\documentclass[11pt,a4paper]{article}
\pdfoutput=1
\usepackage{jheppub}
\usepackage{amsmath,amssymb,amsbsy,amsfonts,latexsym,graphicx,hyperref}
\usepackage{color,array,subfigure}
\usepackage{color,eucal}

\newcommand{\be}{\begin{equation}}
\newcommand{\ee}{\end{equation}}
\newcommand{\bea}{\begin{eqnarray}}
\newcommand{\eea}{\end{eqnarray}}
\newcommand{\beq}{\begin{equation}}
\newcommand{\eeq}{\end{equation}}
\newcommand{\beqa}{\begin{eqnarray}}
\newcommand{\eeqa}{\end{eqnarray}}

\newcommand{\nn}{\nonumber}

\renewcommand{\Im}{{ \rm Im}}

\title{Mott transition with Holographic Spectral function}
\author[a,c]{Yunseok Seo,} 
\author[a]{Geunho Song,}
\author[a,b]{Yong-Hui Qi,}
\author[a]{and Sang-Jin Sin}
\emailAdd{yseo@gist.ac.kr}
\emailAdd{sgh8774@gmail.com}
\emailAdd{qiyh10@gmail.com}
\emailAdd{sangjin.sin@gmail.com}
\affiliation[a]{ Department of Physics, Hanyang University, Seoul 133-791, Korea }
\affiliation[b]{ Research Institute for Natural Science, Hanyang University, Seoul 133-791, Korea }
\affiliation[c]{ {GIST College, Gwangju Institute of Science and Technology, Gwangju 500-712, Korea}}
\abstract{
We show that the Mott transition can be realized  in a holographic model  of a fermion with bulk mass, $m$,  and a dipole interaction of  coupling strength $p$.  
The phase diagram contains gapless, pseudo-gap and gapped phases and the first one can be further divided into four sub-classes.  We  compare the spectral densities of our holographic model  with the Dynamical Mean Field Theory (DMFT)  results for Hubbard model as well as the experimental data of Vanadium Oxide materials. Interestingly, single-site and cluster DMFT results of Hubbard model share some similarities with the holographic model of different parameters, although the spectral functions are quite different due to the asymmetry in the holography part. The theory  can fit the X-ray absorption spectrum (XAS) data quite well, but once the theory parameters are fixed with the former it can fit the photoelectric emission spectrum (PES) data only if we symmetrize the spectral function.  
 }

\subheader{ ~
\today
}

\arxivnumber{}

\begin{document}

\maketitle


\section{Introduction}
The Mott transition, the interaction induced Metal-Insulator transition (MIT)\cite{MOTT:1968aa},  is a challenging subject  and quantitative  understanding  of such phenomena 
is a  necessary virtue  of any successful theory of  for strongly interacting system (SIS).
The physics of Mott transition is usually discussed using the Hubbard model 
$
H = -t (\sum c^{\dagger}_{i \sigma} c_ {j \sigma} + H . c . ) + U  \sum n_{i \uparrow} n_{ i\downarrow} ,
$
which capture the competition between the hopping  and the on-site repulsion. 
However, for 2+1 and higher dimension,   it has not been solved for more than half century. 

Holographic method  has been developed  as a theory of SIS  \cite{Zaanen:2015oix,Hartnoll:2016apf}, and  it is natural  to ask whether  it can describe the Mott transition in terms of  fermion spectral function. 
However, finding the exact gravity dual of a given theory is extremely  difficult  if possible at all. 
Furthermore, Hubbard model is just one  simple model that captures some essence of  the Mott transition, so  finding an exact dual of such model  is not essential either.  
In this situation, instead of  trying to find  the dual of the Hubbard model, it would be more sensible  to  find a holographic model that can  achieve  the same  physics. 

There has been  much effort on  holographic   fermion spectral function starting from \cite{sslee}. 
The marginal non-Fermi liquids in holography is   established \cite{Liu:2009dm,Faulkner:2009wj,Faulkner:2011tm,Faulkner:2013bna}.    
On the one hand,  holographic  gap generation  was discussed in  \cite{Edalati:2010ge,Edalati:2010ww,Vanacore:2014hka,Vanacore:2015poa,Ling:2014bda}  using the  dipole term or Pauli term 
\be 
p \bar{\psi}  F^{\mu\nu}\sigma_{\mu\nu} \psi. \label{pauli}
\ee 
On the other hand,    
the   emergence of free fermion-like point at the bulk mass 1/2 and the nearby  Fermi-liquid-like phase  was found in  \cite{Cubrovic:2009ye,Cubrovic:2010bf,Medvedyeva:2013rpa}. 
Putting these together,  we  might expect that
 Mott transition can be handily described in holography  and Hubbard model
can be replaced  in terms of easily calculable theory. 
However,  the  Hubbard model is  a free  fermion  at $U=0$, while the
holographic theory  is   strongly interacting  even at the absence of the gap generating term, which 
is a generic property of a holographic theory.  Therefore it is not clear how similar and different are the two theories and we need to see the detail  to decide the usefulness of the new theory. 
 
In this paper we study  the phase structure  of the holographic fermion model with the bulk mass term and a gap generating interaction given by  (\ref{pauli}).  We find that rather surprisingly,  for a fixed bulk mass, the model can describe a  transition between the gapless and gapful phases only when we restrict the bulk mass below the critical mass $m_c\simeq 0.35$ \footnote{Notice that the discussion of the 'interaction induced metal insulator transition'  in terms of  conductivity  was already made in ref.'s \cite{Donos:2012js,Andrade:2017ghg} but not in terms of spectral function and also notice that  bosonic Hubbard model in holographic context  was   discussed in \cite{Fujita:2014mqa}.} .
We also find that there is a rather large region of pseudo-gap, a phase where the density of state is depleted near the Fermi sea. 
The phase diagram is  richer than  expected since the gapless phase can be further divided into four subclasses:  the bad metal phases with and without   shoulder peaks and the half-metal phase with a gap between the shoulder peak    and  the central peak  apart from the   Fermi-liquid like phase first discovered in  \cite{Cubrovic:2009ye}. 
The presence of half-metal  is  a surprise at first, but  it can be understood as the 
 proximity effect of the 'free fermion Wall' sitting at $m=1/2$ line  in the phase diagram.  
 
It turns out that there is   a  phenomenologically important difference between the  holographic model discussed here and the Hubbard model: the spectral function of the Hubbard model is symmetric at the half filling, while that in the  present holographic model  is highly asymmetric  for any  non-vanishing charge density without which gap is not generated.  Obviously, the real systems are in between. 

With such differences understood,  it is now meaningful to seek the commonness and similarity between the   different models which realizes Mott transition. 
Since Hubbard model contains essentially one parameter, $U/t$, and the holographic model has two parameters $m$ and $p$, we need to take a path in the phase diagram, which we call embedding.  
We will see that there are two common features: i) transfer of the degree of freedom from the central   to shoulder peaks, ii) smoothness of the transition. 
Our calculation shows that all the transitions are smooth crossover.
Therefore one may wonder why we call a regime as a phase. However, gap and gapless is certainly very different although smoothly connected through the pseudo-gap region, which had been classified as a phase of SIS. We suggest that  such smooth transition with intermediate zone is a general character of SIS. 
For the second feature,  we will see that 
   there are two paths in gap creation: in one path, 
 the central peak  begins to be reduced in height  from  the beginning and the gap is created by such reducing  process. In the middle, pseudo-gap appears in the middle. 
 In the other path,  the central peak remains sharp but its weight and width is getting thinner.  
For the second path or embedding,  the gap creation is done by such thinning process. 
It is really surprising that in the DMFT study of Hubbard model such two different paths for opening the gap were achieved by different approximation scheme. One is called single-site DMFT and the other is cluster DMFT.  It is a bit mysterious how such different features which would be expected from different models could be obtained in the same model in both cases: in DMFT as different approximation schemes
and  in the holographic model as different parameter regimes. 

Finally we  tested  our model with the  experiment using the Vanadium oxide data. 
It  turned out that the X-ray absorption Spectrum (XAS) data  can be fit by our theory but the photoelectric emission (PES)  data can not be unless spectral functions are symmetrized by hand.   
 %
%
  
  \section{Spectral function}
  \subsection{Setup and review}
 We start from the fermion action in the   dual spacetime with non-minimal dipole interaction, 
\begin{align}
S_{D} =\int d^4 x \sqrt{-g} \bar{\psi}\left( \Gamma^M {\cal D}_M -m - i  p\, \Gamma^{MN} F_{MN}  \right)\psi  + S_{\text{bd}},  \label{Eq:S_Dirac}
\end{align}
where the subscript $D$ denotes the Dirac fermion and the covariant derivative is
\begin{align}
{\cal D}_M = \partial_{M} +\frac{1}{4} \omega_{abM} \Gamma^{ab} -i q A_M.
\end{align}
For fermions, the equation of motions are first order and 
we can not fix the values of all the component at the boundary, which make it necessary to 
introduce `Gibbons-Hawking term' 
$S_{\text{bd}}$   to guarantee the equation of motion which
defined as
\beqa
-i S_{\text{bd}} =  \pm \frac{1}{2} \int d^dx \sqrt{h} \bar\psi \psi  = \pm \frac{1}{2} \int d^dx \sqrt{h} (  \bar\psi_- \psi_+ + \bar\psi_+ \psi_- ), \label{Eq:S_bd}
\eeqa
where $h= - g g^{rr}$, $\psi_\pm$ are the spin-up and down components of the bulk spinors.  
The sign is to be chosen such that, when we fix the value of $\psi_+$ at the boundary,  $\delta S_{bd}$ cancel the  terms
 including $\delta { \psi}_- $ that comes from the total derivative of $\delta S_D$. Similar story is true when we fix $\psi_-$. The former defines the standard quantization and the latter does the alternative quantization.
 The background solution  we will use is Reisner-Nordstrom black hole in asymptotic $AdS_4$ spacetime,
\begin{align}
ds^2 &= -\frac{r^2f(r)}{L^2} dt^2 +\frac{L^2}{r^2 f(r)} dr^2 +\frac{r^2}{L^2}d\vec{x}^2  \cr
f(r) &= 1+ \frac{Q^2}{r^4}-\frac{M}{r^3},~~~~~A=\mu \left(1-\frac{r_0}{r}\right),  \label{AdS4}
\end{align} 
where    $L$ is AdS radius, $r_0$ is the radius of the black hole   and   
$
Q= r_0 \,\mu, M= r_0(r_0^2 +\mu^2). 
$
The temperature of the boundary theory is given by
$
T={f'(r_0)}/{4 \pi} 
$
and it can be solved for $r_0$ to give 
$r_0 = ( 2\pi T +\sqrt{(2 \pi T)^2 +3 \mu^2})/3$.
  
Following \cite{Liu:2009dm}, we now introduce  $\phi_{\pm}$ by 
\beqa
\psi_\pm = {(-gg^{rr})}^{-\frac{1}{4}} 
 \phi_\pm, \quad \phi_\pm = 
\left(
\begin{array}{ccc}
  y_\pm    \\
  z_\pm    \\
\end{array}
\right), \label{Eq:psi_pm}
\eeqa 
after Fourier transformation. 
Then  the equations of motion become \cite{Liu:2009dm},
\begin{align}
\sqrt{\frac{g_{xx}}{g_{rr}}} z_{+}'(r) -m L \, \sqrt{g_{xx}} z_{+}(r) +i [u(r)+k- p \sqrt{g_{xx}} A_{t}'(r)] y_{-}(r) &= 0\cr
\sqrt{\frac{g_{xx}}{g_{rr}}} y_{-}'(r) + m L \, \sqrt{g_{xx}} y_{-}(r) +i [u(r)-k+ p \sqrt{g_{xx}} A_{t}'(r)]z_{+}(r) &= 0, \label{eom01}
\end{align}
where
$u(r) = \sqrt{\frac{g_{xx}}{-g_{tt}}} ( \omega + q A_t (r) ).  \label{ur}
$
Here, the  momentum is along $x$ direction. 
The corresponding equations for $ y_{+}, z_{-}$ are obtained from the above by $(A_t, \omega) \to (-A_t,-\omega)$.

At the boundary region($r \rightarrow \infty$), the geometry becomes $AdS_4$ and the equations of motion (\ref{eom01}) have analytic solution as
\beqa
z_+ &= A_1 \chi_{1} (r) + B_1 \chi_2 (r) ,   \quad 
y_- &= C_1 \chi_{3} (r) + D_1 \chi_{4}(r)  ,\\
y_+ &= A_2 \chi_{1} (r) + B_2 \chi_2 (r)  ,  \quad
z_- &= C_2 \chi_{3} (r) + D_2 \chi_{4}(r)  ,
\eeqa
where
\beqa
\chi_{1} (r) &=& r^m \, _0F_1\left(\frac{1}{2}-m;-\frac{W}{4 r^2}\right), \quad \chi_{2} (r) =r^{-m-1} \, _0F_1\left(m+\frac{3}{2};-\frac{W}{4 r^2}\right), \nn\\
\chi_{3} (r) &=&r^{m-1} \, _0F_1\left(\frac{3}{2}-m;-\frac{W}{4 r^2}\right), \quad \chi_{4} (r) =r^{-m} \, _0F_1\left(m+\frac{1}{2};-\frac{W}{4 r^2}\right), \label{chis}
\eeqa
with  $W = (\omega + q\, \mu)^2- k^2$. 
The asymptotic behaviors of (\ref{chis}) are manifest if we notice $_0F_1 \to 1$ in $r\to \infty$. 
The equation of motion produces the relations of coefficients: 
\begin{align} \label{coeff:CA-BD-1}
 C_1 = \frac{i A_1 \left( k -(\omega + q\,\mu) \right)}{2m -1},~~~~~B_1=\frac{i D_1 \left( k +(\omega + q\,\mu) \right)}{2m +1}, \\
\label{coeff:CA-BD-2}
 C_2 = \frac{i A_2 \left( k +(\omega + q\,\mu) \right)}{2m -1},~~~~~B_2=\frac{i D_2 \left( k -(\omega + q\,\mu) \right)}{2m +1}. 
\end{align}
Here, we 
made an abbreviation for $mL$ with $m$,  which we will restore at  the end. 
 
The boundary term in Eq.(\ref{Eq:S_bd}) becomes
\beqa
-i S_{\text{bd}} =&  y_- z_+ - y_+ z_-  = (A_1 D_1 - A_2 D_2)  + \sum_\pm E_{\pm} r^{\pm 2m-1}+E_{2}r^{-2}, \label{ADAD}
\eeqa
using the asymptotic behavior of wave functions $\chi_{i}$. 
Here, $E_{\pm}$ and $E_{2}$ are   functions of the coefficients of $\chi_{i}$. 
A few remarks are in order. 
First, for   $m>1/2$, the second term($E_{\pm}$) dominates but it can be cancelled by counter terms \cite{Ammon:2010pg}, which do not contribute any finite terms to the effective action. 
Second, 
in the standard quantization where we fix $\psi_+$ at the boundary, $A$'s are the source terms. 
While in the alternative quantizaton where we fix $\psi_-$ at the boundary,  
$D_i$ is taken to be the source. 
 Therefore, if  variables with index 1 and those with index 2 are separable, the retarded Green's function in standard quantization, 
 is given by  
\begin{align} 
{\cal G}  = \text{diag} \bigg( i \frac{D_1}{A_1}, -i \frac{D_2}{A_2} \bigg) \equiv \text{diag} (G_+^R, G_-^R ), \quad m>0,  
\end{align}
while that in alternative quantization is given by
\begin{align} 
{\tilde {\cal G} } = \text{diag} \bigg( i \frac{A_1}{D_1}, -i \frac{A_2}{D_2} \bigg) &\equiv  \text{diag} (\tilde{G}_+^R, \tilde{G}_-^R ) \cr
&=-\text{diag} (1/G_{+}^{R}, 1/G_{-}^{R} ),\quad m>0.
\end{align}
Since  $G_{R}$ for  $m<0$ case, can be also obtained by $G_{R}\to -1/G_{R}$, 
   ${\tilde G_R}$, the Green function for the alternative quantization for $m>0$, is the same as  that for   $-m$ in the standard quantization: 
   \begin{align}
  {\tilde G_{\pm}^R(\omega,k;m)}= -1/G_{\pm}^R(\omega,k; m) =G_{\mp}^R (\omega,k;-m).
\end{align}  
  Introducing the $\xi_\pm $ by 
  \be 
  \xi_+ =i \frac {y_-}{z_+}, \hbox{ and } \quad \xi_- = - i\frac{ z_-}{y_+},
  \ee 
  the   equations of motion Eqs.(\ref{eom01})  can be recast into two independent   equations for $\xi_\pm$: 
 \beqa
\sqrt\frac{g_{xx}}{g_{rr}} \xi_\pm^\prime = -2m \sqrt{g_{xx}}\xi_\pm  + [  u(r)   +p \sqrt{g_{xx}} A_{t}'(r) \mp k]  + [ u(r)-p \sqrt{g_{xx}} A_{t}'(r) \pm k ]\xi_\pm^2. \label{eomxipm}
\eeqa
and the Green functions    for  $m<1/2$  can be written  as  
\beqa
G^R_\pm (\omega,k) = \lim_{r\to\infty} r^{2m}\xi_\pm(r,\omega,k). \label{greenless}
\eeqa 
Notice that   two components of the Green function are not independent: 
$ 
G^R_- (\omega,k)=G^R_+(\omega,-k).
$
The spectral function is defined as the  imaginary part of the Green function. There are two of them 
$\Im[G^R_+]$ and $\Im[G^R_-]$ and we can define the spectral function for each of them:
\be  
A_\pm(\omega,k)=\Im[G^R_\pm(\omega,k)].
\ee

There is an issue on the finiteness of the spectrum: it was pointed out \cite{Gursoy:2011gz} that 
  the high frequency behavior of the spectral function  diverges like $\omega^{2m}$ so that the sum of the degree of freedom  over   frequency is infinite if $m$ is positive.  
  Therefore we need to take the negative bulk mass only in the standard quantization. For the ease of discussion  we want to maintain the positivity of the mass  which can be done simply by going to the alternative quantization.  
  Even in this case spectral function is $\sim \omega^{-2m}$ which does not decay fast enough to guarantee the finite integration. The sum rule can be still an issue and we do not treat this problem here.  
  Summarizing,  we work in  alternative quantization with positive mass and 
we treat the fermion as a probe and do not consider its back reaction.

\subsection{Non-relativistic system in terms of  relativistic formulation}
Now how do we define the physical spectral function that can be compared with experimental data? 
For relativistic system like Dirac or Weyl semi-metal,  it is natural to define it as the traced object which is the sum of the two: 
$\Im[G^R_+ +G^R_-]$.
One expect that the chemical potential is small so that the Fermi level is near the Dirac point. 
In fact we have a few  experiences that such Dirac material with small Fermi sea can be well described by the RN black hole physics \cite{Seo:2016vks,Seo:2017oyh,Seo:2017yux}.  

However, if we want to describe a non-relativistic system, the problem is more subtle as we describe below.  
Notice that in the presence of chemical potential, the dispersion relation near the Fermi sea is linear so that we expect that dynamical aspect  are not much different between relativistic and non-relativistic cases. However, the relativistic spectrum is a double of non-relativistic case in the former handle the 
negative and positive frequency at equal footing. 
Therefore when  compare  the two,  half of the relativistic  spectrum should be   projected out.   
Indeed,  the relativistic case  has a serious  problem in describing real  non-relativistic system because  it has unphysical spectrum  far below fermi surface. 
This can be seen by considering weakly interacting system with chemical potential. 
Notice that the fermi level is defined by 
$G_-^{-1}\sim \omega+\mu - k -\Sigma =0 $ with $\omega=0$. 
For a weakly interacting system, the self energy $\Sigma\simeq0$, then $k_c$, the momentum at which we consider the spectral function can be taken as $k_c=\mu:=k_F$. Then the spectrum or  the pole of the $G_+^{-1}$ would be at  $\omega= -2\mu$, because $G_+^{-1}\sim \omega+\mu + k_c -\Sigma =0 $. 
For large chemical potential  the spectral function has high peak deep under the Fermi sea, which is certainly unphysical.
\begin{figure}[ht!]
\centering
      \subfigure[Relativistic Spectrum(RS)]
       {\includegraphics[width=5cm]{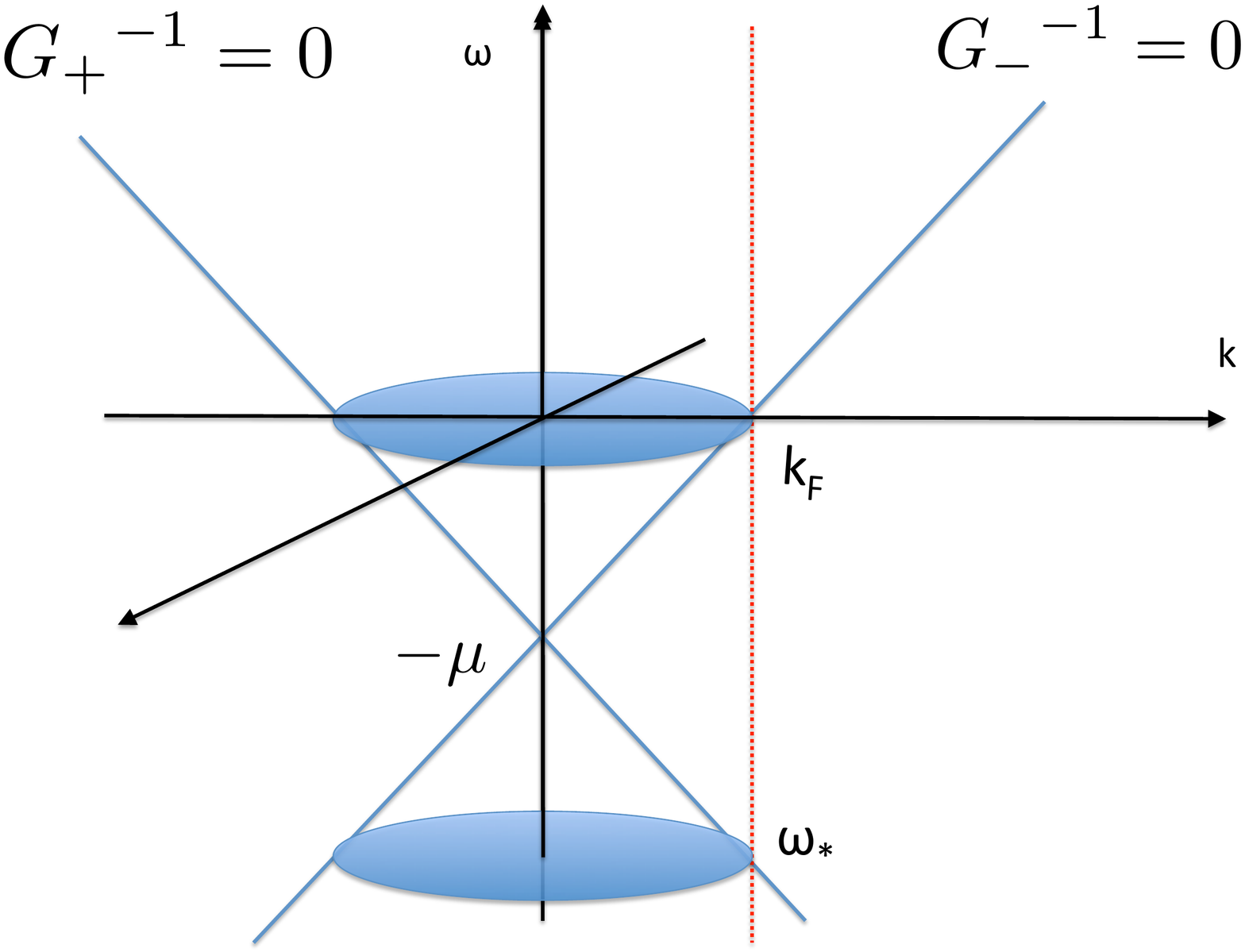}  }  
     \subfigure[NRS approximated by RS]
       {\includegraphics[width=5cm]{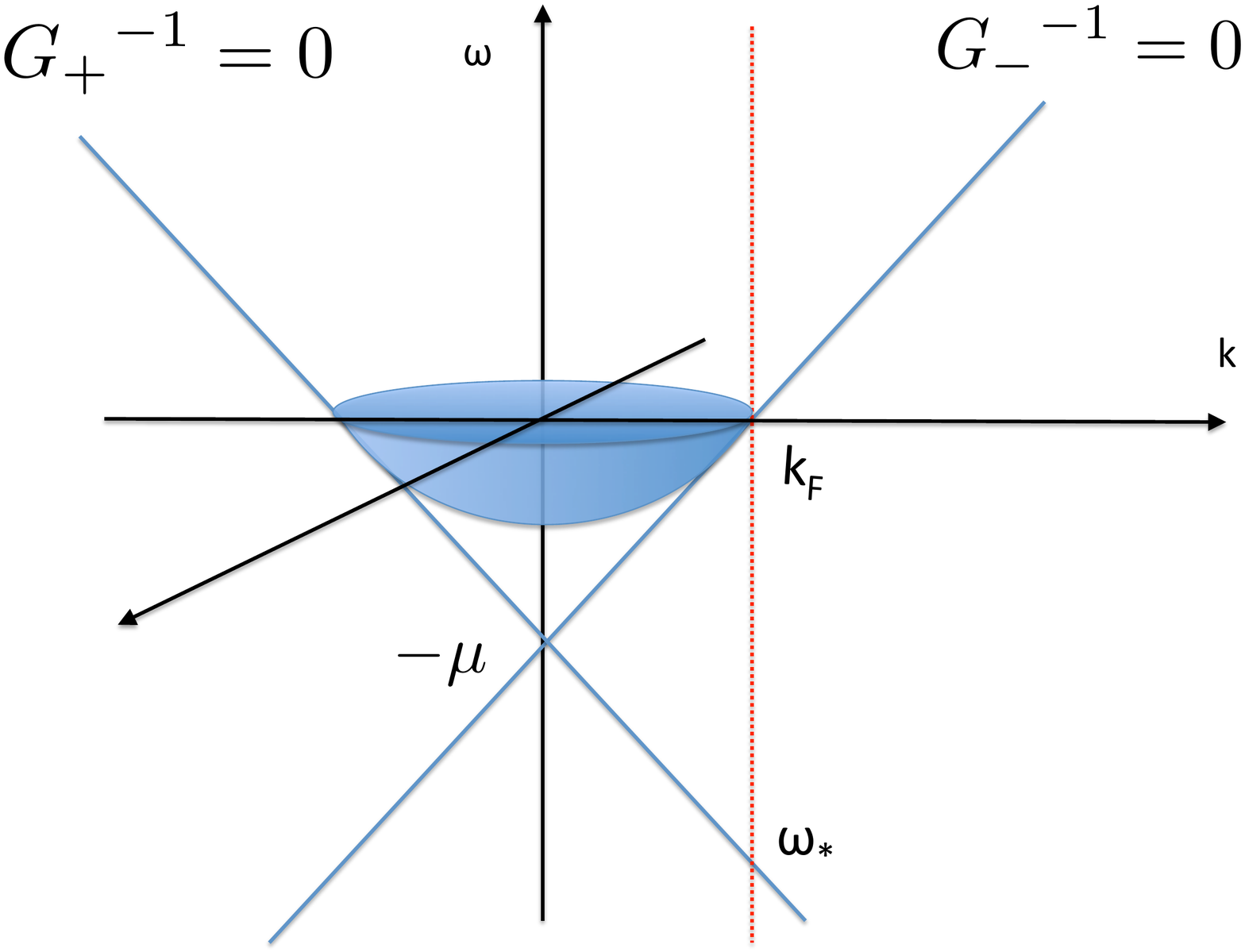}  }  
     \subfigure[Half filling]
       {\includegraphics[width=5cm]{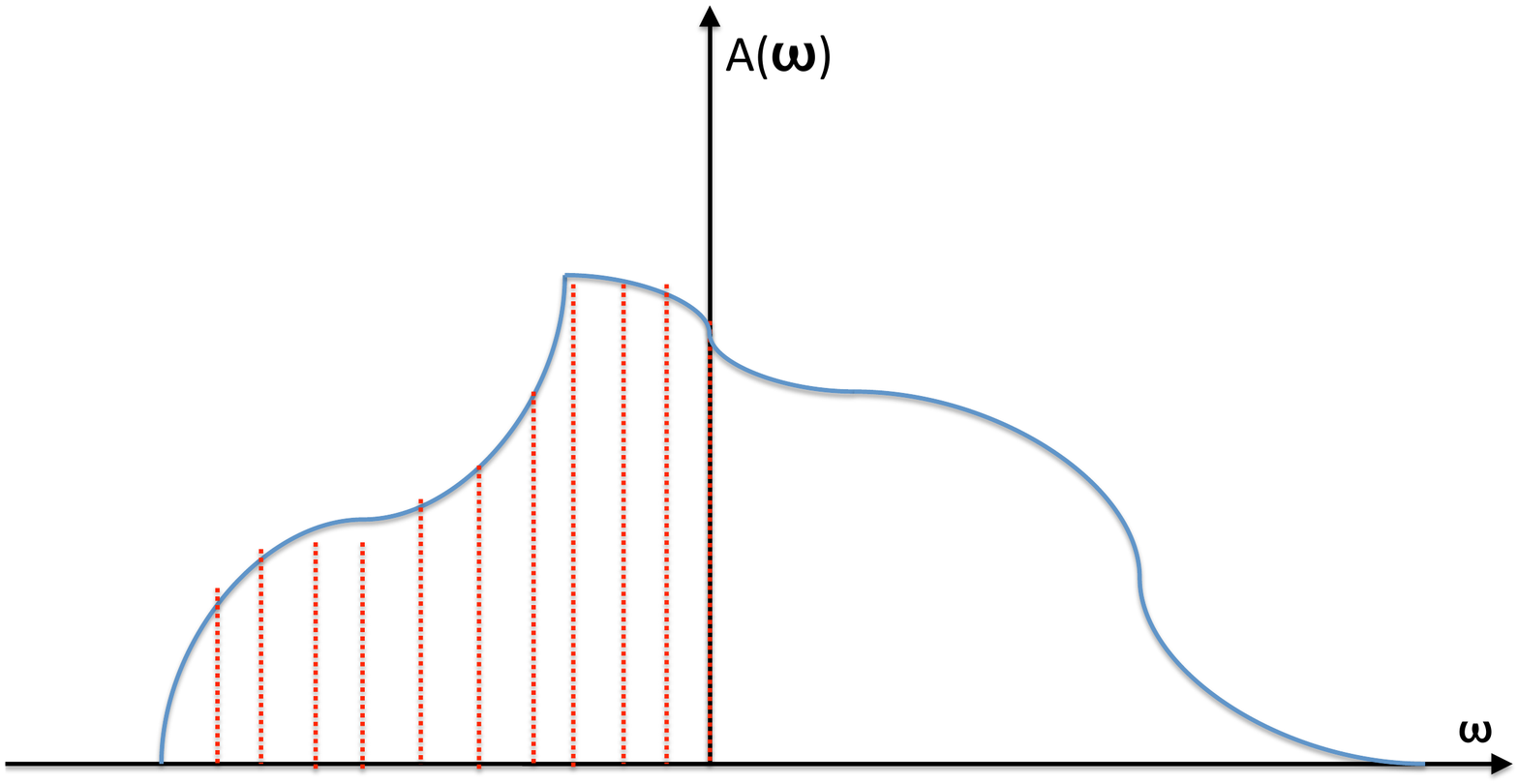}  }  
            \caption{ (a)spectrum in relativistic formulation. (b) Unphysical spectrum appear in the $G_+$ in hole sector  when we approximate the non-relativistic spectrum (NRS) in terms of the relativistic theory near Fermi level.    To elliminate such pole we define the spectral function by eliminating the spectrum deep under the Fermi sea: $\omega<-\mu$.          
(c) Definition of half filling. 
             }   \label{fig:relativistic}
\end{figure}
See  the figure \ref{fig:relativistic},   where the spectral function as function of $\omega$   along the $k_c=k_F$ line has two peaks: one at the fermi surface and the other at the 
 $\omega=\omega_*=-2\mu$, deep in the fermi sea.   
Such feature  is attributed to the relativistic formulation of the fermion  and it continues to exist  for strongly interacting system. 
When we describe a non-relativistic  system in terms of relativistic fermions,  
we have to exclude such spectrum. 
Therefore we identify our spectral function as $\Im[G^R_-]$ rather than the traced one. 
 \be
 A(\omega,k_c)=A_-=\Im[G^R_-(\omega,k_c)]
 \ee
 This is fine for practical purpose where we set $k$ non zero only along x-direction 
 but  it is not a rigorous definition, because  
 in more than 1+1 dimension $G_+$ and $G_-$ are not separable, as the lightcone structure in figure 1 suggests.  
  More proper way to state it is to discard the spectrum under $\omega<-\mu$ if $\mu>>T$.  
 
 \subsection{definition of the half filling  in the absence of the lattice}
 In most practical calculation of holography, one does not encode the presence of the 
 lattice. However, the starting point of Mott transition is the having half filled band which is certainly 
 band conductor, which should become insulator under the growth of the coupling strength.
 Therefore we need to ask what is the definition of the   the half filling  when we do not  encode the lattice. 
 One may think this is not a serous question in holography since generic phase of fermion in the absence of extra interaction is a bad metal and we have at least one  a gap generating  interaction. 
 However, it is rather confusing in understanding the definition of the doping which  is necessary to  calculate physical quantity in terms of the doping rate.  
 Here we propose that the system is half filled if the Fermi level  $\omega=0$ divide the density of state 
 by half, namely if the   area under the density of state (DOS)  graph is divided into two equal areas by $\omega=0$.   
 See figure \ref{fig:relativistic}(c). If we denote the general density by  $Q$ and the half filling density by  $Q_0$,  then the doping rate $x$ is expressed by 
$
 x={(Q-Q_0)}/{Q_0}.
 $ 

\section{The phases of holographic fermion} 
We study the phases diagram of the model given by Eq. (\ref{Eq:S_Dirac})  
as function of   $p$ and  $m$.
There are two  self evident  phases: gapless,  gapped phases.  The pseudo-gap appears as an interpolating zone of these two phases. 
The phase diagram is  richer than  expected, because   the gapless phase can be subdivided into four subclasses: Fermi liquid  like (FL), bad metal(BM), bad metal prime(BM') and half-metal(hM) phases. 

The most typical phase of the gapless phase is bad metal phase. 
Since  $A(\omega,k)\sim \omega^{-2m}$,    it is not well localized near Fermi surface (FS). 
The peak at the free fermion point $(p,m)=(0,1/2)$ is singularly  sharp. For the continuity of the 
phase diagram, we have to install a Fermi liquid like phase near that point and   
we take the phase boundary value to be $m\sim 0.35$.   However this phase  can  not be a real Fermi liquid in two aspects: i) the central peak is not really localized  and 
decays too slowly as mentioned above. 
ii) the width of the central peak does not follow $\Gamma \sim T^2$ law. Instead,  we find that they follow $\Gamma\sim T$ law for small chemical potential $\mu/T\ll 1$. 
The reason we call it Fermi liquid like phase is due to the sharp linear dispersion relation $\omega +\mu=\pm k$.  Also for large chemical potential 
$\mu/T\gg1$,  we find  the half width of the central peak $\Gamma\sim T^{\alpha}$ with $\alpha\simeq 2$ so that it resembles the true  Fermi liquid.  See the figure \ref{fig:GFF}. 

\begin{figure}[ht!]
\centering
    \subfigure[Fermi liquid like (FL)  ]
    {\includegraphics[width=43mm]{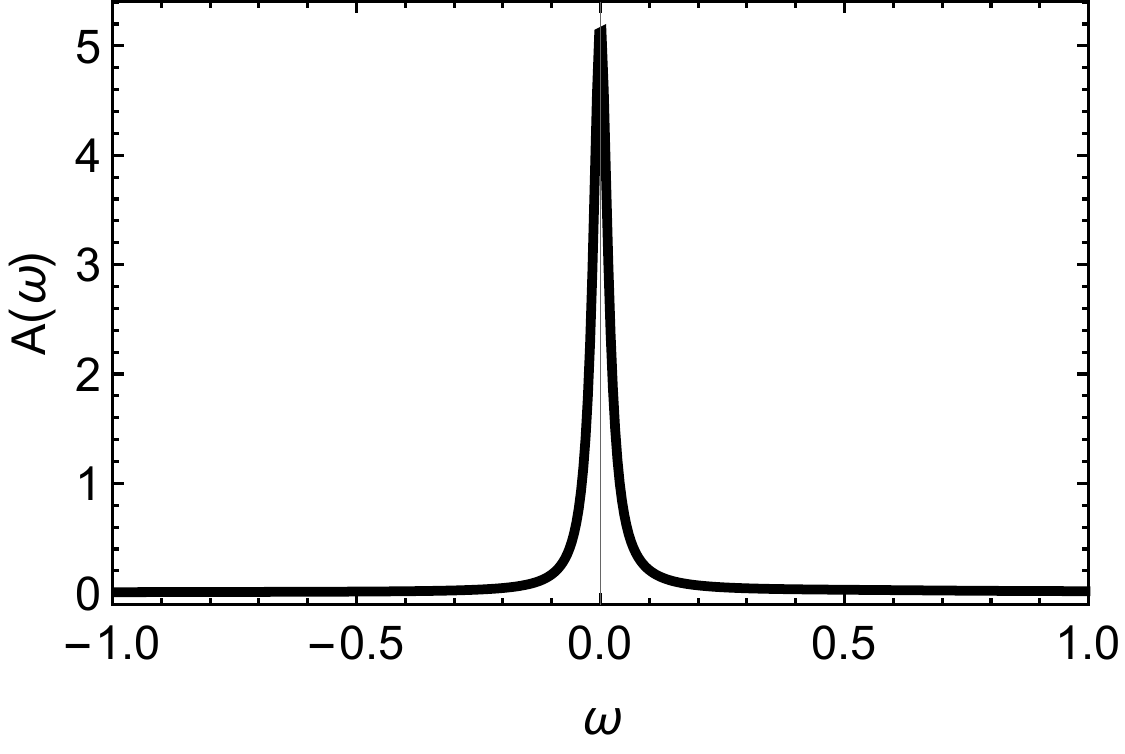}  }
        \subfigure[ bad metal prime (BM')]
   {\includegraphics[width=4.5cm]{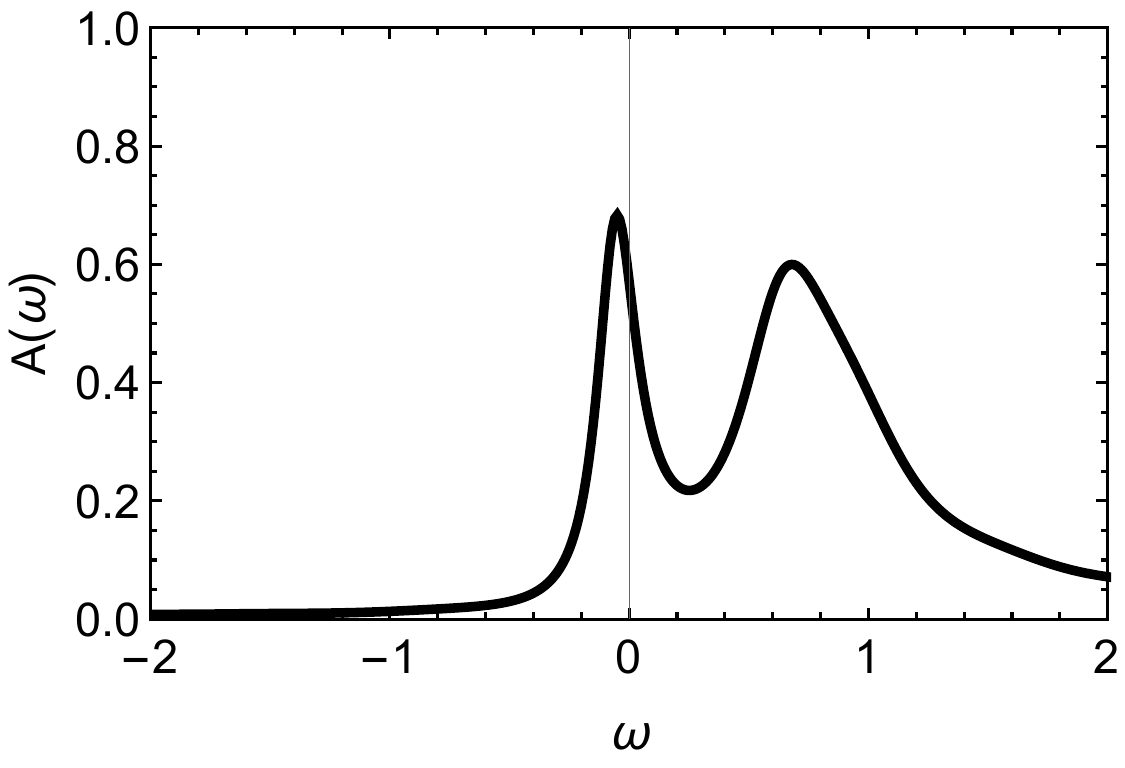}  }
       \subfigure[half-metal  (hM)]
   {\includegraphics[width=4.3cm]{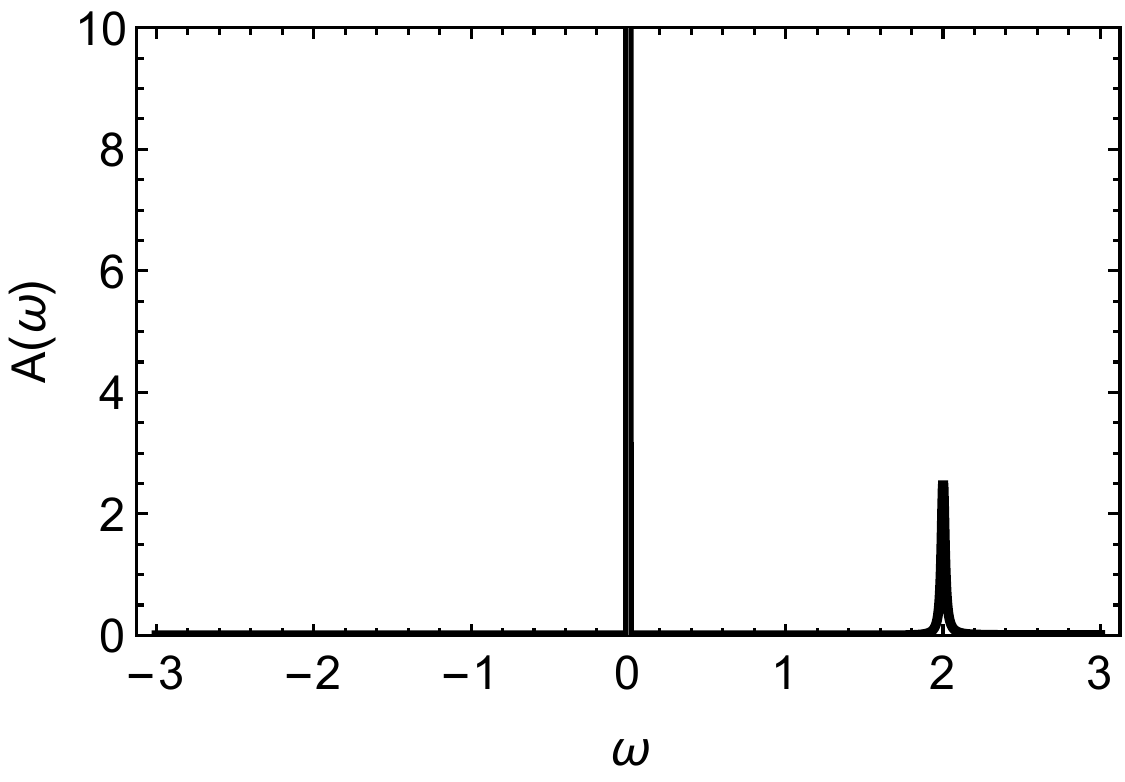}  }
         \subfigure[ bad metal (BM)]
   {\includegraphics[width=4.3cm]{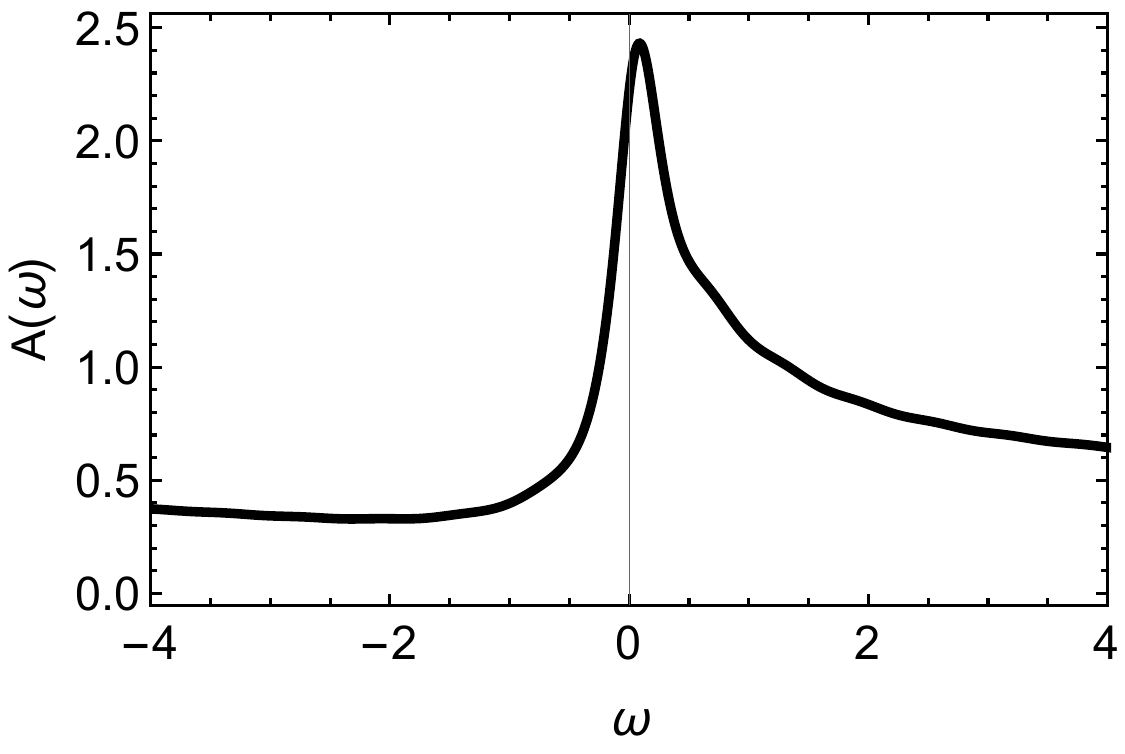}  }
       \subfigure[pseudogap (PG)]
   {\includegraphics[width=43mm]{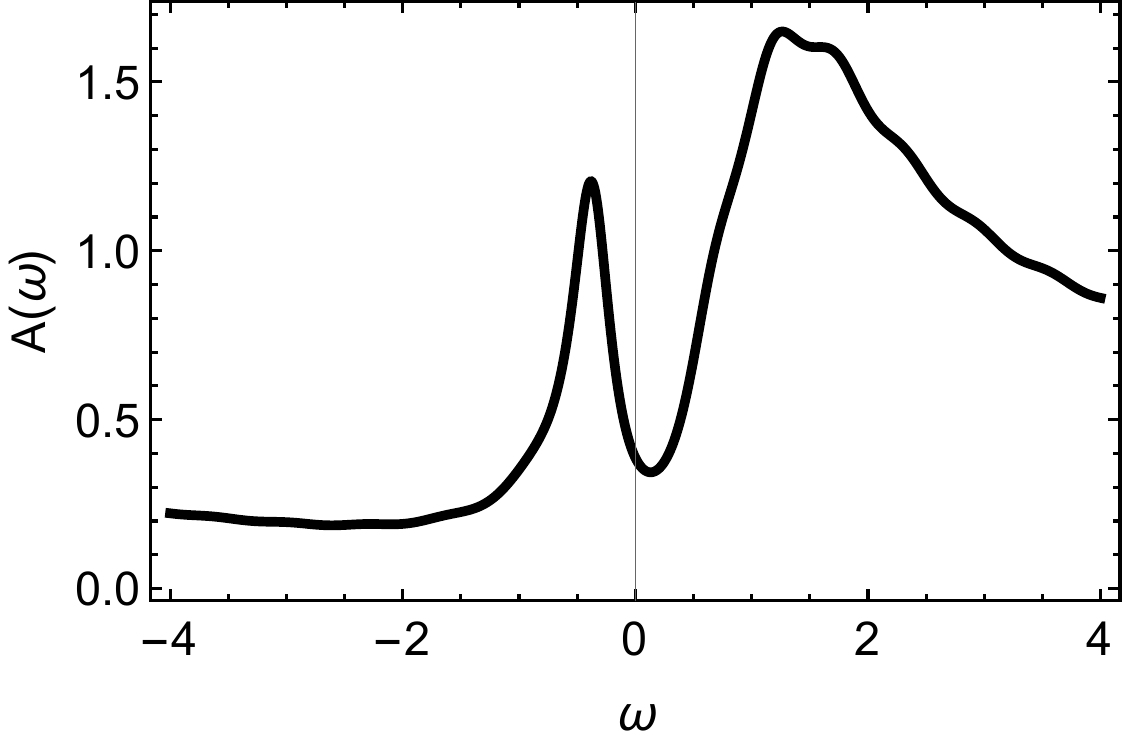}  }
            \subfigure[ gapped (G)]
   {\includegraphics[width=45mm]{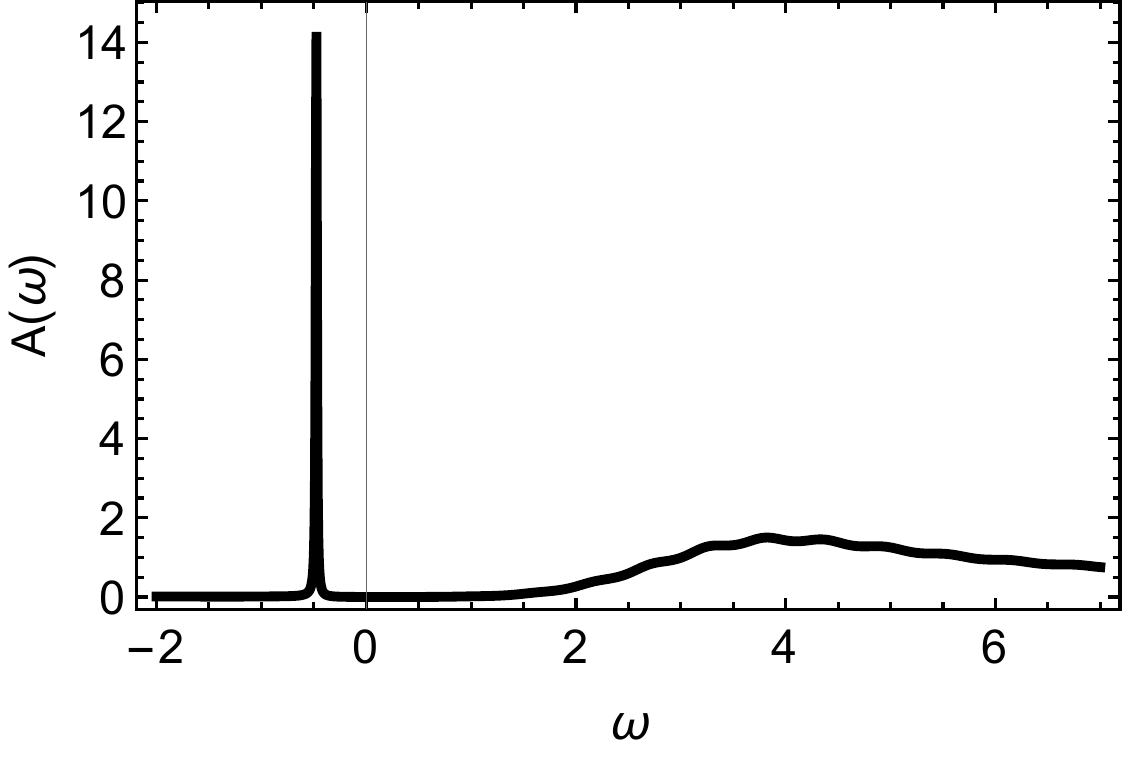}  }
        \caption{(a)-(f):Typical fermion phases.  
        (a) FL with (p=0.5, m=0.45,$k_c=1.65$), (b) BM' with shoulder  (p=2, m=0.4, $k_c=2.30$), (c) hM (p=6, m=0.45, $k_c=2.48$), (d) BM without shoulder (p=0.5, m=0.1, $k_c=1.20$), (e) PG (p=2,m=0.1,$k_c=2.89$): notice the position of the $\omega=0$ compared with the BM' phase in (b),  (f) Gapped phase  (p=6, m=0.15, $k_c=5.68$)
     } \label{spec06}
\end{figure}

The reason of free fermion point at $(p,m)=(0,1/2)$ is well understood \cite{Cubrovic:2009ye,Faulkner:2009wj}.
$\psi$ is the dual of the operator with dimension $\Delta=d/2-m$ which is dimension of free fermion when $m = 1/2$ so that  the fermion with $m=1/2$  in  alternate quantization is dual to the free fermion.  
 Along the $m=1/2$ line  in the phase diagram, the  spectral function is also sharp although the spectrum can be more diversified.  We call that line as   `free fermion wall'.  
The bad metal prime is the bad metal with  shoulder peak(s). See Figure \ref{spec06}(d) and (b). 
The  half metal is the bad metal prime with a gap between the central peak and the shoulder peak. 
See Figure \ref{spec06}(c). 
   
 Since the transitions are smooth everywhere,  
one may wonder whether we can classify the phases. 
However,  it would be more strange if we say that SIS has just one phase since even the gap and gapless phases are smoothly connected.  
With this understanding,   the phase boundary naturally depends on the choice of the criterion:  we choose the onset of pseudo-gap by $R=0.9$ where $R$  is the ratio of the spectral function at the central dent, $A(\omega=0,k_c)$, to that at the Hubbard peak. Here $k_c=k_F$ if $k_F$ exists, otherwise it is the momentum at which one of the dispersion curve branch  just touches  the fermi level $\omega=0$ which happens at $m=0.35$. See figure \ref{fig:spec02}(b). 
 For the gapped phase we choose $R=0.01$.
 
 \begin{figure}[ht!]
\centering
      \subfigure[Phase diagram]
       {\includegraphics[width=7cm]{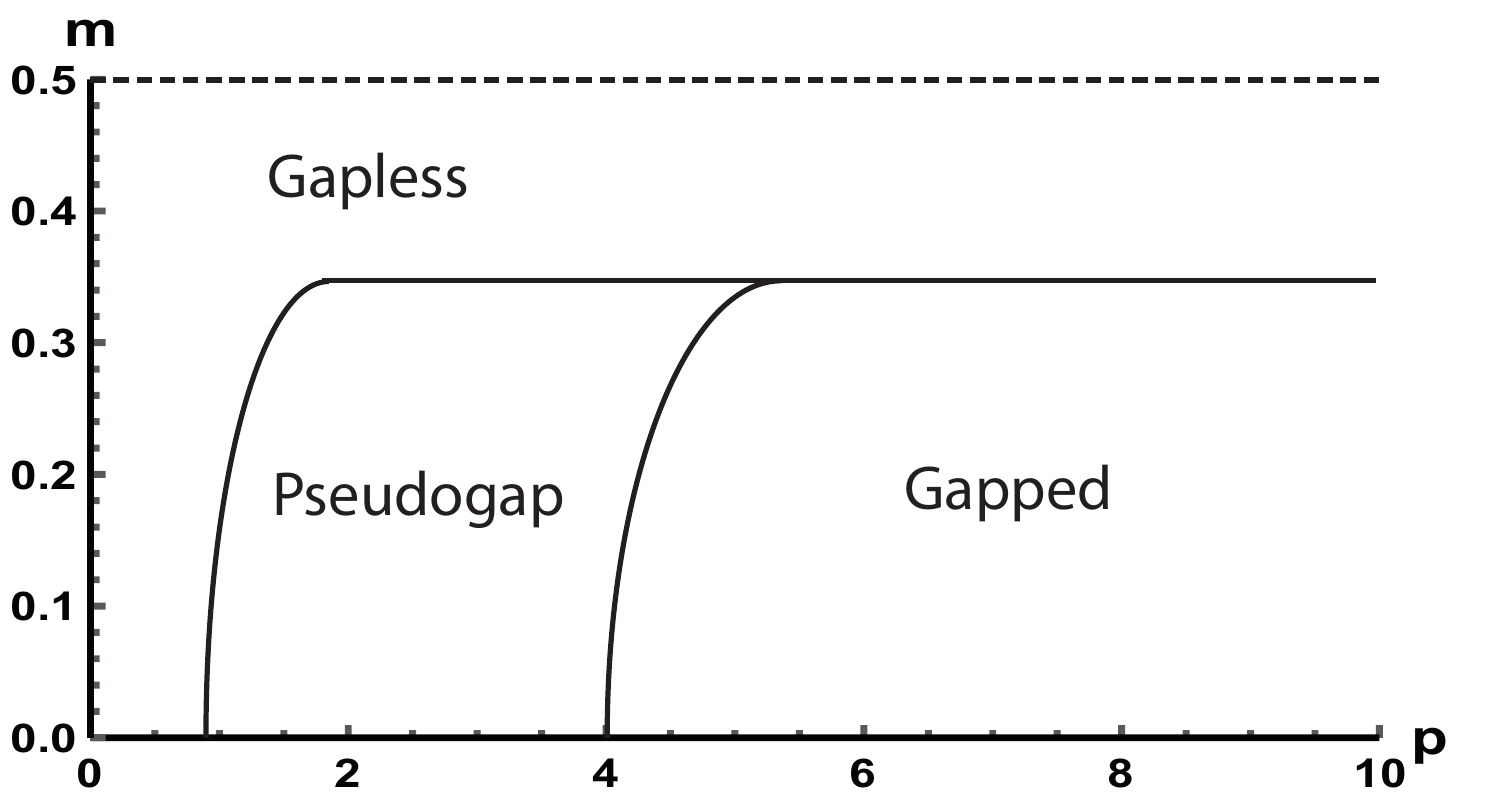}  }  
     \subfigure[Substructure in gapless phase ]
       {\includegraphics[width=7cm]{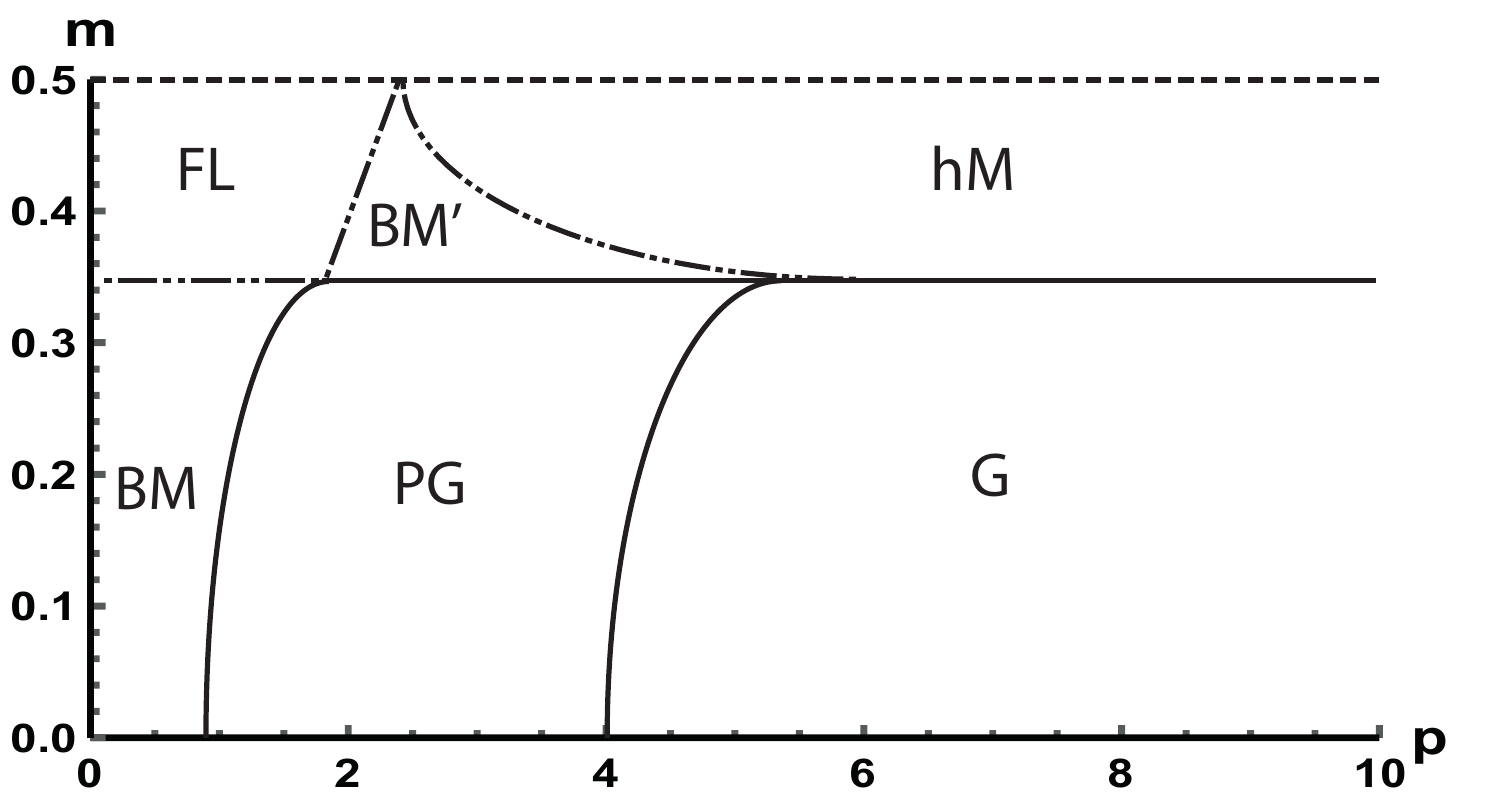}  }  
         \caption{          
          Phase diagram in $(p,m)$ space. (a) Three phases which are gapless, psuedo-gap and Gapped phases, appear and all  transitions are smooth crossover. The gapless phase can be further divided into four   subregions:  Fermi liquid like (FL), bad metals (BM), bad metal prime (BM'), half metal (hM).}   \label{fig:PhD}
\end{figure}

%
The result of the detailed study of phases are summarized by the phase diagram given in
 Figure \ref{fig:PhD}. The dashed line along $m=0.5$  represents the free fermion wall, the FL phase is located at the upper-left corner and gapped phase is at the  lower-right corner. All other phases are in between the two and  can be understood as   proximity and competition of  the two phases. 
Notice also that the phase diagram is divided by  the line of  $m \simeq 0.35$:   the lower half region is where   gap-generation phenomena is observed  as we expected from the presence of the dipole term. 
However,  in the upper  region,    a  new metallic phase appears instead of gapped one. 
We call it   half-metal phase,  because    significant fraction of density of state is depleted from the  quasi-particle  peak near the Fermi level and  moved  to the shoulder region.    
The emergence of this new metallic phase in the strongly coupled system was unexpected.
To understand its appearance, we    study the effect of the dipole term on the spectral density near $m=0.5$. See figure \ref{fig:GFF}. 
\begin{figure}[ht!]
\centering
       \subfigure[ $m=0.5, p=0$]
   {\includegraphics[width=4cm]{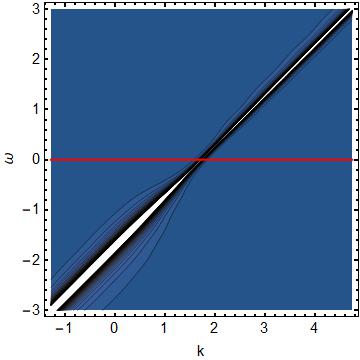}  }
       \subfigure[$m=0.5, p=5$ ]
   {\includegraphics[width=4cm]{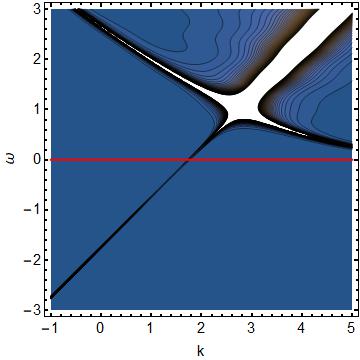}  }
        \subfigure[ $m=0.47, p=5$]
  {\includegraphics[width=4cm]{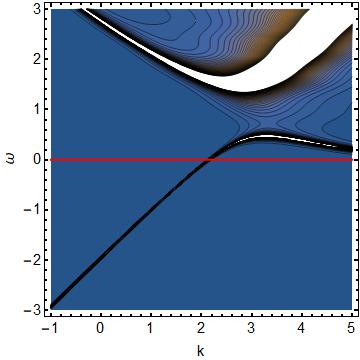}  }
 \caption{Contour plot of spectral density for $m\sim 0.5$.  Red   line indicates the Fermi level.  
    (a) At $p=0$, the degree of freedom  follows the dispersion relation $\omega+\mu=k$,   (b) At $p=5$,  
    new branch of dispersion curve appears. 
    (c)  Lowering $m$ from 0.5, the  spectral curves  are reconnected  to avoid the 'level crossing'. 
     } 
     \label{fig:GFF}
\end{figure}

It turns out that the peak along the dispersion curve $\omega+\mu=k$    exists  along $m=0.5$ line although more and more degrees of freedom are  depleted from the central peak and moved to the shoulder as we increase $p$ \footnote{
Previously, the free fermion phase near the $m=0.5$ was noticed by Leiden group \cite{Cubrovic:2009ye} at $p=0$ and here we  study  it in the presence of the gap generating dipole term. }. 
We call the line  $m=0.5$ `the free fermion wall'.  
One  important  effect of the   dipole term is  {\it the creation of new   band}. See figure \ref{fig:GFF}(b).  
As $p$ increases, it push down the new  band below the Fermi level so that a gap is created 
and  will be increased. 
The third  effect of increasing $p$  is  to make the  new  band sharper which means it keep transferring the spectral density from the  central peak to the shoulder peak. 
This is  similar to the effect of   $U$ in the DMFT calculation of Hubbard model.    
 Now lowering $m$ from 0.5, the  spectral curves  are reconnected  to avoid the 'level crossing'.  Consequently, the density profile moves from Figure \ref{fig:GFF}(b) to (c). 
  
We can now understand the    
the  role of mass  in creating the half-metal phase:  increasing the $m$ pushes up the   new  band created by  $p$ so that  the band  can cross the Fermi level.  See figure \ref{fig:spec02}. 
This effect  competes with  that of increasing $p$, but the effect of mass is  stronger.  
For $m>0.35$ the new   band  always crosses  Fermi sea and 
this is the mechanism of the hM phase. 
  Notice that the new band touch the Fermi  level  at $m=0.35$    for all $p$.

\begin{figure}[ht!]
\centering
    \subfigure[$m=0.25$ ]
   {\includegraphics[width=4cm]{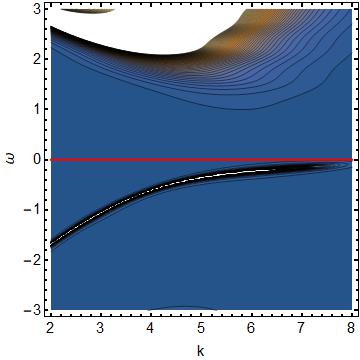}  }
       \subfigure[ $m=0.35$]
   {\includegraphics[width=4cm]{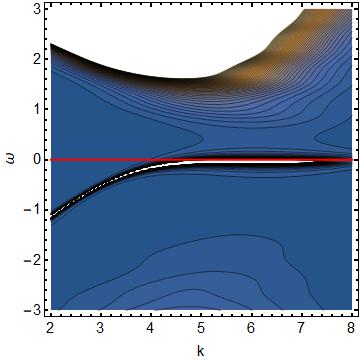}  }
         \subfigure[ $m=0.45$]
   {\includegraphics[width=4cm]{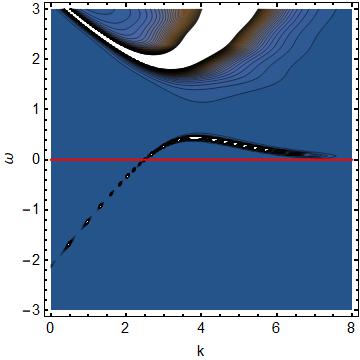}  }
    \caption{$m$-evolution at $p=6$  shows the  origin of the  the half-metalic phase: increasing the $m$ pushes up the new  band, making the Fermi sea. Large $p$ is responsible for the sharpness of new band. } 
    \label{fig:spec02}
\end{figure}
 
 For small mass  $m < 0.35$, the dipole interaction leads to metal-insulator transition as $p$ increases.  For $p>4$ and $m=0$,  gap is dynamically generated  as it was shown by Phillips et.al \cite{Edalati:2010ge}. For larger mass, the gap generation requests slightly higher values of dipole strength.    The pseudo-gap is nothing but the intermediate zone of this smooth transition,  namely $0.8<p<4$ for $m=0$. Notice that in the phase diagram Figure \ref{fig:PhD}, there is a rather large territory of PG.   

 Similarly, for    $m > 0.35$, the dipole term drives  BM' -hM transition because     the new band   always  crosses the Fermi level.  This is why   strong  dipole interaction leads to   the half metal rather then a Mott insulator for  in this regime.   As $p$ increases the new band is narrowed and sharpened    but  it never disappears even at very large $p$.  In the appendix, we study the evolution in m for fixed $p$ and evolution in $p$ for fixed $m$ in more detail. 

\vskip .5cm
%

\begin{figure}[ht!]
\centering
   \subfigure[ ]
   {\includegraphics[width=60mm]{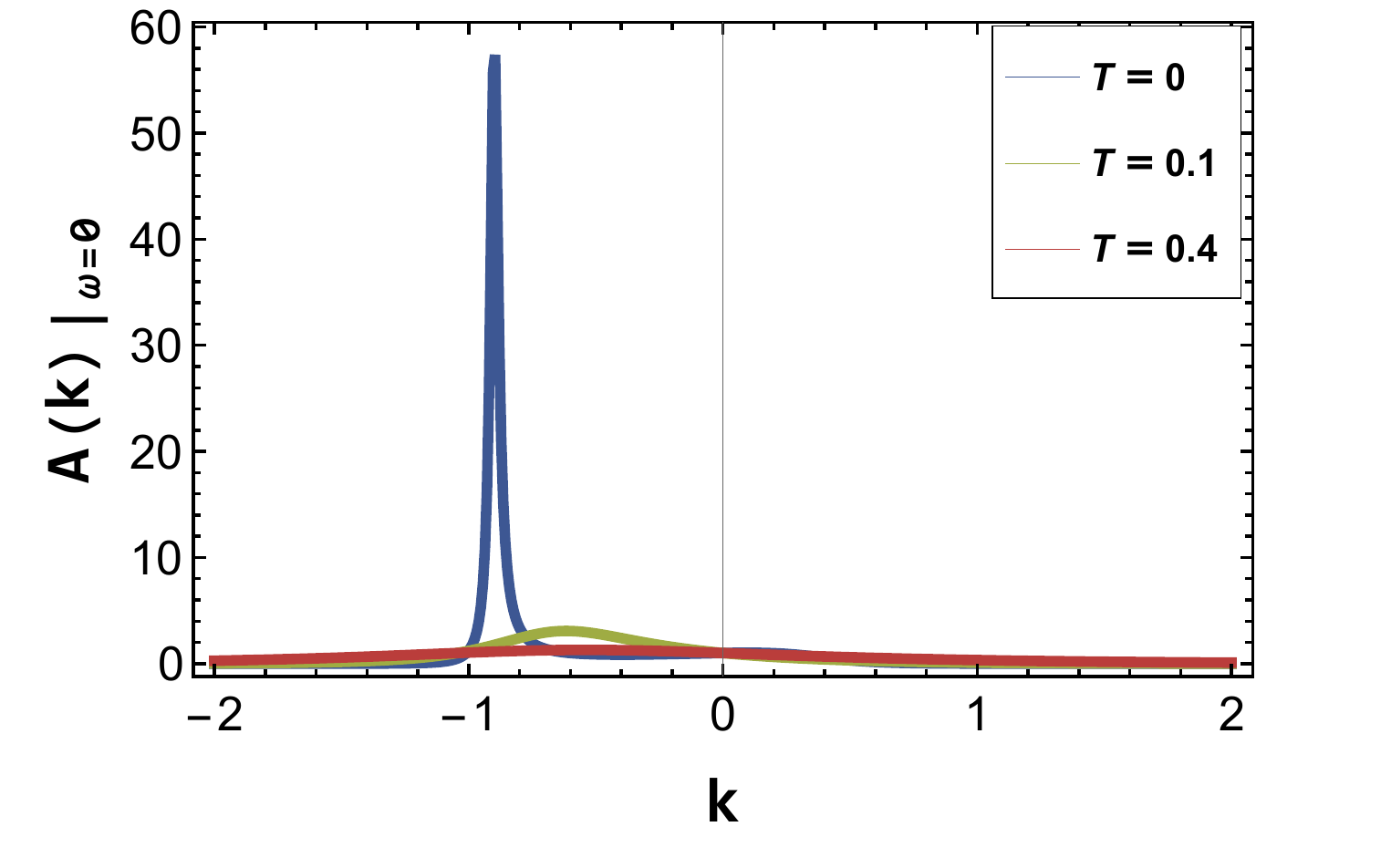} }
       \subfigure[]
   {\includegraphics[width=60mm]{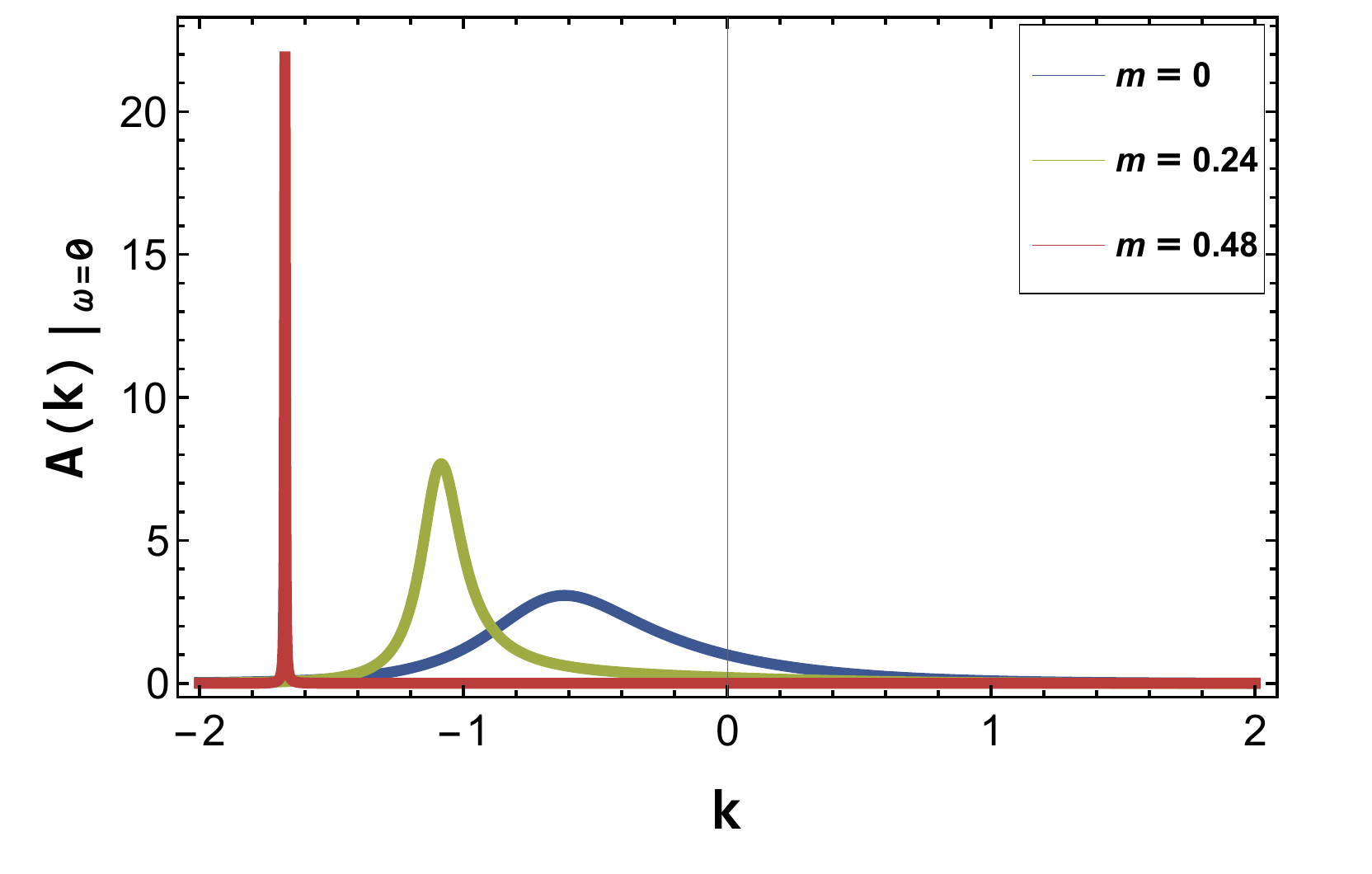}  } 
           \caption{Role of mass in stabilizing the Fermi surface (FS): (a)Fast decay of FS  for nonzero Temperature $T=0,  0.1, 0.4$   at $m=p=0$. 
           (b) FS stabilized by bulk mass: $m=0,0.24,0.48$   at $T=0.1$ and $p=0$.        
           } 
           \label{mrole} 
\end{figure}

The  figure \ref{mrole}(a) shows that  in the absence of the bulk mass,   peak in  the spectral density 
$k$-plot  goes away very rapidly as soon as we turn on  temperature. That is, quasi-particles are  fragile at finite temperature, which is the character of non-Fermi liquids. 
 On the other hand, if we turn on the mass, the Fermi surface peak  becomes  sharper as we can see 
 in the Figure   \ref{mrole}(b). 
 %
%
We can see the role of mass is the stabilizer of the quasi-particle nature in holographic matter.
As  $m\to 1/2$, such `quasi-particle stabilizing  tendency' increases singularly so that 
the system is a Fermi liquid  like whatever   is the strength of dipole term.  
In fact, the spectral function shows that the dispersion curve is straight line   as if  it is a free fermion.   For applications to the  realistic material, having  such a dial to make the system Fermi liquid like in a limit is  very useful   because  in the real experiments, one tunes the coupling by applying pressure or doping rate.  
In the presence of the dipole coupling whose role is to introduce a gap which break the conformal symmetry  dynamically, there is no guarantee that  the 'free fermion'  continues to exists.  
Our observation is that, nevertheless,  such free fermion nature at $m=1/2$ persists   in the presence of the dipole interaction regardless of its strength. We call it {\it free fermion wall}  in $m$-$p$ phase diagram.
We found that if $m>0.35$,   metalic phase exist always as a consequence.  
  
  \section{Comparison with other studies}
  \subsection{Comparison with Hubbard Model}
\begin{figure}[ht!]
\centering
  	{\includegraphics[width=7cm]{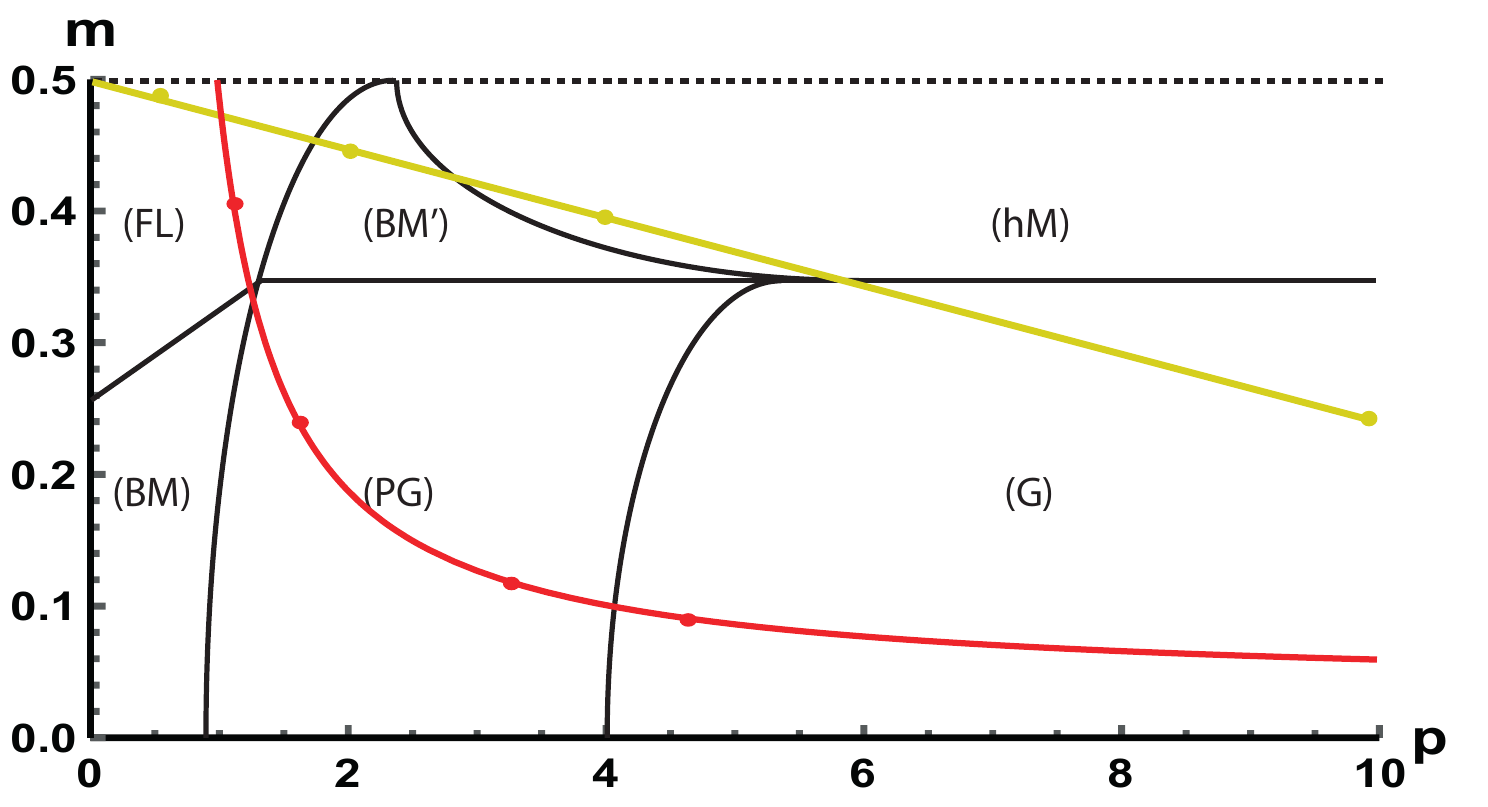}}
	\caption{Phase diagram with two class of embeddings:  Linear embedding mimics single site DMFT and the Hyperbolic one resembles two site DMFT. $\alpha,\beta$ in eq. (4.2) and (4.4) are $1/40$ and $1/20$ respectively. 
	}
         \label{fig:emb3}
\end{figure}

Usually a theoretical study for Mott transition has been done using the  Hubbard model. Therefore 
it is inevitable to compare our result with the previous study of Hubbard model. 
We emphasize that our model is not the holographic dual of Hubbard model but a replacement of it for the Mott transition purpose. In fact the model studied here has a notable difference from the Hubbard model. First, $U=0$ in the latter is free fermion while $p=0$ in the former is  not  unless the bulk mass is fine tuned. Second, at the half filling, the Hubbard model has symmetric spectral function while the holographic theory does not.    Third but most vividly, we have two parameters $m,p$ while the Hubbard model has only one, $U/t$. 
In order to compare our model to the result of Hubbard model, we have to restrict to an 1 dimensional 
subspace of the phase space, which we call embedding:  namely,   we associate a line in the parameter space $(p,m)$ in the holographic model that gives qualitatively the same spectral density. 
Any line connecting the  free fermion point and the gapped phase can realize a Mott transition and  define   an embedding.  Here we consider  two simplest choices:  the first one start from the free fermion point and reach at the gapped phase following  a straight line
\be
m+\alpha p=1/2,  \quad p=U/t.
\ee  
This defines a  linear  embedding given in Figure \ref{fig:emb3}.    
The second    starts from a point in the  'free fermion wall', the line $m=1/2$,  and rapidly goes down to small bulk mass regime  and and turn there to reach the gapped regime following a curve
\be
p=\beta(\frac{1}{2m}+1), \hbox{ with } \beta=constant.
\ee
 We call it  'hyperbolic  embedding' and it is the red line in Figure \ref{fig:emb3}. 
 
\begin{figure}[ht!]
\centering
  \subfigure[single site DMFT result]
   {\includegraphics[width=4.5cm]{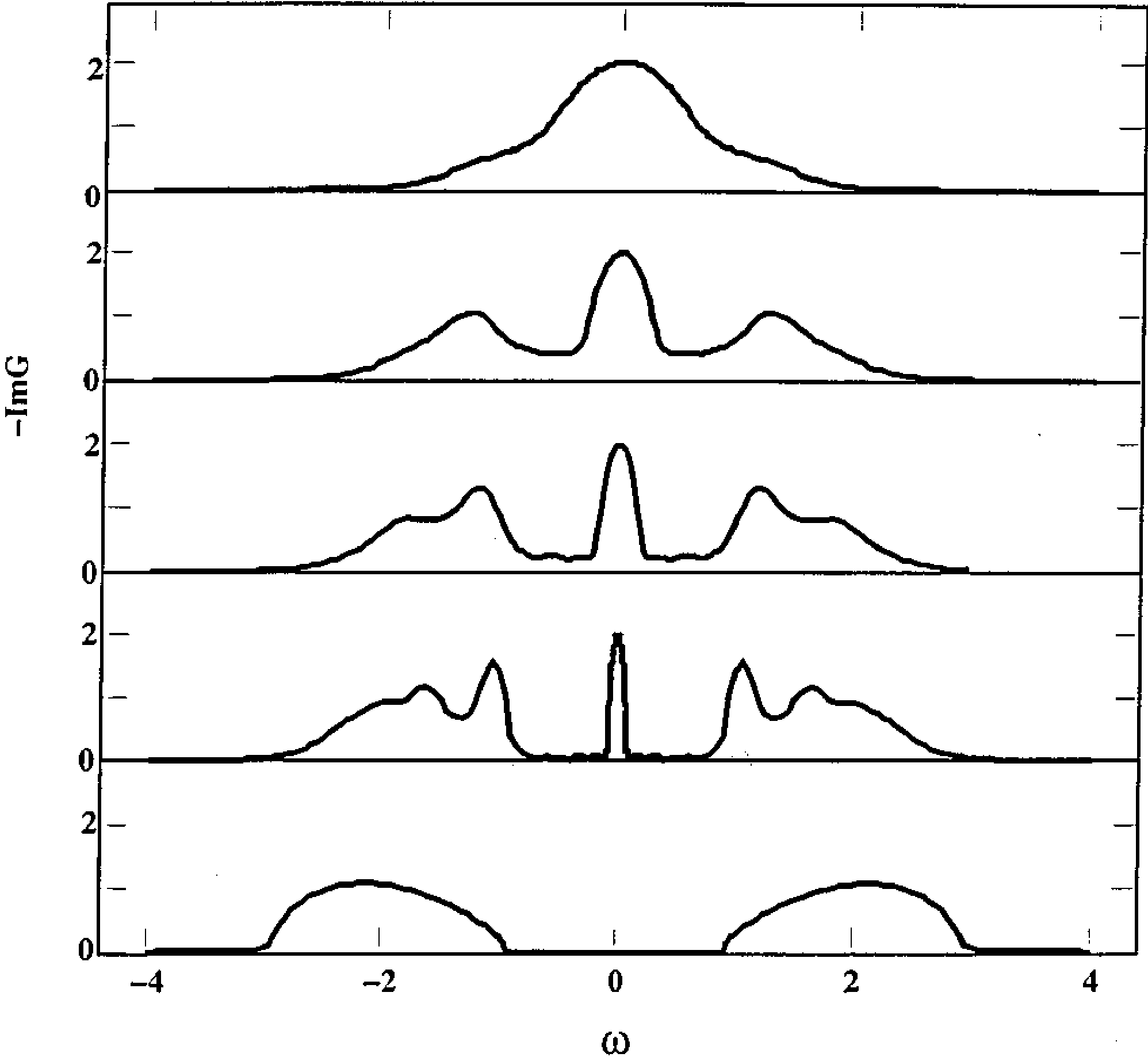}  }
   \hskip 1cm
    \subfigure[Linear embedding]
   {\includegraphics[width=5cm]{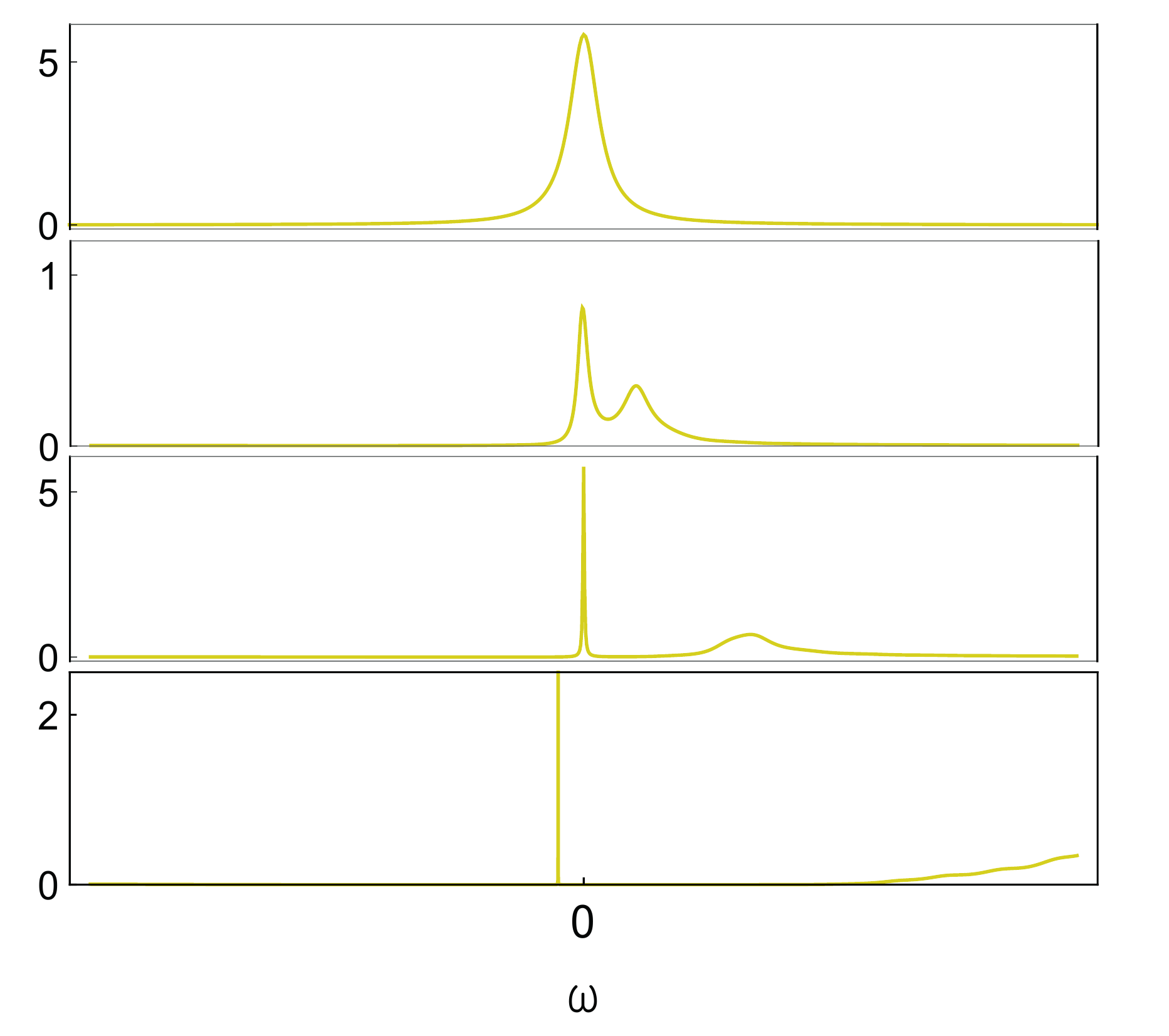}  }
      \caption{Comparision of DMFT results and Linear embedding. 
 (a) Hubbard model  in single site DMFT.  The figure is from  \cite{RevModPhys.68.13}. 
(b)  Evolution of of Linear embedding.  Transfer of DOS, persistence of central peaks till the gap formation are common with DMFT. But due to the heavy asymmetry in spectral function, 
the  three peak structure of single site DMFT is not manifest here.   $\alpha\simeq 1/40$. } 
     \label{fig:motive}  
\end{figure}

There is no deep reason why we choose these two but it turns out to be of some interests. 
The linear  embedding  i) has central and shoulder peak structure ii) does not have pseudo-gap,  and iii) as we increase the coupling $p$  (iii) the degree of freedom transfer  from central peak to  shoulder peak so that  
the central peak becomes thinner and thinner  until gap is created.
These three are the characteristic property of the single site DMFT result for Hubbard model\cite{RevModPhys.68.13}. 
 However, the holographic model has too much asymmetry in spectral function so that the 
and  three peak structure which is one of the property of single site DMFT is not manifest since one of the shoulder is too weak.  
See Figure \ref{fig:motive} (a) and (b). 

\begin{figure}[ht!]
\centering
   \hskip -1cm
   \hskip.5cm
   \subfigure[2-site DMFT result ]
   {\includegraphics[width=4.5cm]{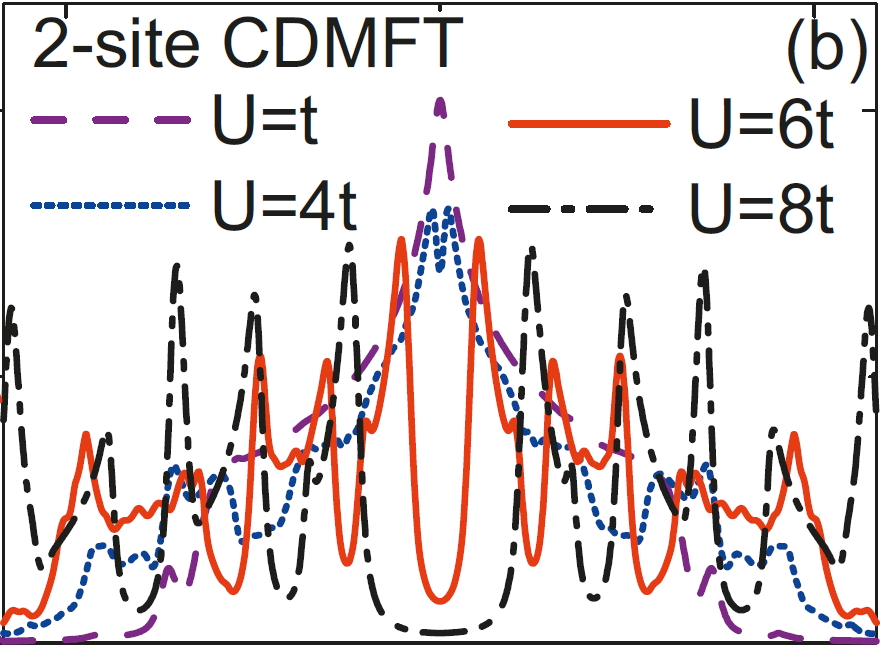}  }
   \hskip.5cm
     \subfigure[a hyperbolic embedding]
   {\includegraphics[width=7cm]{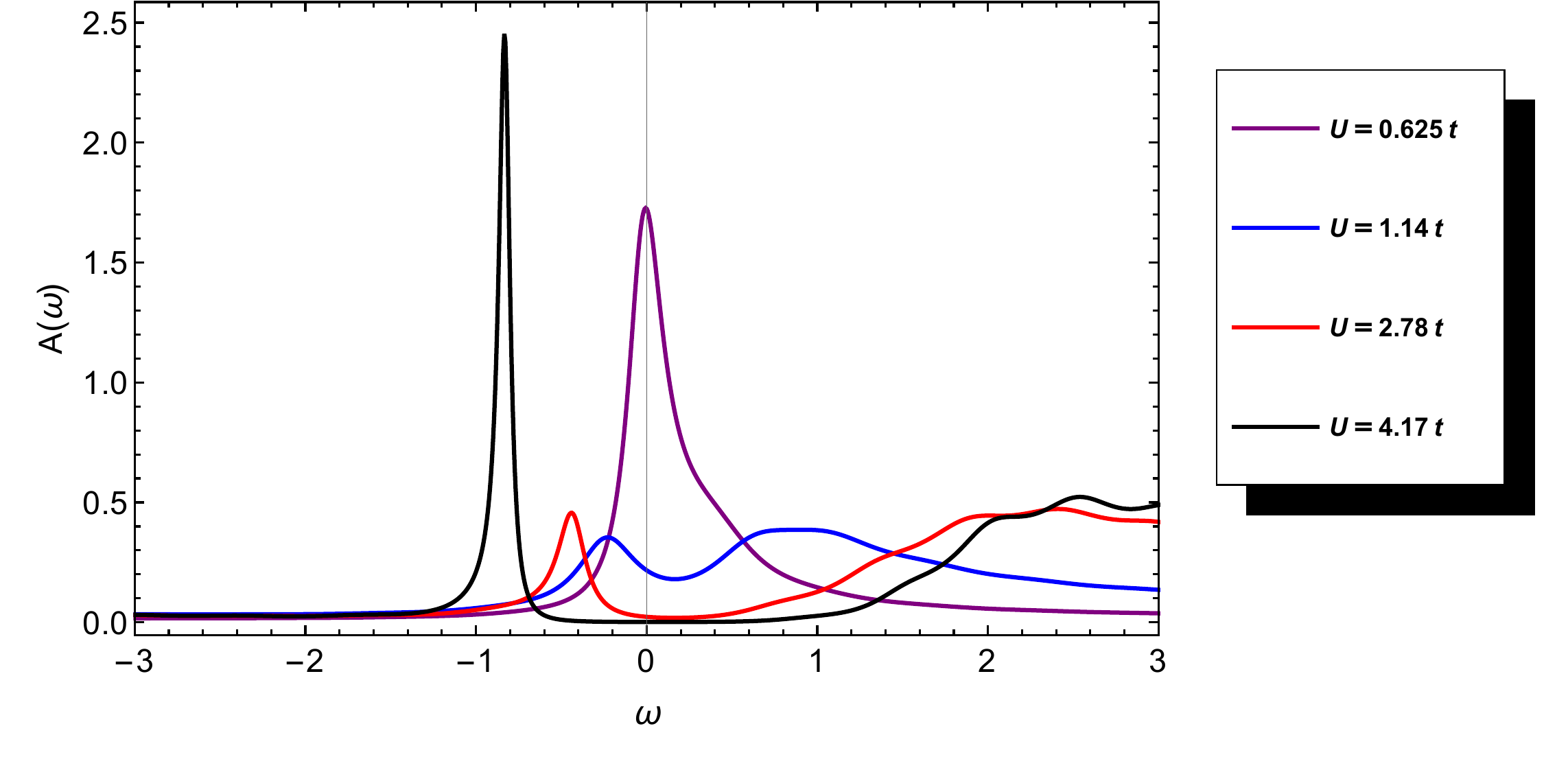}  }    
      \caption{Comparision of cluster DMFT results:  
(a) Hubbard model   in 2-site DMFT.   The figure is from    \cite{zhang2007pseudogap}.
(b)  Evolution along the hyperbolic embedding in holographic model.    Appearance  of the pseudo-gap resembles  cluster DMFT, but  here again, due to the large asymmetry, the details look different. 
} 
     \label{fig:cluster}  
\end{figure}

 The  second embedding has pseudo-gap without central-peaks which is analogous to cluster DMFT results \cite{zhang2007pseudogap}. 
 The `transfer' of the  spectral density  from the quasi-particle peak to the Hubbard side peaks are common to both embeddings. 
The apparent similarity between the two should be coming due 
to sharing the Mott transition.  However,  due to the large asymmetry again, the detail is different.
The comparison of 2-site DMFT  and its holographic analogue, the hyperbolic embedding,  is given in figure \ref{fig:cluster} with  $\beta=0.5$.

It is rather surprising that two different approximation scheme of DMFT for the same model
 behave as if they are different models and yet the holography can accommodate them with
  different parameter regime. 
 Since we are comparing different theories, the similarity is overall one and they are different in detail. 
 The difference in gap creation is worthwhile to emphasize. 
 The single site DMFT   \cite{RevModPhys.68.13}    shows that the gap creation is `sudden' since 
 it is created  with a finite size.  
On the other hand, linear embedding  opens gap  starting with zero size.    

 \subsection{Comparing with experiment}
\begin{figure}[ht!]
\centering
       {\includegraphics[width=5cm]{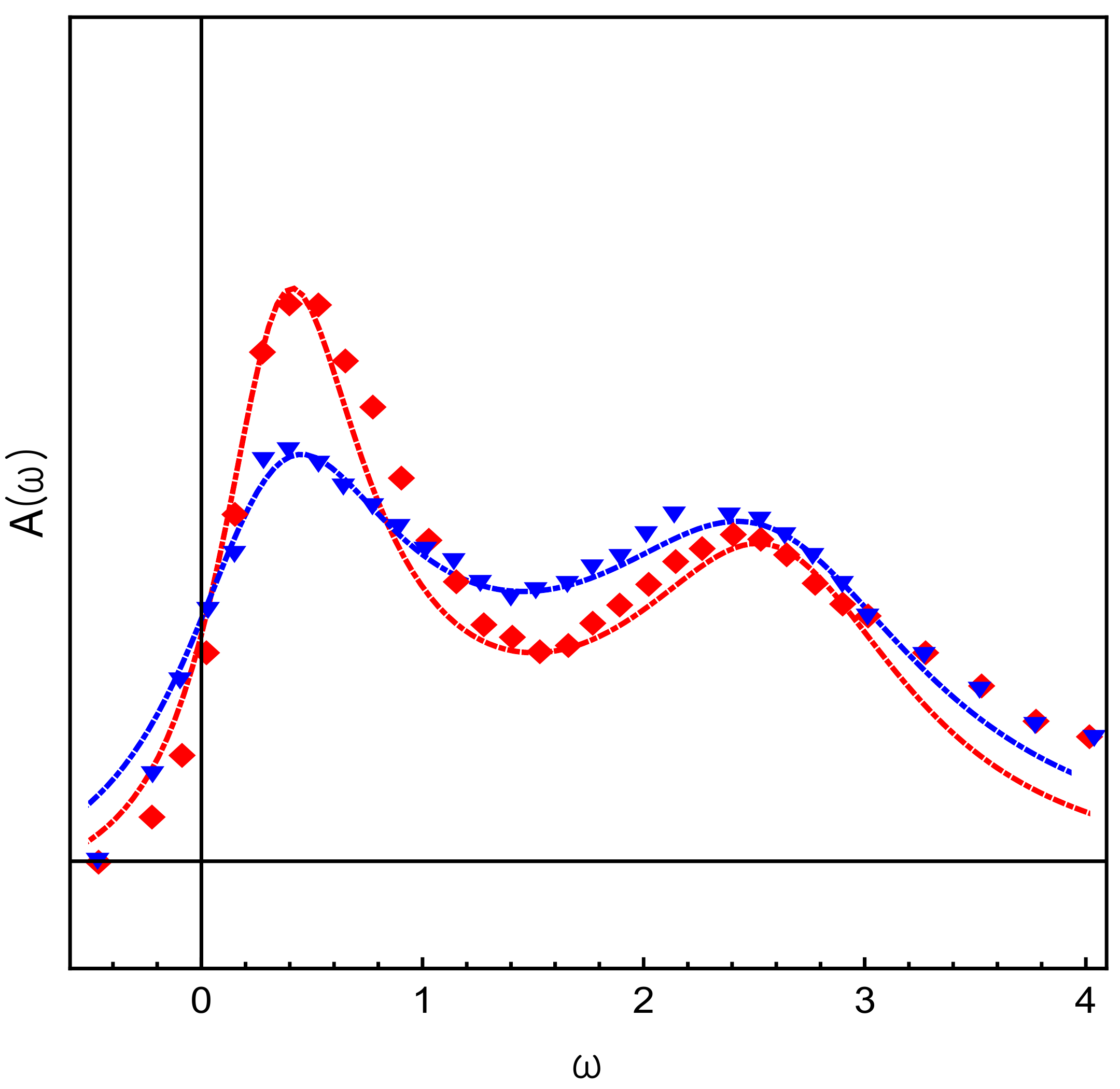}  }  
         \caption{Experimenal data  vs holographic theory:  
           XAS data ;  color red is for SrVO$_{3}$ and (color blue) is for  CaVO$_{3}$. The data for  SrVO$_{3}$ is from   \cite{sekiyama2002genuine},  and  that for  CaVO$_{3}$ is from \cite{inoue1995systematic}.   The parameters values  we used 
 $(m,p,k_c,\mu)=(0.47,1.9, 2.05,1.732)$ for red line and $(0.455,1.7, 2.07,1.732)$ for blue line. 
           } \label{fig:fitting} 
\end{figure}

The ultimate test of a physical model is  the capability of its explaining the data. 
Here we take Vanadium Oxides data and see how the theory fits data.   
It turns out that the  X-ray absorpsion spectroscopy data for  SrVO$_{3}$ (red circles and diamonds) \cite{sekiyama2002genuine} and Ca$_{0.9}$Sr$_{0.1}$VO$_{3}$ (blue boxes and triangles) \cite{inoue1995systematic} can be well fit with our theory. However the Photoemission (PES) data  can not. The parameters   taken to fit the XAS data create too much asymmetry in 
spectral function so that unless we symmetrize by hand, we do not have a room to accommodate the PES data. We do not have a good reason to perform such symmetrization although in some literature it is   practiced \cite{norman1998phenomenology,lee2007abrupt,kondo2009competition}.  Once   symmetrized, can the data can be fit very well by our model. See the figure \ref{fig:PES} in the appendix.    
We adapted the data presented in the lecture note of Vollhardt in \cite{pavarini2014dmft} and DMFT study in \cite{sekiyama2004mutual,nekrasov2005comparative} and fit those with our theory. 
The result is given in Figure \ref{fig:fitting}.  
 
 \section{Discussion}
In this paper we studied the phase diagram of a holographic model which can accommodate the physics of Mott transition. The key feature is the presence of gapped phase  and the  Free fermion point. 
The competition  of the two generate pseudo-gap phase as an intermediate zone.   
  Any line connecting the gapped and free fermion point  in the phase space can be regarded as a   
   an analogue of the  Hubbard model. We   report that  all the phase change is smooth and we  did not find  any signal of instability in the spectral density within the unitarity bound $m<1/2$.
Comparing the DMFT result on the Hubbard model with ours, three features agree with single site DMFT: the appearance of  shoulder peak and transfer of the DOS to the shoulder peak and the maintenance of the central peak till the 
gap creation.  However, due to the large asymmetry created by the chemical potential, three peak structure is not manifest since one of the shoulder is too weak. 
For the cluster DMFT and the hyperbolic embedding, there is an agreement in the appearance of the pseudo-gap.  But again due to the asymmetry, the details are different. 

What is the origin of the spectral asymmetry? If spatial dimension is bigger than 1, the momentum space light-cone
$(\omega+\mu)^2-k^2=0$ is asymmetric because  the region below  the chemical potential is closer to the tip of the cone.  In 1+1 dimensional theory we do not expect such asymmetry.  The charge dependence of the interaction term enhanced this phenomena. 

Apart from the asymmetry there is one more problem in this model. 
 It turns out that for the model with dipole interaction,  the filling fraction  changes the interaction strength $p$, which is odd at first sight. 
 This can be understood   the if we note that $p$ always comes with $Q$, the charge density of RN black hole, so that   increasing $p$ has the effect of increasing $Q$. Increasing $Q$ has the effect of increasing $\mu$ so that in the presence of the Fermi sea, the fermi level should look as if it is increased. 
In real material changing the coupling strength should not involve the effect of  changing the particle number. 
Therefore we should conclude that the dipole term is not proper to model the Mott transition in a system where particle number is preserved.

We describe some future interests below. 
First, we need to find other gap creation mechanism which can realize the Mott transition such that  spectral function maintains  particle-hole symmetry at least approximately, lack of which is the most 
serious defect of present model in practical application. 
Second, we need to consider the back reaction of the background geometry. 
This is especially interesting due to the parallelism of holographic theory and the DMFT calculation near quantum critical point\cite{PhysRevLett.107.026401,2016arXiv160200131D}. 
Third,  when the gap is generated, the conformal invariance is also broken therefore the conformal unitarity condition that restricted us $m \leq 1/2$ is not much meaningful. 
Then,  we should investigate beyond  $m=1/2$. 
Next, we should study the fermions  in the presence of the complex scalar, the superconducting order parameter. Also  we did not investigated temperature and chemical potential  evolutions much. It will gives most practical results to data fit.  Many interesting questions are waiting analysis  to accommodate the reality in the holographic model.     

 \vskip 1cm
\appendix {\bf \Large Appendix}

\section{phase space of T=0}
 The phase diagram depends on temperature. We   examined the phase space in zero temperature: Bad metal and its prime regions are reduced but not go away at $T=0$ because the presence of sharp peak itself does not guarantee that it is Fermi liquid. 
At $p=0$, the spectral function depends on the frequency $\sim \omega^{-2m}$ in alternative quantization we use. If $m$ is not large enough,  localizing the spectral function 
 to the Fermi surface can not be achieved  so that it is far from Fermi liquid behavior. Therefore 
 we need to distinguish the phase near the $m=0.5, p=0$, the free fermion point, from the phase near  $m=0,p=0$, which is dubbed as bad metal region.  
Notice also that  the zero temperature Fermi surface   
  disappears  very quickly (if not on   m=1/2 line)  as soon as we turn on  T just a bit.   
  
Bad-metal    prime   region
 shrinks to $m>0.45$ but it still exists.  
See figure \ref{fig:p0mevol2} below. 

\begin{figure}[ht!]
\centering
        \subfigure[$k=k_F$]
   {\includegraphics[width=6cm]{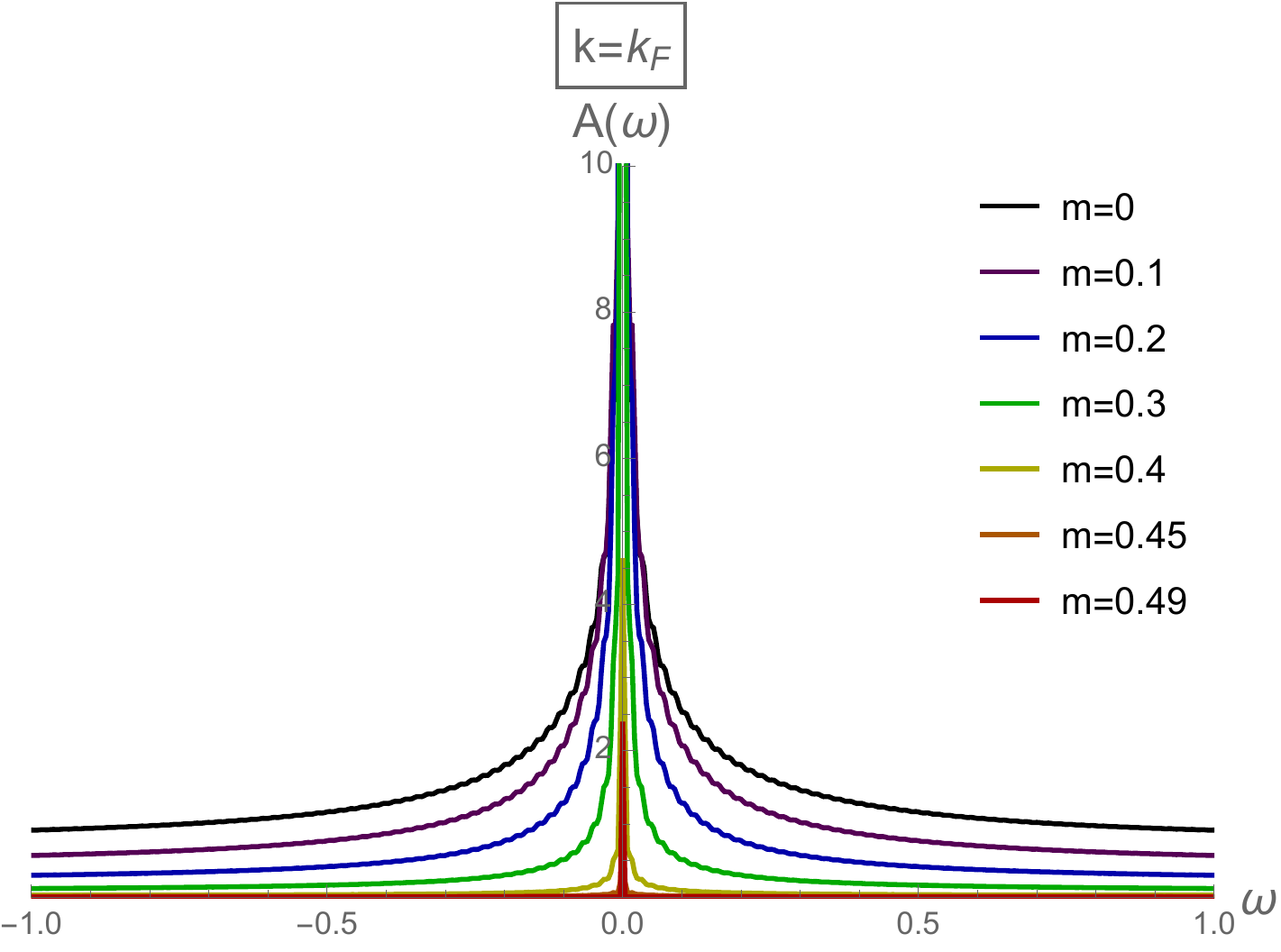}  }
   \hskip1cm    
       \subfigure[$k=-3$ ]
    {\includegraphics[width=6cm]{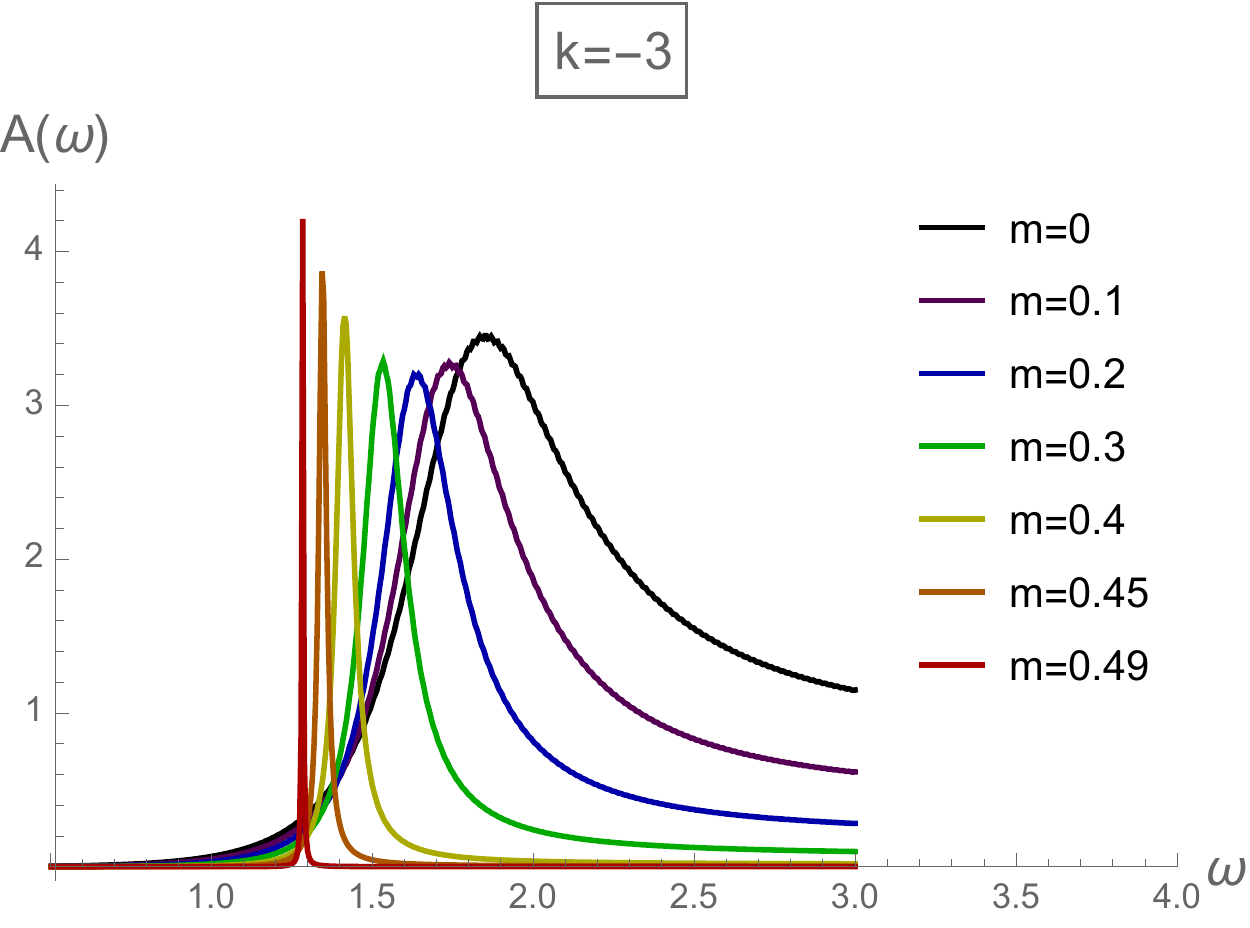}  }
     \caption{The spectral density at $T=0$ and $p=0$ at fixed momentum $k_{F}$. 
    \small     (a) Spectral density at $k=k_F$. There is sharp peak at $k_F$. As mass decreases,  it is not localized at the fermi-surface and can not be classified as a Fermi liquid.  (b) Spectral density at $k=-3$ also shows that it is   broad for small mass.  The hight decreases up to $m\sim 0.25$ and increases.  }
      \label{fig:p0mevol2}
\end{figure}
\begin{figure}[ht!]
\centering
    \subfigure[$m=0$ ]
    {\includegraphics[width=4cm]{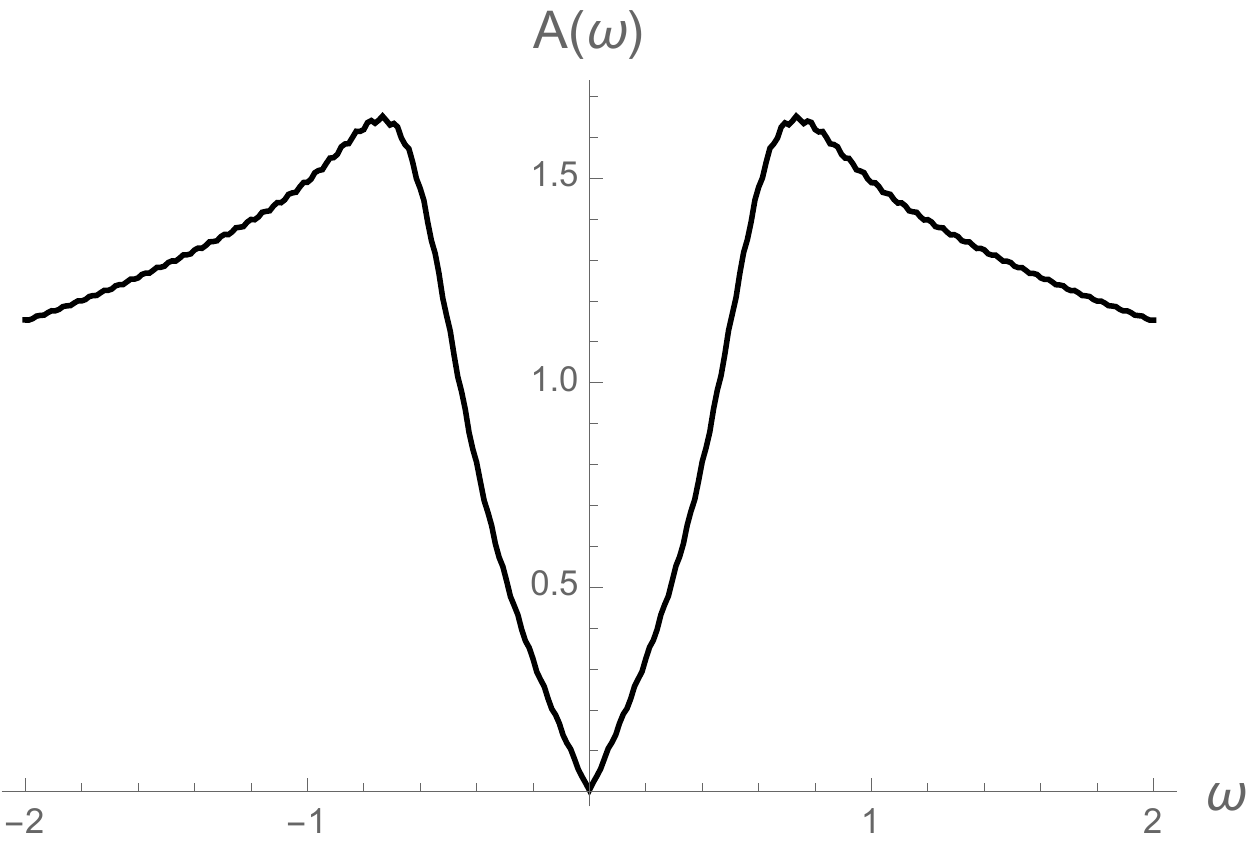}  }
     \subfigure[$m=0.45$]
   {\includegraphics[width=4cm]{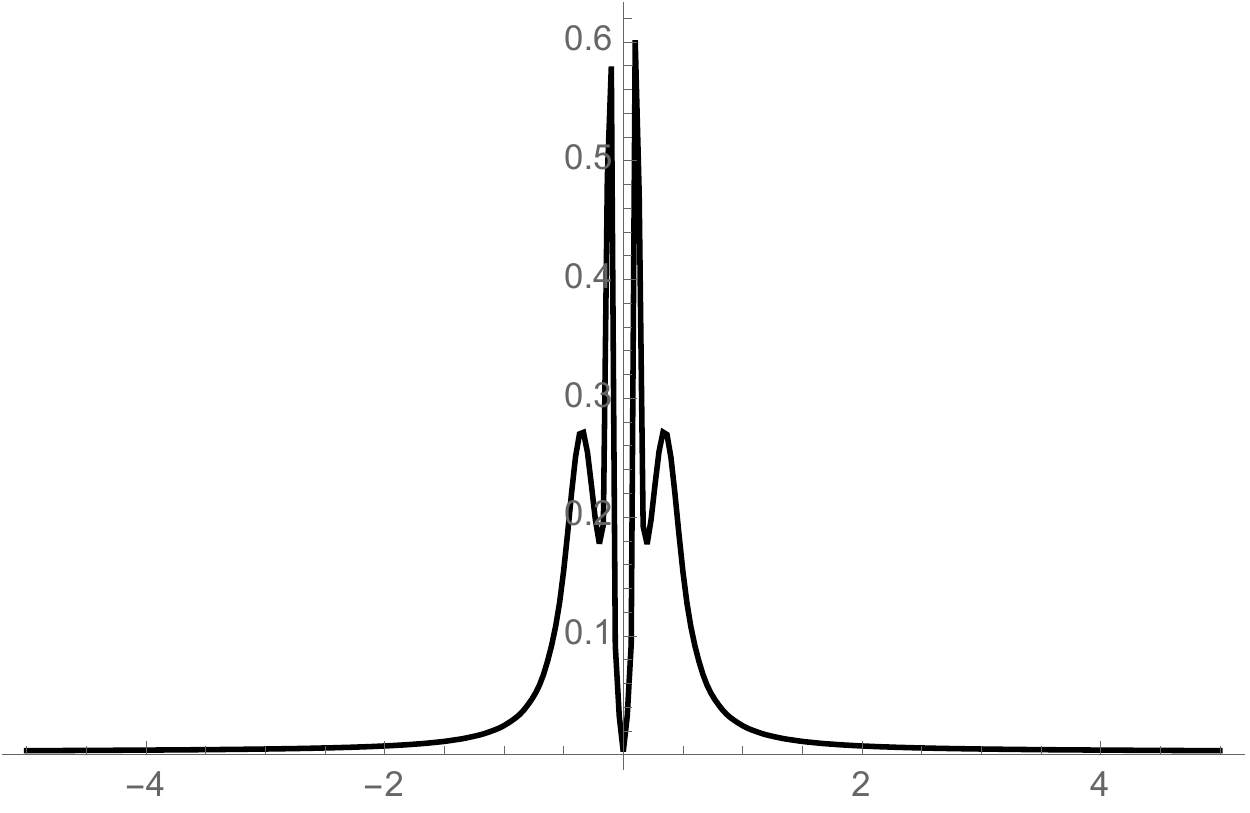}  }
     \subfigure[$m=0.49$]
   {\includegraphics[width=4cm]{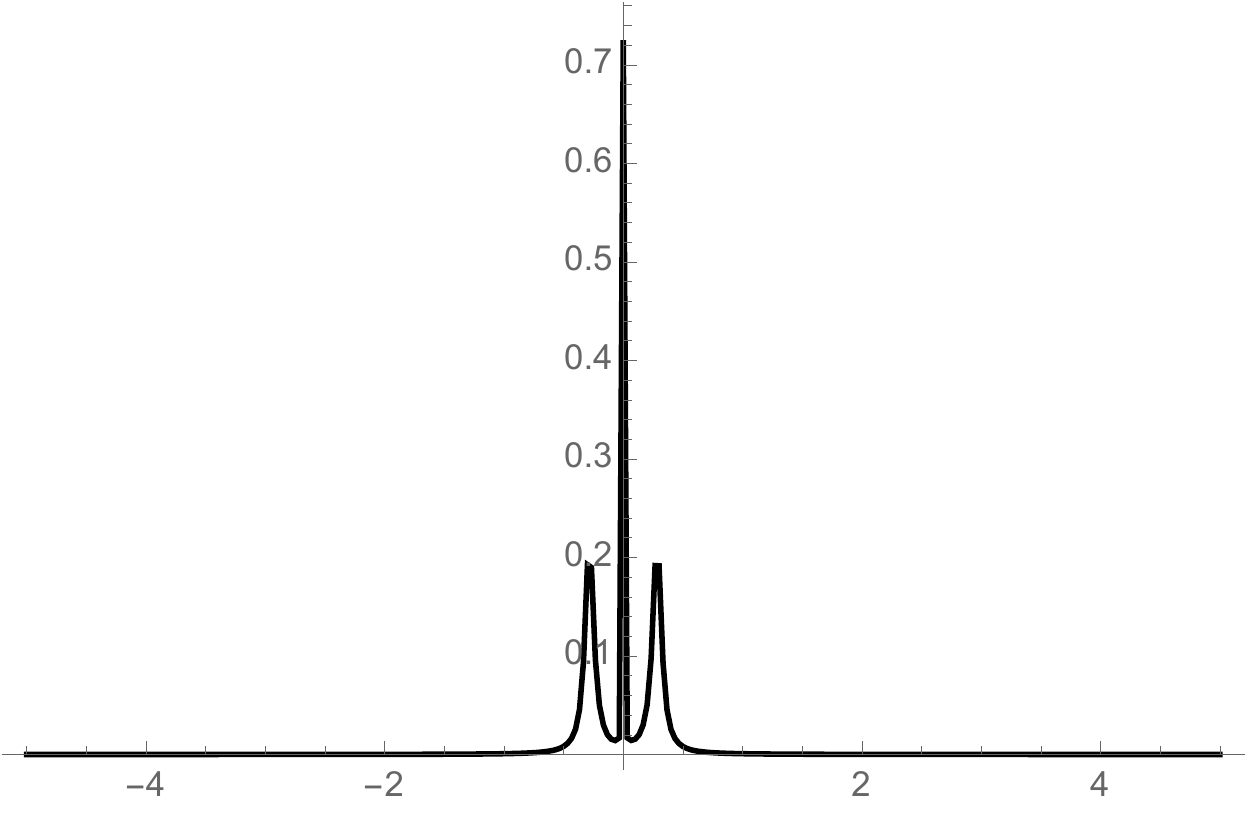}  }
    \caption{Evolution of the spectral density at $T=0$ and $p=2$. BM' region is much reduced at $T=0$ but still exist.}
      \label{fig:p2mevol}
\end{figure}
\begin{figure}[ht!]
\centering
    {\includegraphics[width=6cm]{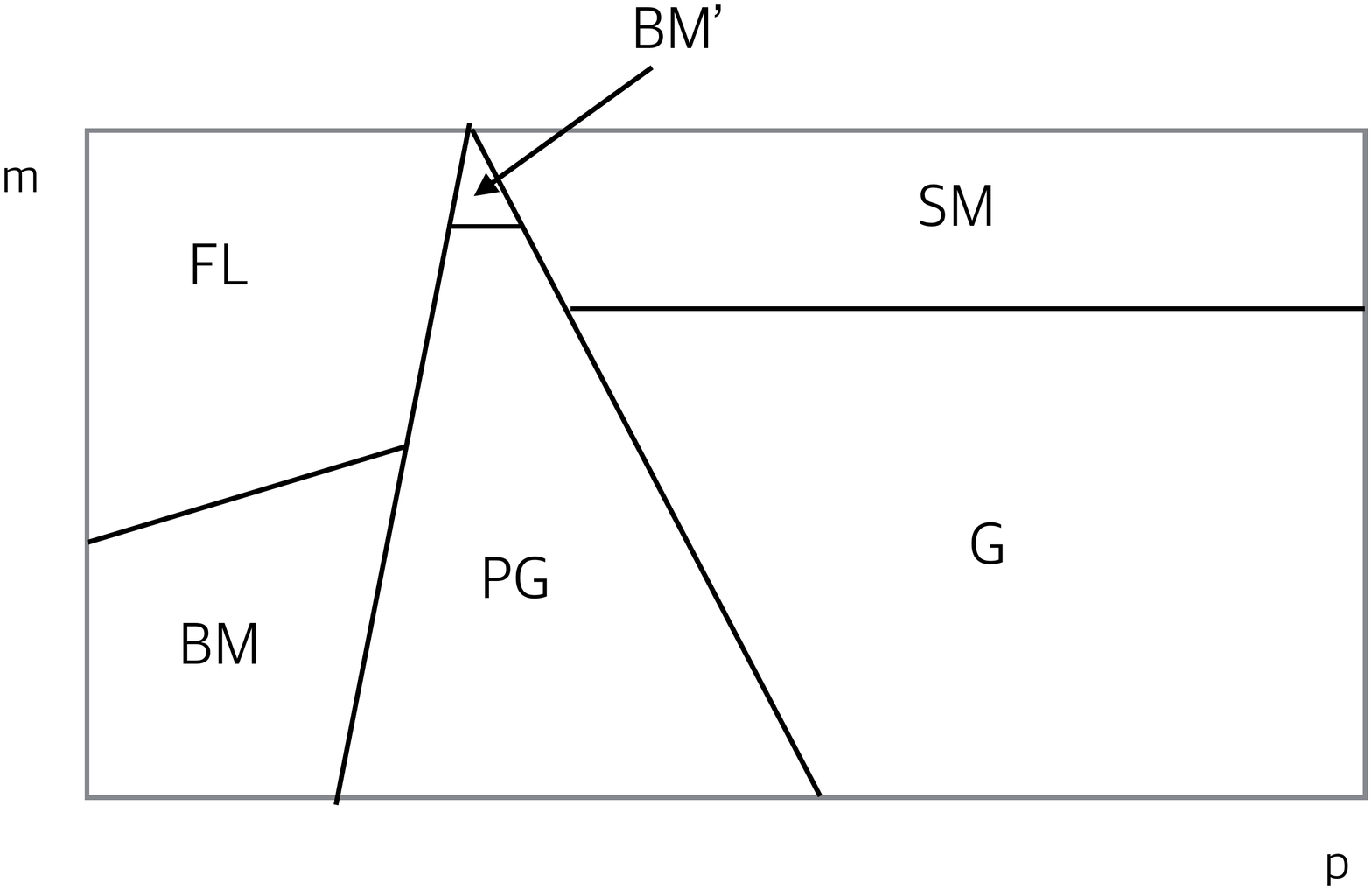}  }
     \caption{ Schematic  phase diagram at $T=0$. }
      \label{fig:zeroTpd}
\end{figure}
Gapped phase and pseudo-gap phase expand such that much of the bad metal prime region become pseudo-gap at zero temperature. 
The phase boundary of gapped phase  and pseudo-gap is moved to near p=3 at m=0 according to our criterion. The phase diagram is drawn schematically  in the figure \ref{fig:zeroTpd}, where we do not find any qualitative change. 
However, the  zero temperature phase diagram is not
very useful to fit the data of the transition metal oxides. 
This is because typical data  belong to bad metal prime phase, and at zero temperature, this relevant 
regime is tiny,   therefore there is not much  room to adjust the  parameter   to  fit data.     
In this paper,  we have drawn it at   $T=0.1$,  which gives  a typical phase diagram and useful to us for data fitting.  
 
\section{The role of $m$ and $p$} 
 \subsection{ Gap generation versus Appearance of half-metal phase}
\begin{figure}[ht!]
\centering
       \subfigure[$p=1$ ]
   {\includegraphics[width=3.7cm]{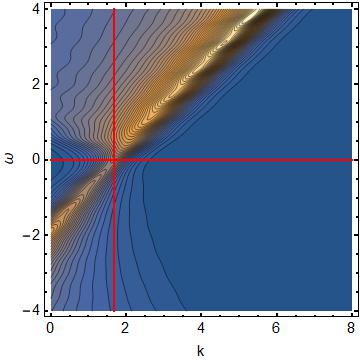}  }
       \subfigure[ $p=2$]
   {\includegraphics[width=3.7cm]{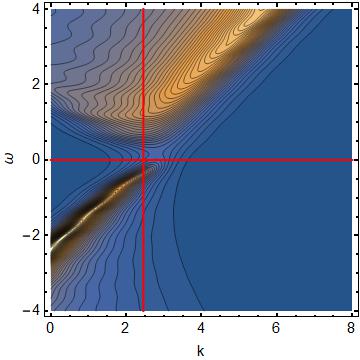}  }
         \subfigure[ $p=6$]
   {\includegraphics[width=3.7cm]{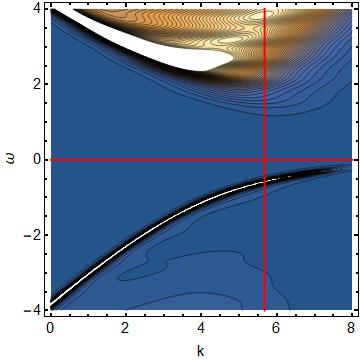}  }
         \subfigure[ $p=8$]
   {\includegraphics[width=3.7cm]{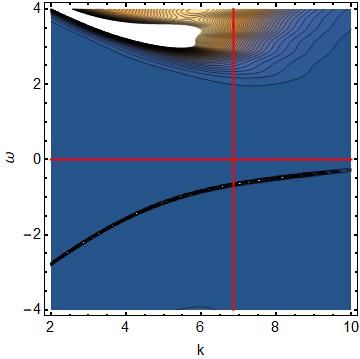}  }
       \subfigure[$p=1$ ]
   {\includegraphics[width=3.7cm]{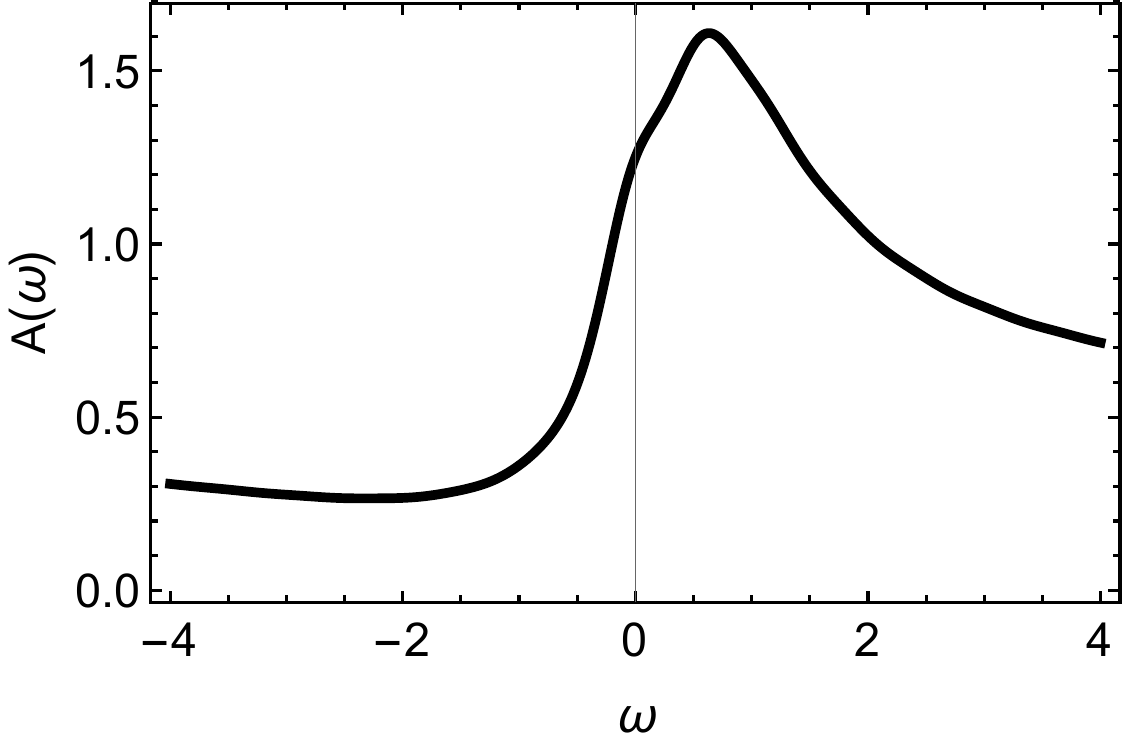}  }
       \subfigure[ $p=2$]
   {\includegraphics[width=3.7cm]{specm01p20wplot.pdf}  }
         \subfigure[ $p=6$]
   {\includegraphics[width=3.7cm]{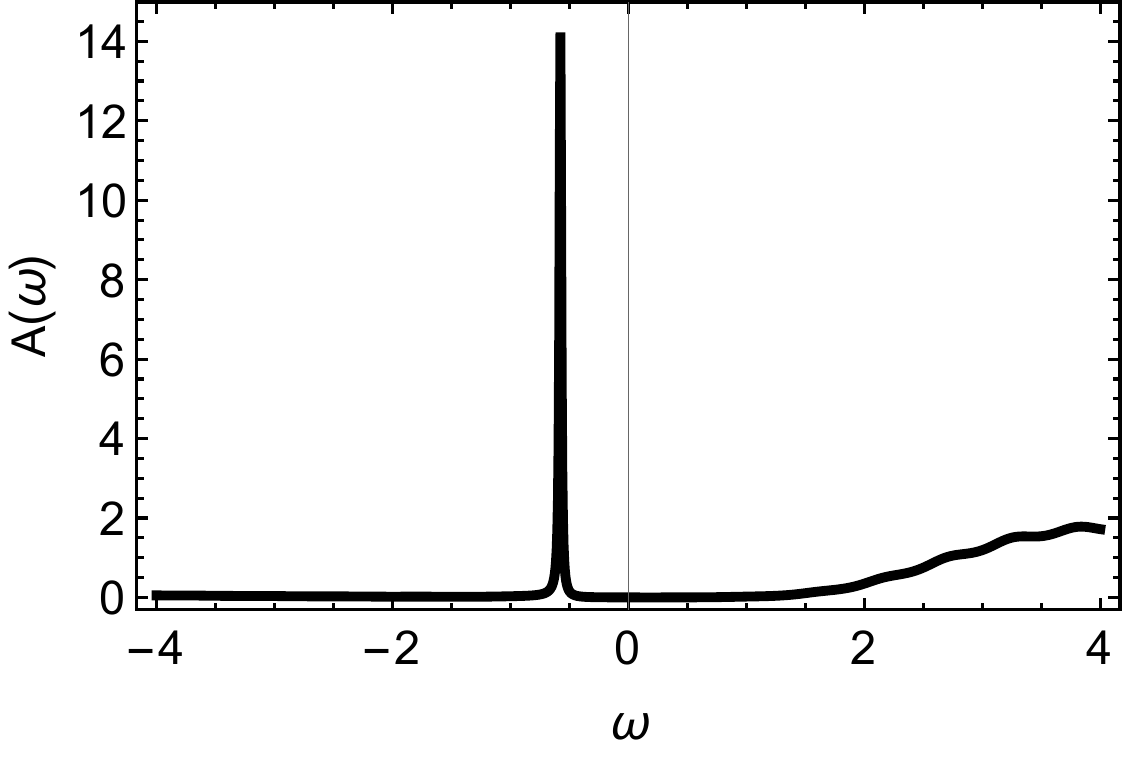}  }
         \subfigure[ $p=8$]
   {\includegraphics[width=3.7cm]{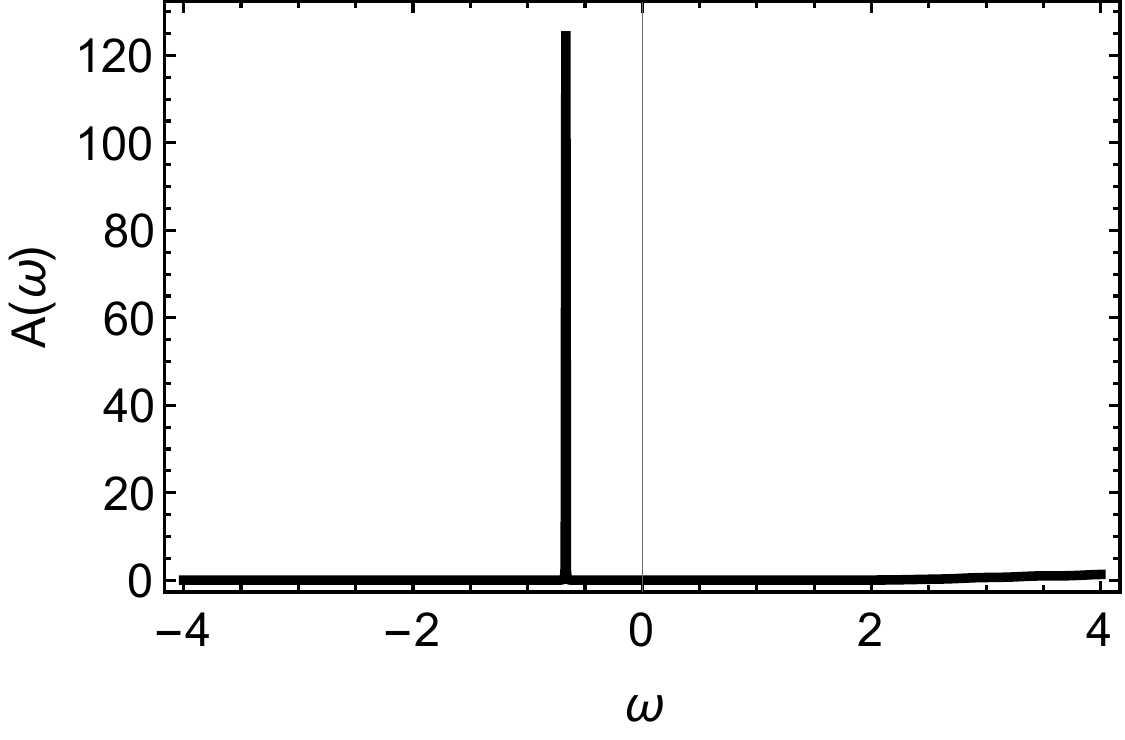}  }
    \caption{  (a)-(d) on p-evolution of the DOS along the line  m = 0.1 show appearance of and pseudo-gap and
gapped phase; (e)-(h) spectral function along vertical red line in each figure (a)-(d).   $k_c=1.65, 2.46, 5.68, 6.86$ respectively from (e) to (h).
       } 
       \label{fig:spec03}
\end{figure}

 Here we follow   fixed mass line in the parameter space to see the evolution as we increase the $p$. 
We first   study the  lower half   of the phase diagram
  by calculating  its evolution along the line $m=0.1$  with increasing the dipole strength. 
  The result is  is drawn in  Figures \ref{fig:spec03} (a)-(h), where  three different phases appear: 
  \begin{enumerate}
\item Fig. \ref{fig:spec03}  (a,e) $p=1$, bad metal phase with broadened  peak with  low height at Fermi level, 
\item Fig. \ref{fig:spec03}  (b,f) $p=2$, psuedo-gap phase with incomplete depletion of DOS at Fermi  level. 
\item Fig. \ref{fig:spec03}  (c,g)  $p=6$, gapped phase; Fig. \ref{fig:spec03}  (d,h)  $p=8$, gapped phase with increased gap size.  
 
\end{enumerate}
The overall feature of the evolution is from metalic to the insulating phase with pseudo gap phase in the middle and   it agrees    with our expectation. 

\begin{figure}[ht!]
\centering
       \subfigure[$p=1$ ]
   {\includegraphics[width=3.7cm]{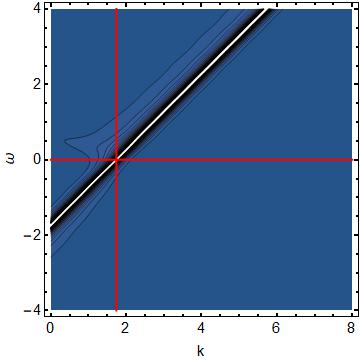}  }
       \subfigure[ $p=2$]
   {\includegraphics[width=3.7cm]{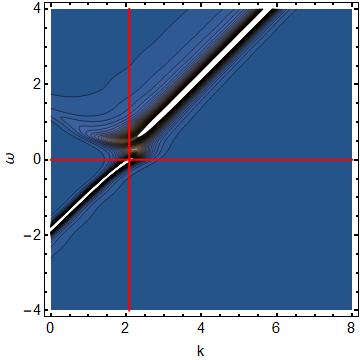}  }
         \subfigure[ $p=3$]
   {\includegraphics[width=3.7cm]{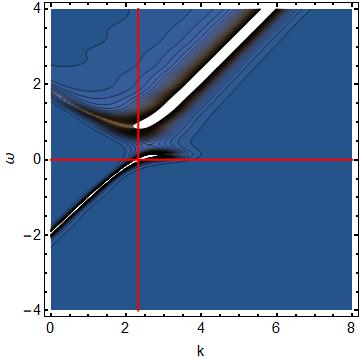}  }
         \subfigure[ $p=6$]
   {\includegraphics[width=3.7cm]{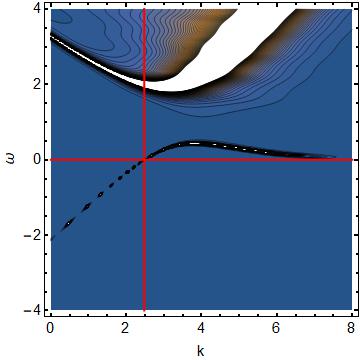}  }
       \subfigure[$p=1$ ]
   {\includegraphics[width=3.7cm]{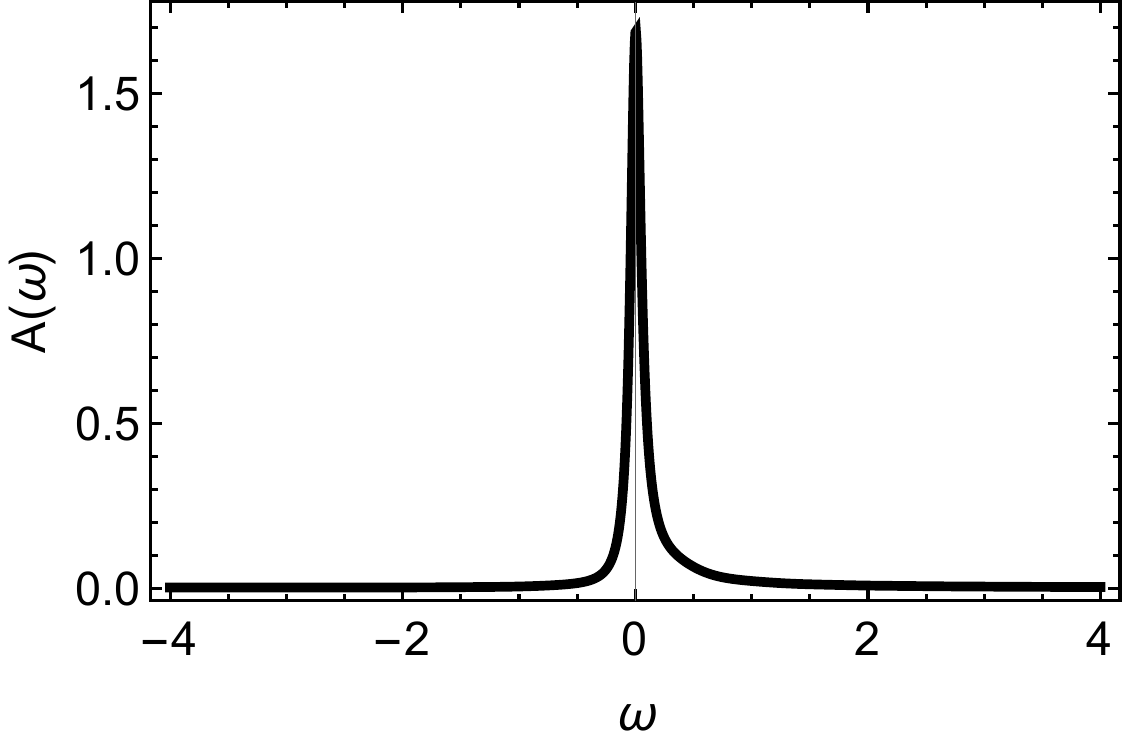}  }
       \subfigure[ $p=2$]
   {\includegraphics[width=3.7cm]{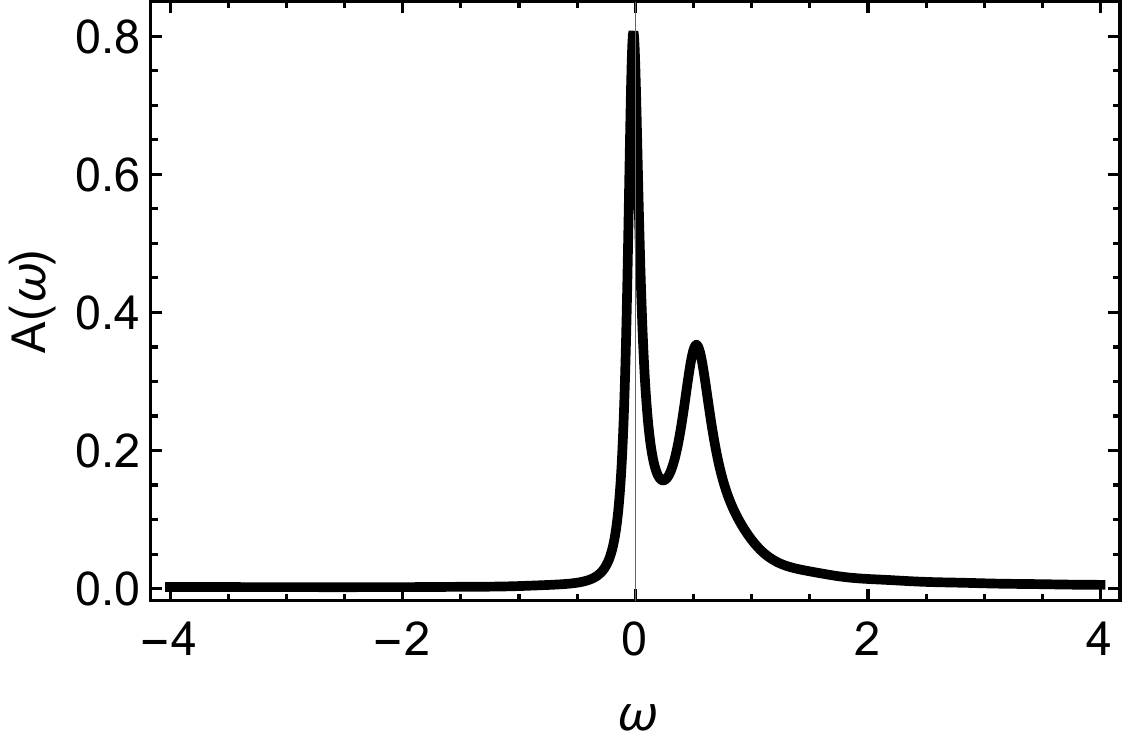}  }
         \subfigure[ $p=3$]
   {\includegraphics[width=3.7cm]{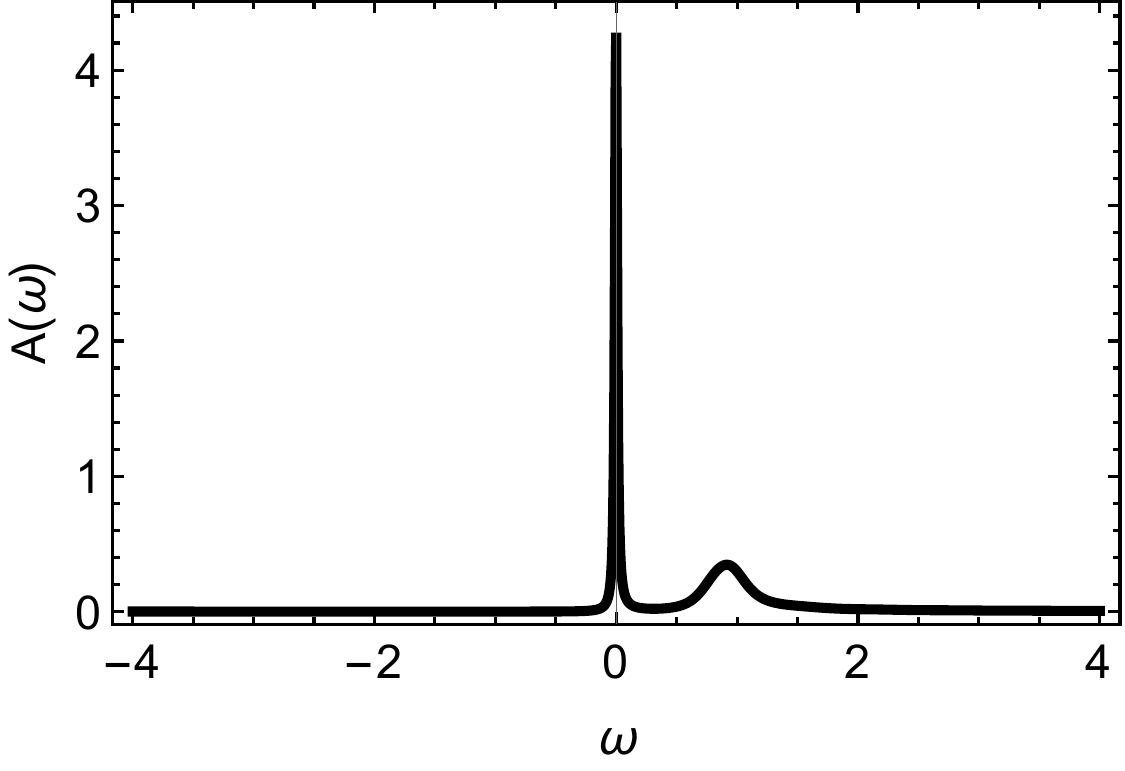}  }
         \subfigure[ $p=6$]
   {\includegraphics[width=3.7cm]{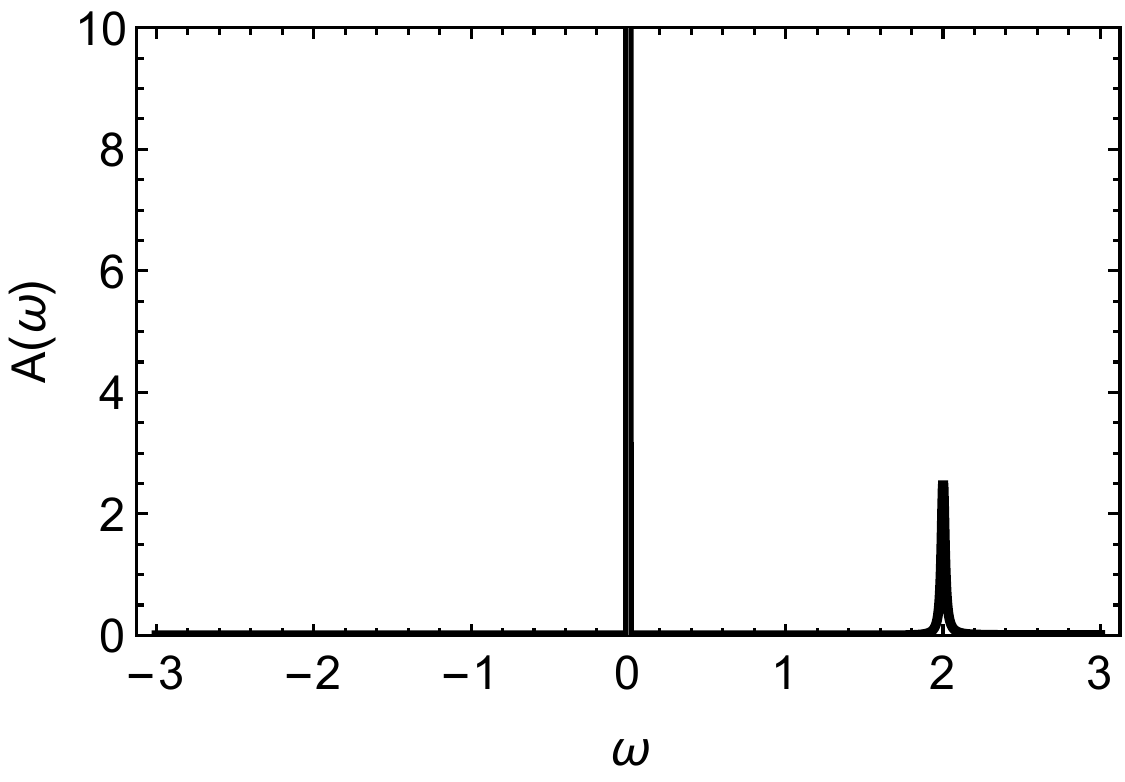}  }
    \caption{  (a)-(d):$p$-evolution of  the DOS   at  $m=0.45$   shows appearance of   and bad metal prime and half-metal;  
     (e)-(h) spectral function along vertical red line in each figure (a)-(d). We choose $k_c$ as $(1.74, 2.08, 2.32, 2.48)$ respectively.} \label{fig:spec01}
\end{figure}

We  now show the evolution along the line of   $m=0.45$  with  increasing 
$p$ to  demonstrate  the changes of phase in the upper  half part of the phase diagram.
The  $\omega$-plots in  Figures \ref{fig:spec01}(e)-(h) are is along the  red vertical redlines  
in Figure \ref{fig:spec01}(a)-(d) respectively. We can see three different phases: 
\begin{enumerate}
\item Gapless metalic phase with linear dispersion:  it is a Fermi liquid (FL) regime.   
\item  The   bad metal phase due to development of  incomplete generation of conduction band. 
\item  New metalic phase which we call  half-metal due to the development of the  conduction band.
half-metal because half of the DOS at the Fermi sea is depleted and moved to shoulder region. 
\end{enumerate}

\subsection{Bulk-mass evolution at fixed $p$}

\begin{figure}[ht!]
\centering
       \subfigure[$m=0$ $p=2.5$]
   {\includegraphics[width=3.7cm]{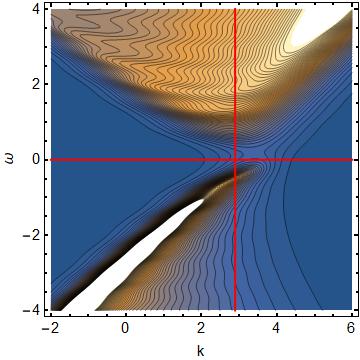}  }
       \subfigure[ $m=0.15$ $p=2.5$]
   {\includegraphics[width=3.7cm]{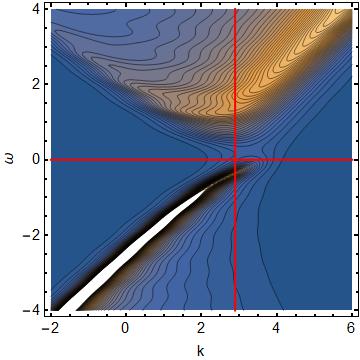}  }
         \subfigure[ $m=0.35$ $p=2.5$]
   {\includegraphics[width=3.7cm]{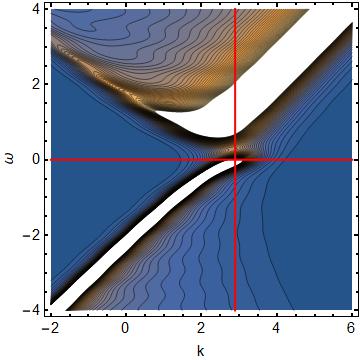}  }
         \subfigure[ $m=0.45$ $p=2.5$]
   {\includegraphics[width=3.7cm]{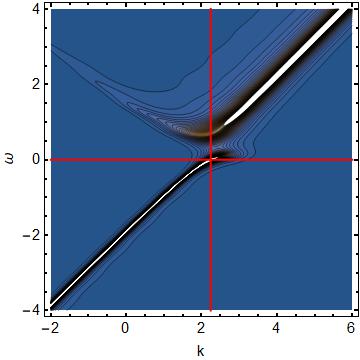}  }
        \subfigure[$m=0$ $p=2.5$]
   {\includegraphics[width=3.7cm]{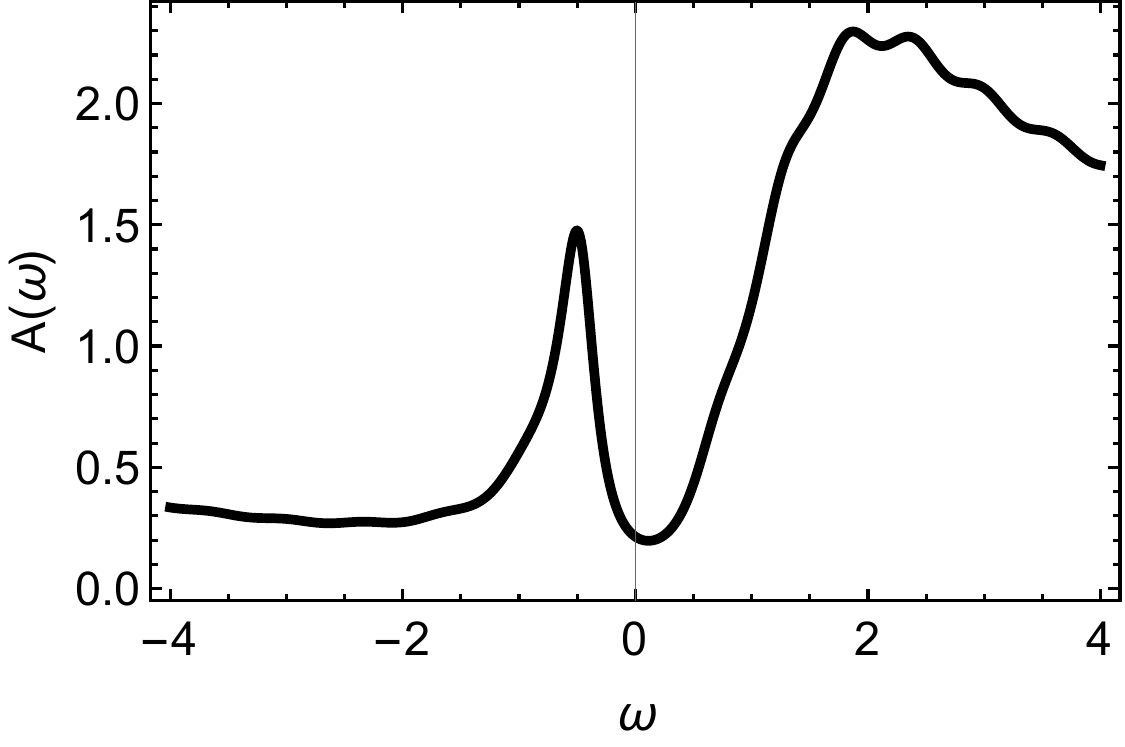}  }
       \subfigure[ $m=0.15$ $p=2.5$]
   {\includegraphics[width=3.7cm]{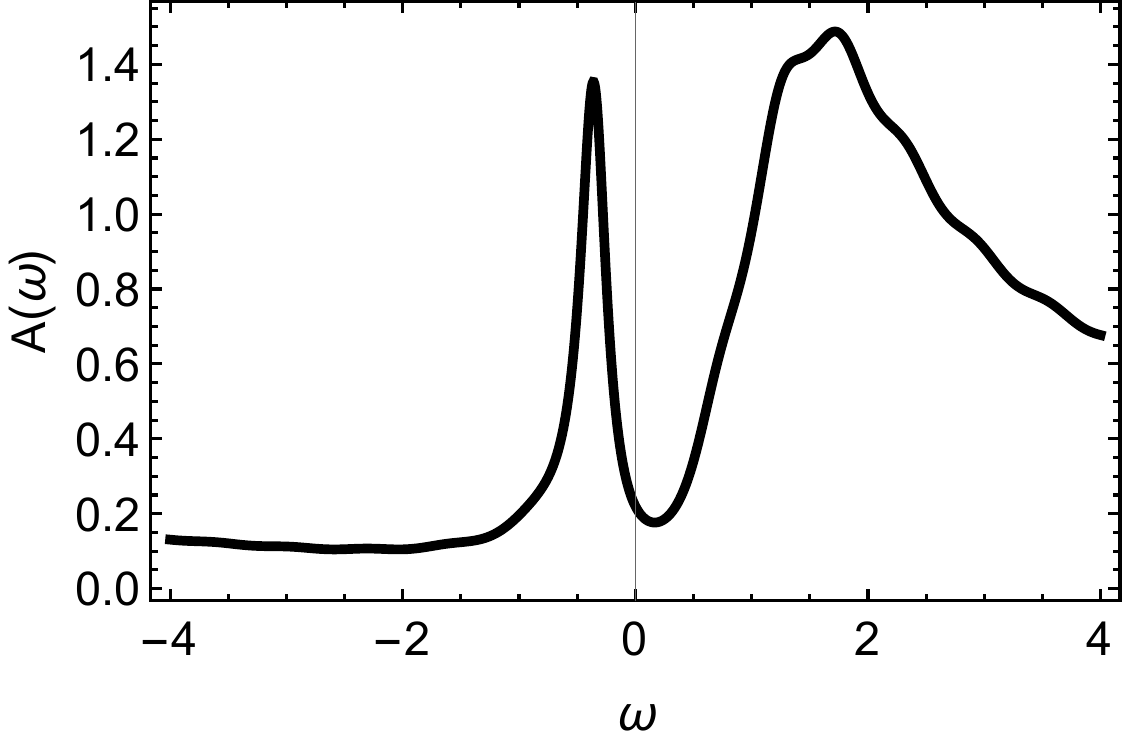}  }
         \subfigure[ $m=0.35$ $p=2.5$]
   {\includegraphics[width=3.7cm]{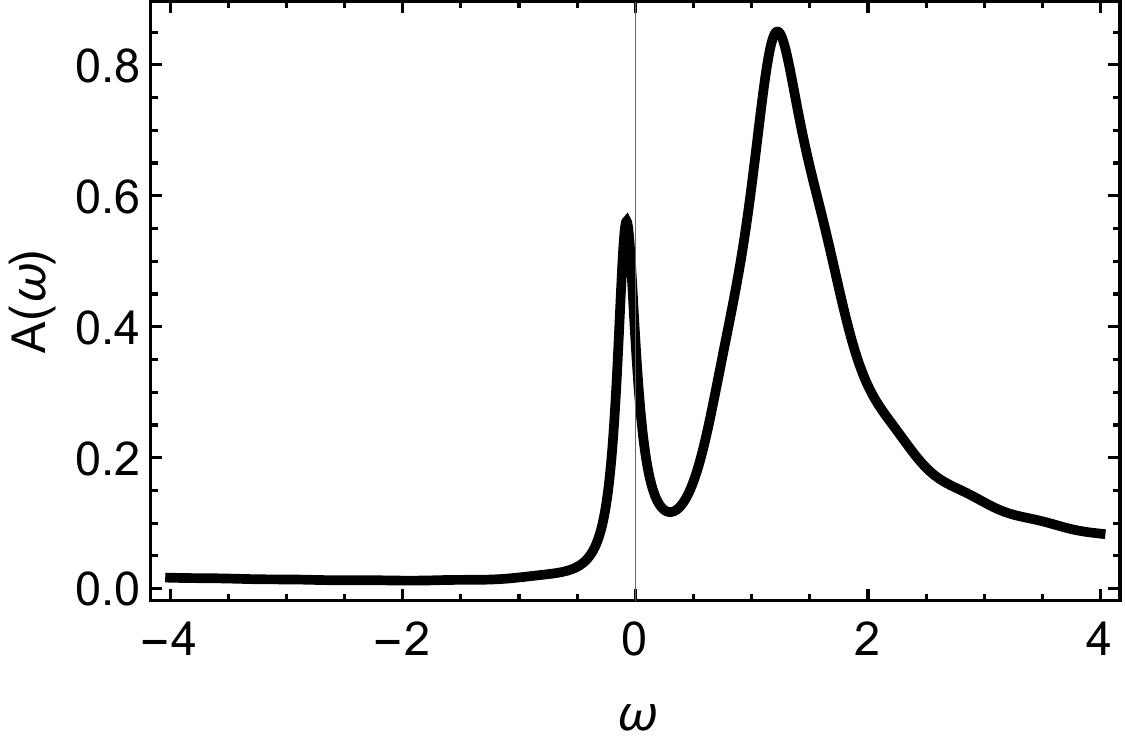}  }
         \subfigure[ $m=0.45$ $p=2.5$]
   {\includegraphics[width=3.7cm]{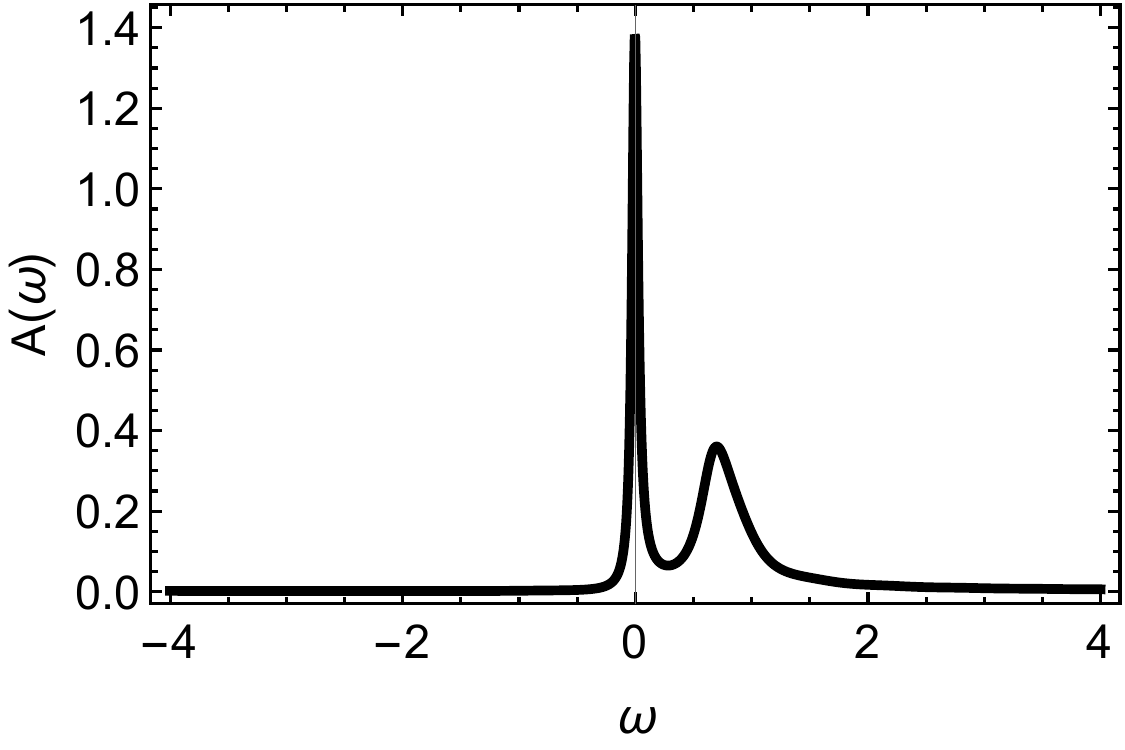}  }
   \subfigure[$m=0$ $p=6$ ]
   {\includegraphics[width=3.7cm]{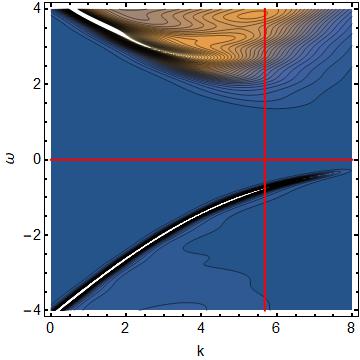}  }
     \subfigure[ $m=0.15$ $p=6$]
   {\includegraphics[width=3.7cm]{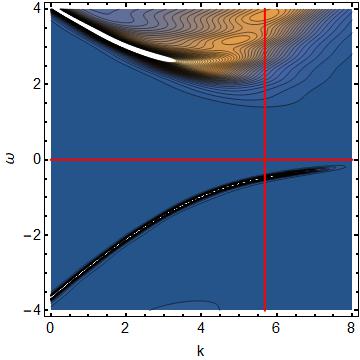}  }
         \subfigure[ $m=0.35$ $p=6$]
   {\includegraphics[width=3.7cm]{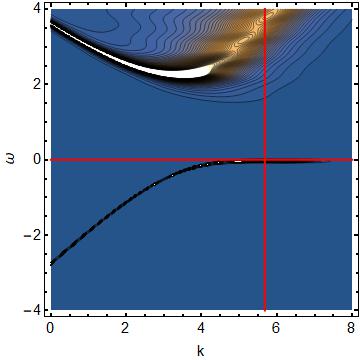}  }
         \subfigure[ $m=0.45$ $p=6$]
   {\includegraphics[width=3.7cm]{contourm045p60r.jpg}  }
        \subfigure[$m=0$ $p=6$]
   {\includegraphics[width=3.7cm]{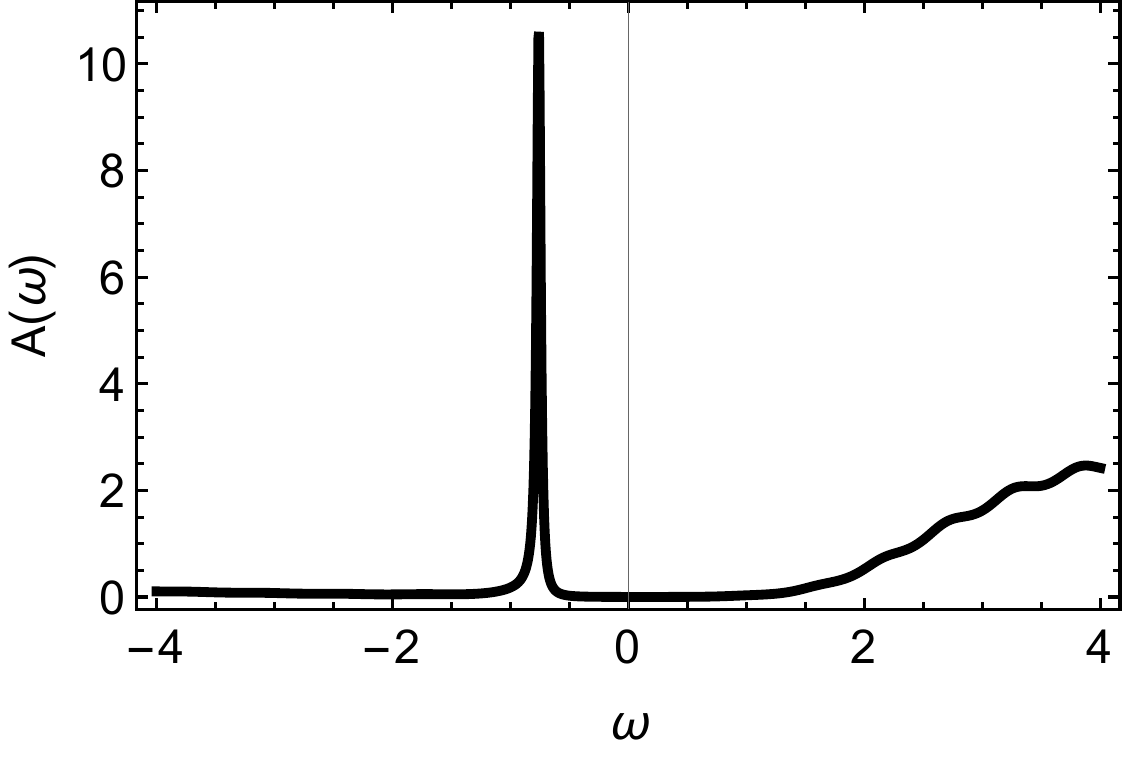}  }
        \subfigure[ $m=0.15$ $p=6$]
   {\includegraphics[width=3.7cm]{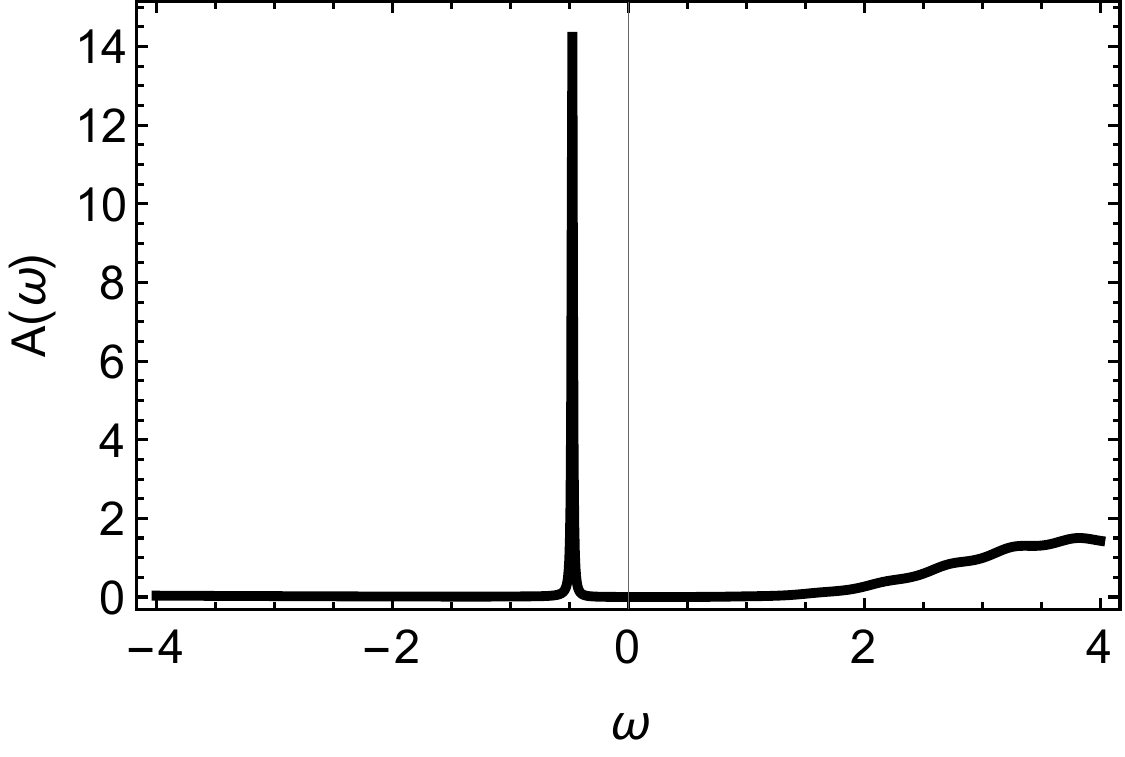}  }
         \subfigure[ $m=0.35$ $p=6$]
   {\includegraphics[width=3.7cm]{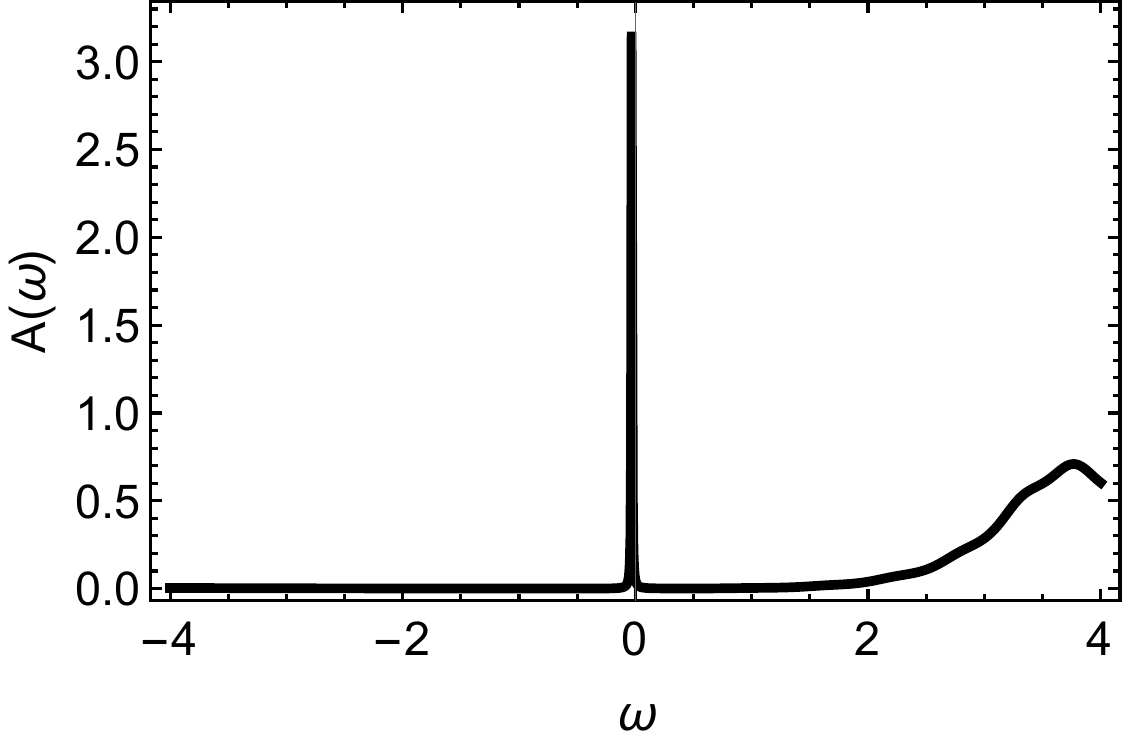}  }
         \subfigure[ $m=0.45$ $p=6$]
   {\includegraphics[width=3.7cm]{specm045p60wplot.pdf}  }
    \caption{(a)-(h): $m$- evolution of  the spectral density with increasing $m$ with $p=2.5$. We choose $k_c$ as $(2.89, 2.89, 2.89, 2.25)$ respectively. 
       (i)-(p):Evolution of  the spectral density at $p=6.0$.   For both cases the final stage is  hM phase.  We choose $k_c$ as $(5.68, 5.68, 5.68, 2.48)$ respectively.    }\label{fig:spec05}
\end{figure}

We now study the role of mass more systematically by calculating the evolution of the DOS at two nonzero  fixed values of $p$, that is   along two vertical lines  $p=2.5$ and  $p=6.0$ in phase dragram.
In the Figure  \ref{fig:spec05}(a)-(h), the $m$-evolution along $p=0.2$ line is drawn, where a few physically interesting phases appear.  From the figures \ref{fig:spec05}(i)-(p), we can   see  that the bulk  mass sharpens the peak at the Fermi surface consistently  regardless of the value of $p$. 
One can see that increasing $m$ pushes up the new band so that gap is reduced. When the middle band crosses the Fermi level,  central peak appears signaling the creation of the half-metalic phase.     For both cases the final stage is hM phase. 
\vfill

\section{Symmetrized spectral function}
\begin{figure}[ht!]
\centering
      \subfigure[BM' $(m=0.4,p=2)$ ]
   {\includegraphics[width=60mm]{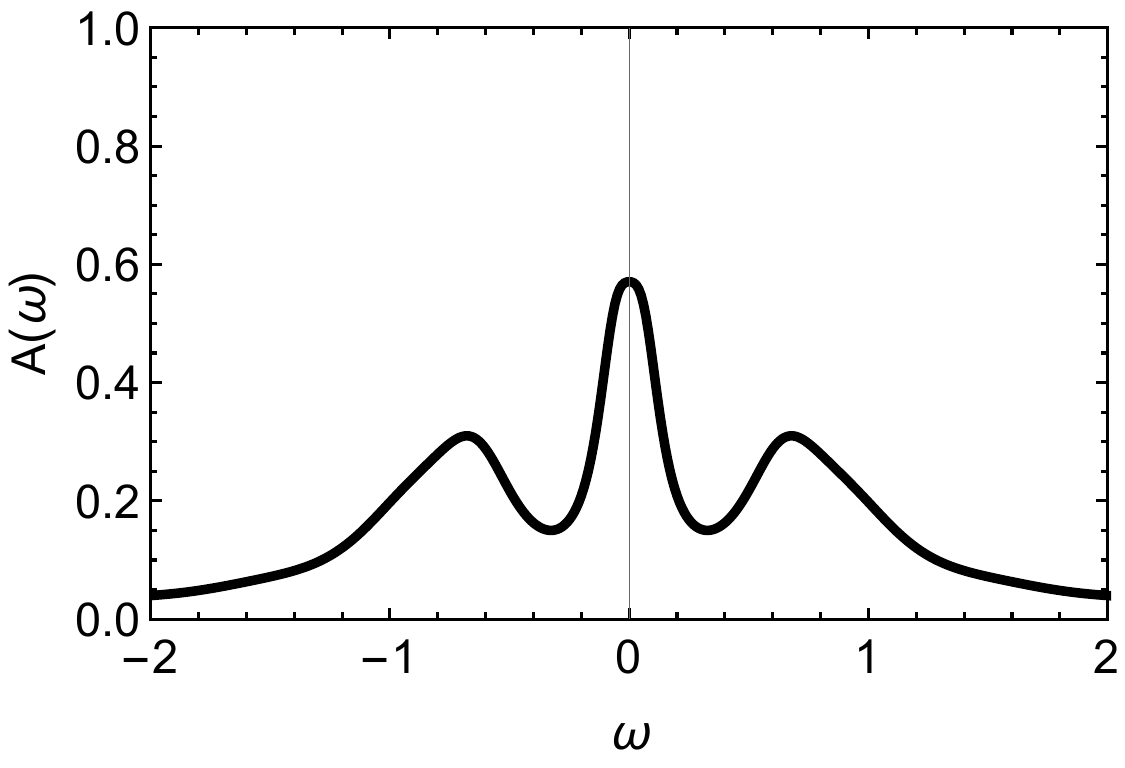} }
    \subfigure[PG $(m=0.1,p=2)$ ]
   {\includegraphics[width=60mm]{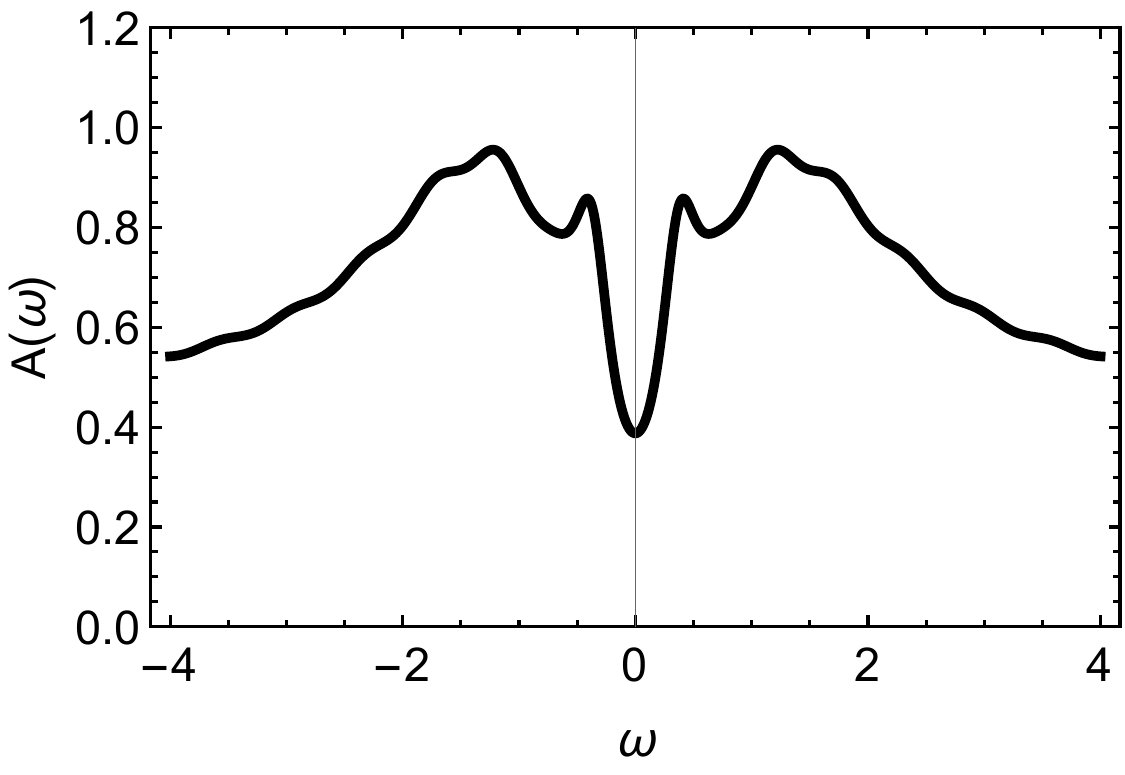} }
           \caption{Symmetrized spectral functions for     Bad metal prime and psuedo-gap.}    \label{fig:SSF} 
\end{figure}

Pseudo-gap data in the context of High-Tc superconductor theory is usually presented using symmetrized spectral function (SSF) \cite{norman1998phenomenology,lee2007abrupt,kondo2009competition}.  
We present the  result  it in figure \ref{fig:SSF}
for those who are already familiar to  condensed matter literature. 
 
\subsection{PES data with symmetrized spectral function}
\begin{figure}[ht!]
\centering
  {\includegraphics[width=5cm]{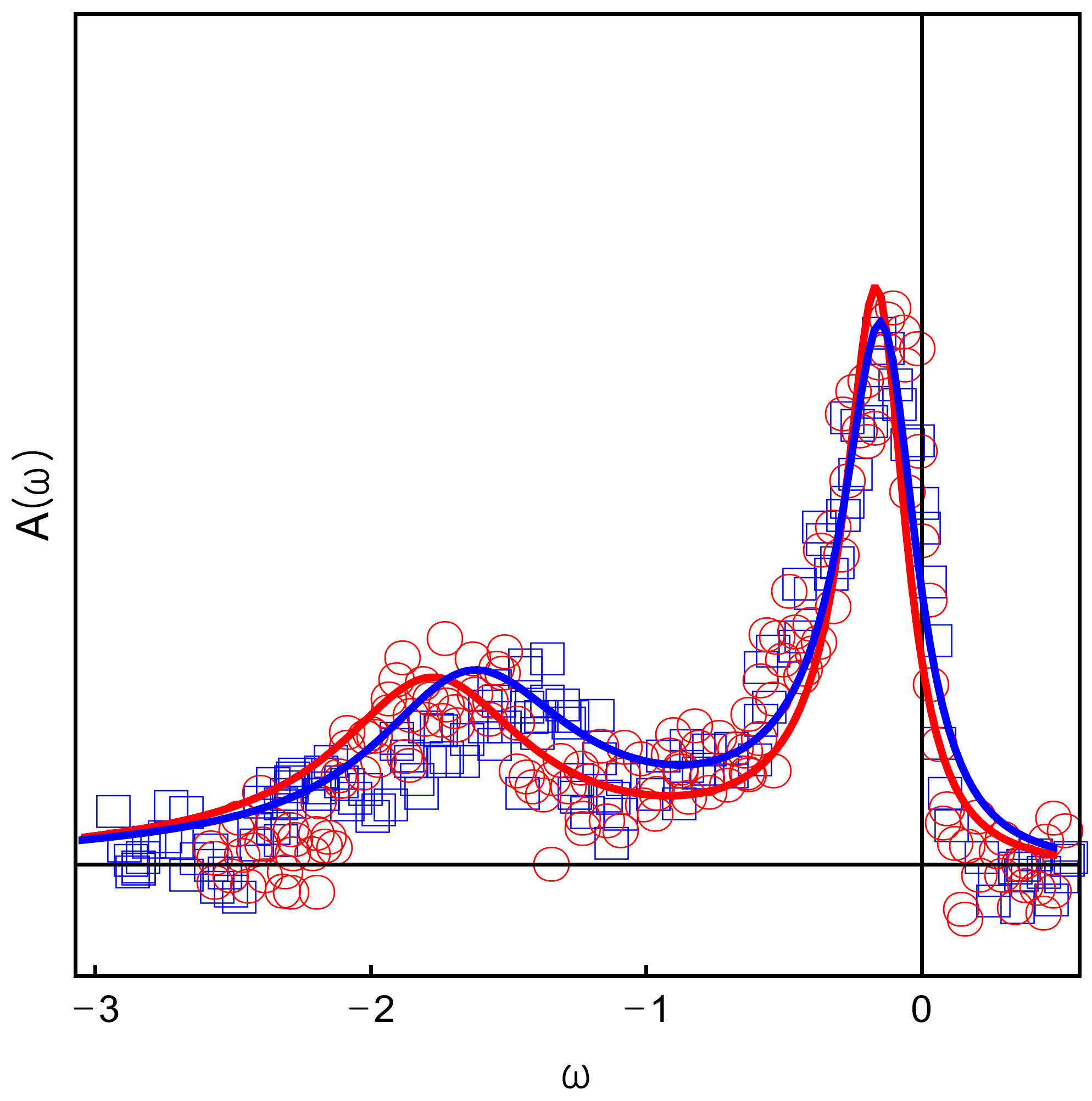}  }
         \caption{PES data  with symmetrized spectral function:  
         color red is for SrVO$_{3}$ and (color blue) is for  CaVO$_{3}$. The data for  SrVO$_{3}$ is from   \cite{sekiyama2002genuine},  and  that for  CaVO$_{3}$ is from \cite{inoue1995systematic}.   The parameters values  we used are 
            $(m,p,k_c,\mu)=(0.47,2.2, 2.08,1.732)$  for  red line and  $(0.47,2.05, 2.04,1.732)$ for   blue line. 
           } \label{fig:PES} 
\end{figure}

As we mentioned in the main text, the photoemission data can be fit by the holographic theory only when we symmetrize the spectral function in $\omega$: $A(\omega,k)\to f(A(\omega,k)+A(-\omega,k))$   fermion distribution function $f=1/(1+e^{E/kT})$. 
 Although we do not have good reason to do it, 
the result is fantastic. In figure \ref{fig:PES} we record the result for possible use in the future.  

\subsection{Evolution along two embeddings}
\begin{figure}[ht!]
\centering
   {\includegraphics[width=6cm]{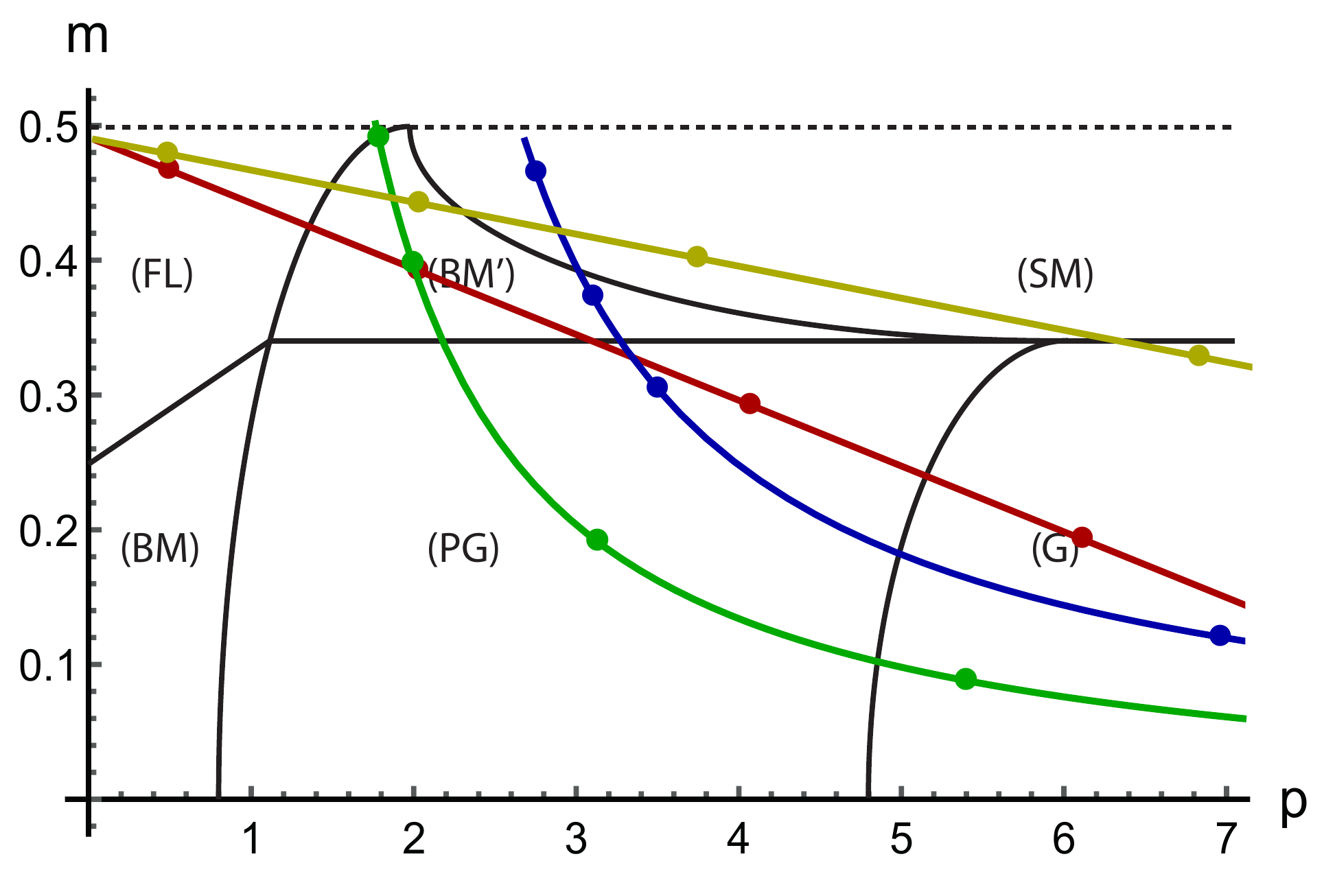}  }
         \caption{          
           An embedding defines a path from gapless to gapped phase in holographic model: 
         $\alpha=1/40$ (1/20)  for yellow (red) line in linear embedding.  
         $\beta=$8/9 (4/3) for blue (green) curve in  hyperbolic embedding. 
     }   \label{fig:PhD2}
\end{figure}

 Here we give four embeddings corresponding to the four colored lines in Figure \ref{fig:PhD2} 
 and corresponding spectral functions using the symmetrized embedding which was used in the early stage of the work. 
\begin{figure}[ht!]
\centering
      \subfigure[]
   {\includegraphics[width=3.7cm]{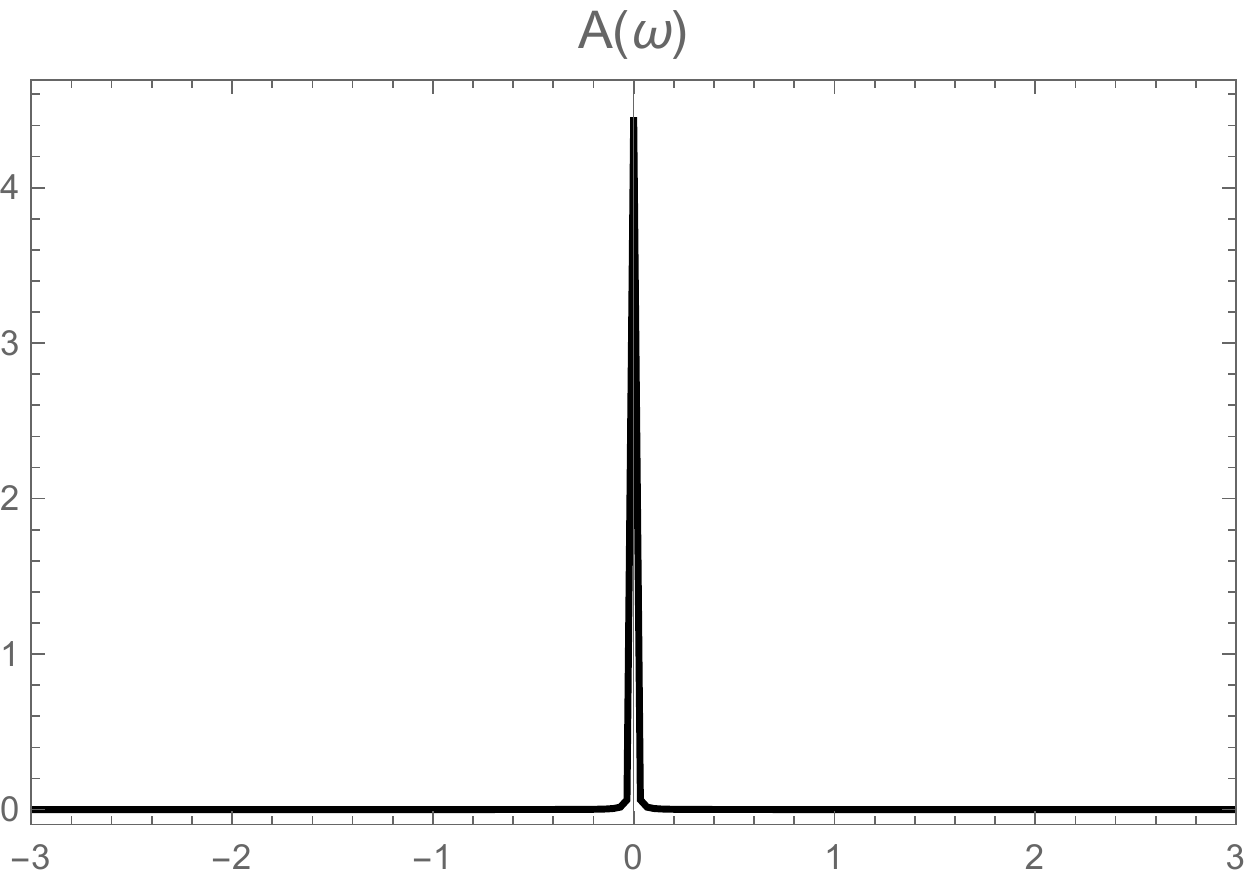}  }
       \subfigure[ ]
   {\includegraphics[width=3.7cm]{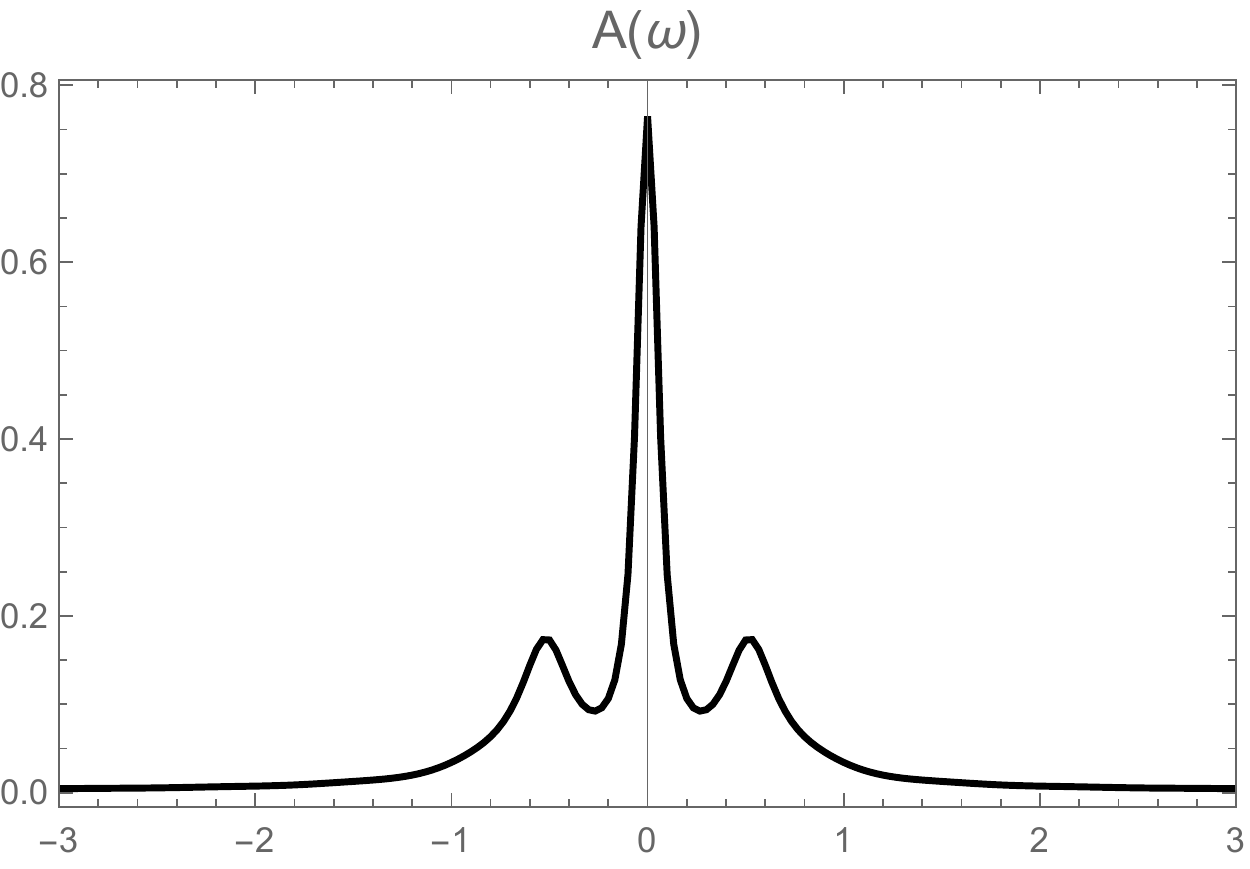}  }
         \subfigure[]
   {\includegraphics[width=3.7cm ]{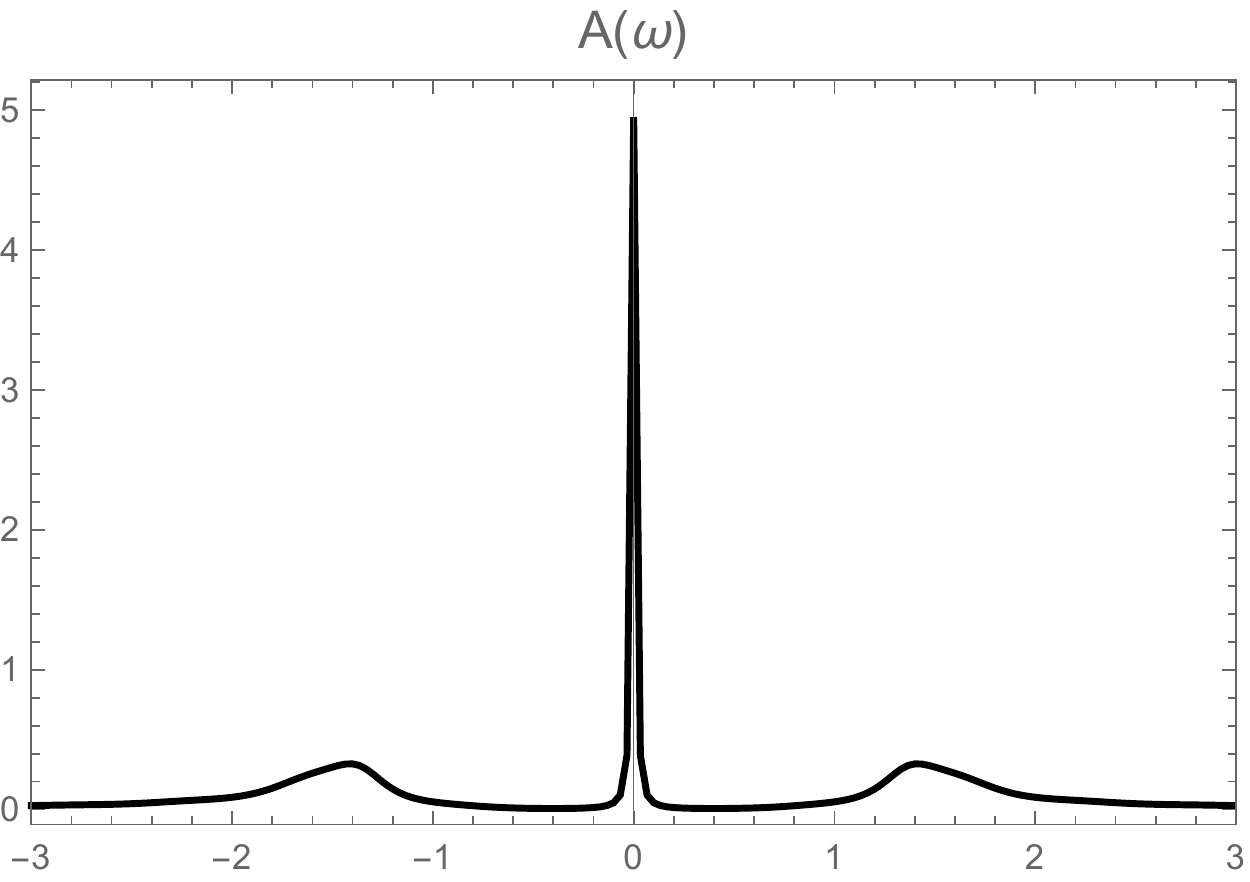}  }
         \subfigure[ ]
   {\includegraphics[width=3.7cm]{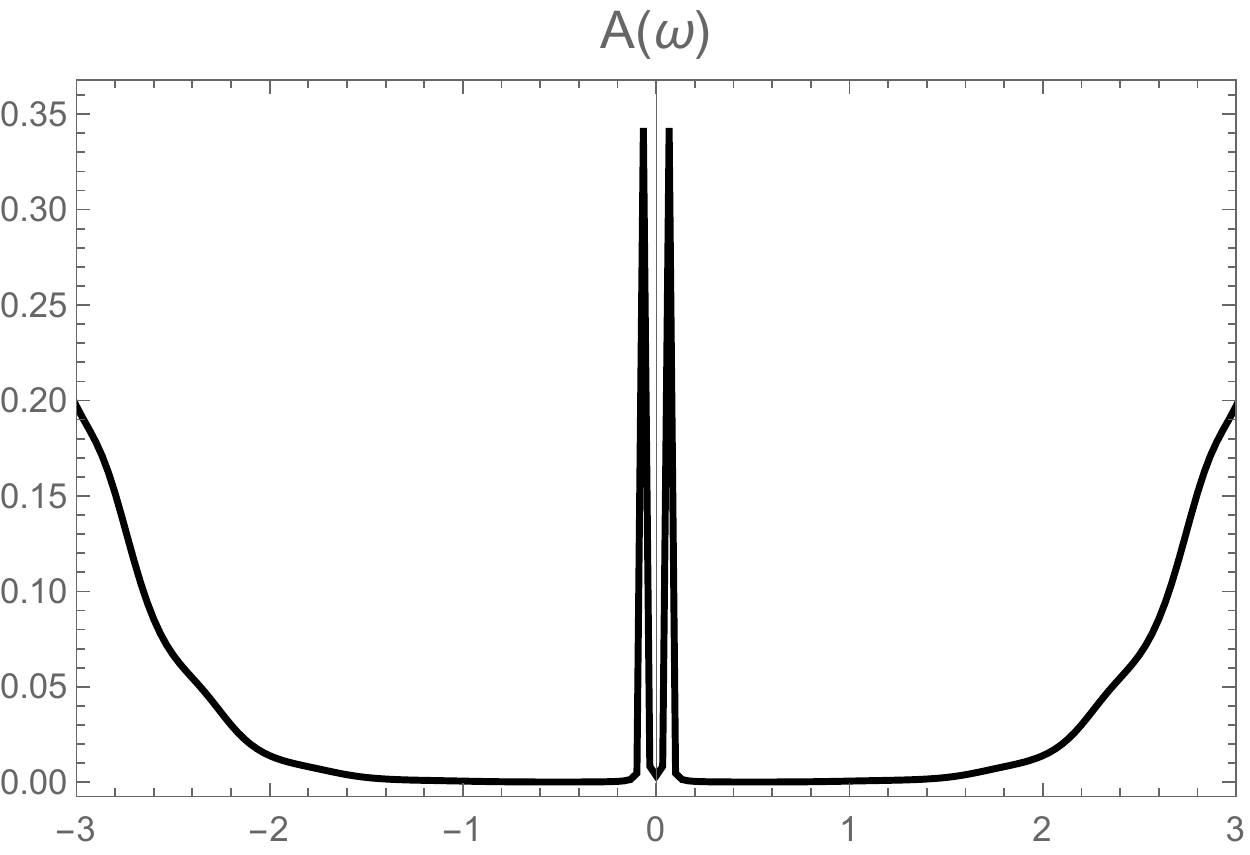}  }
    \subfigure[]
   {\includegraphics[width=3.7cm]{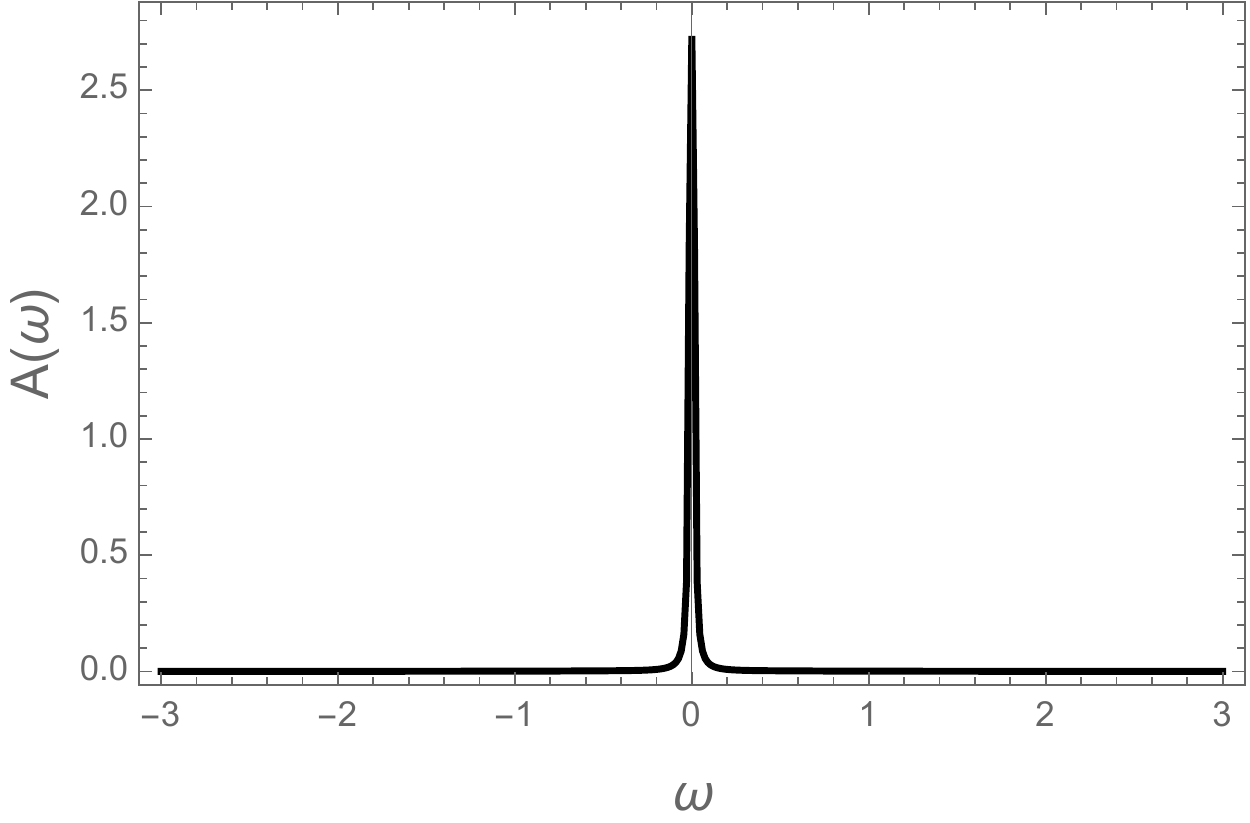}  }
       \subfigure[ ]
   {\includegraphics[width=3.7cm]{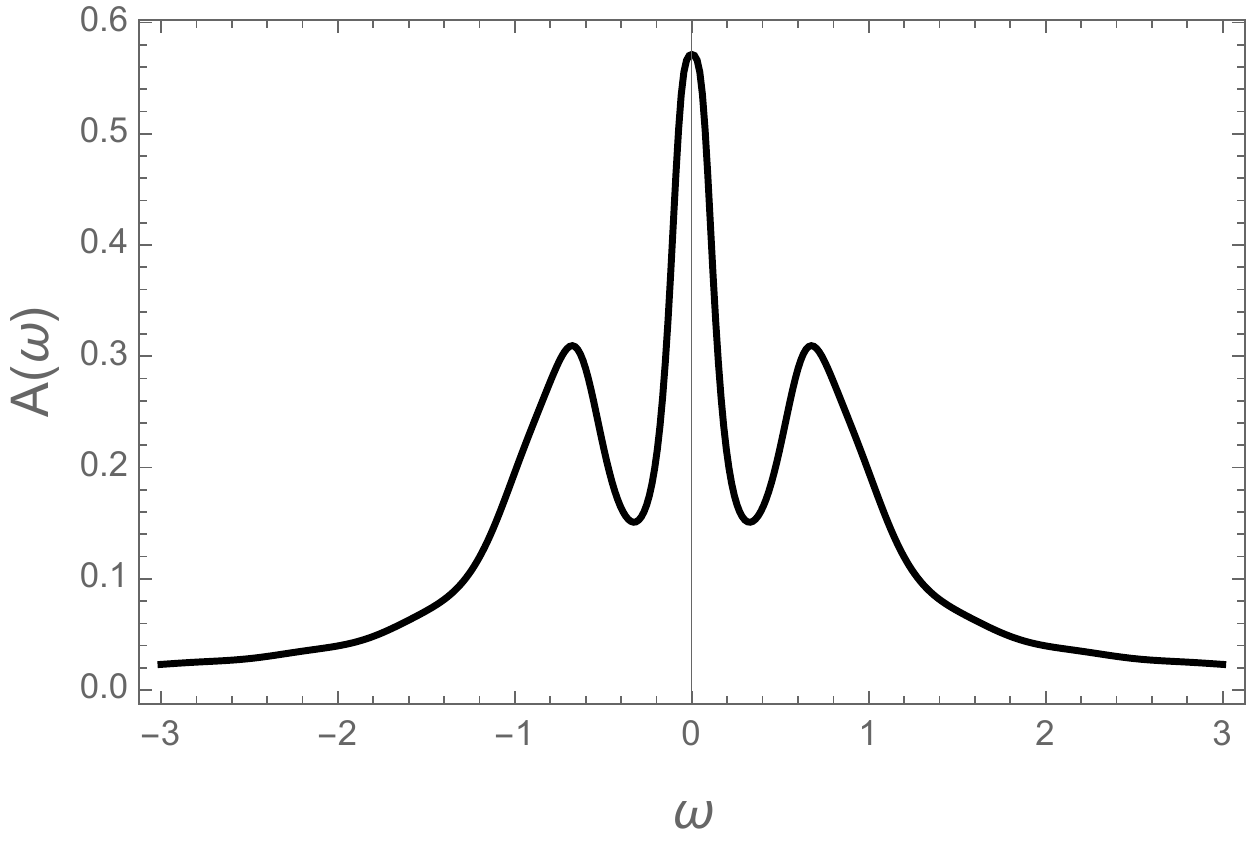}  }
         \subfigure[]
   {\includegraphics[width=3.7cm]{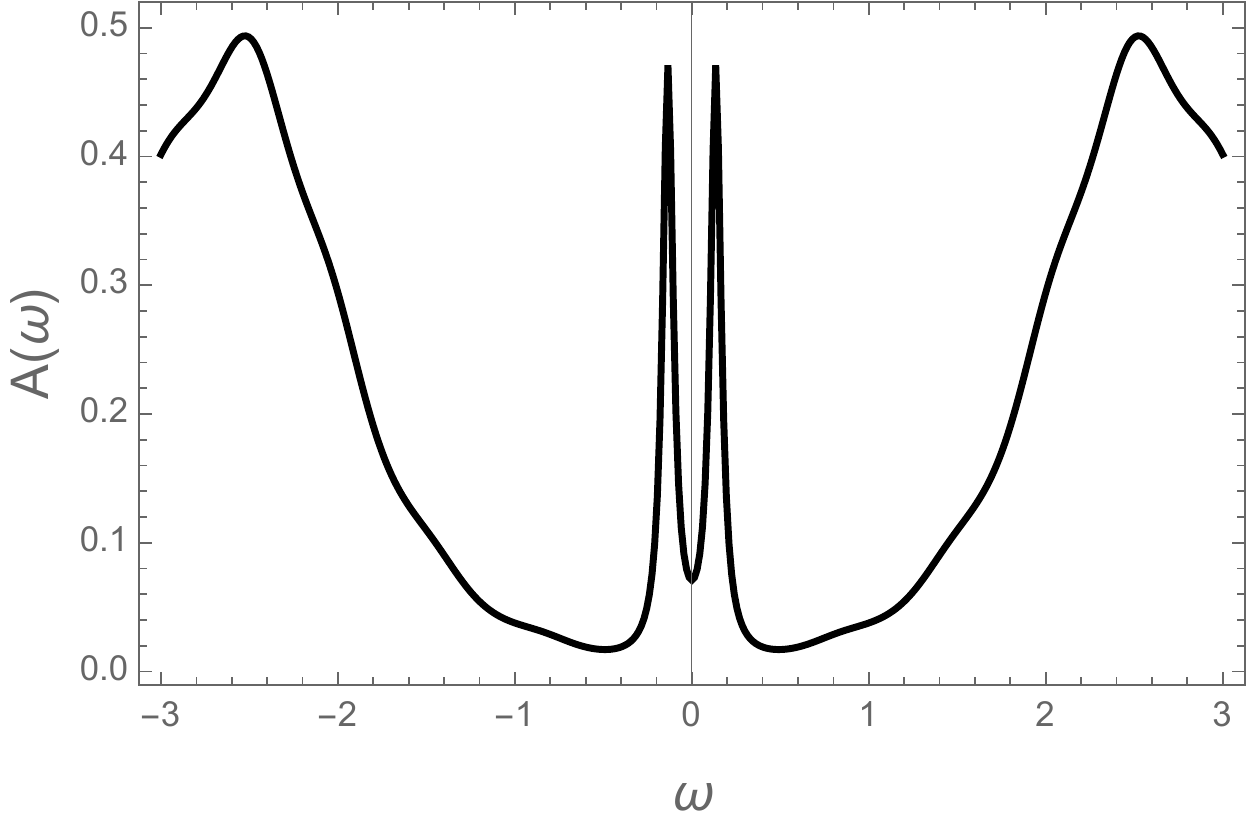}  }
         \subfigure[ ]
   {\includegraphics[width=3.7cm]{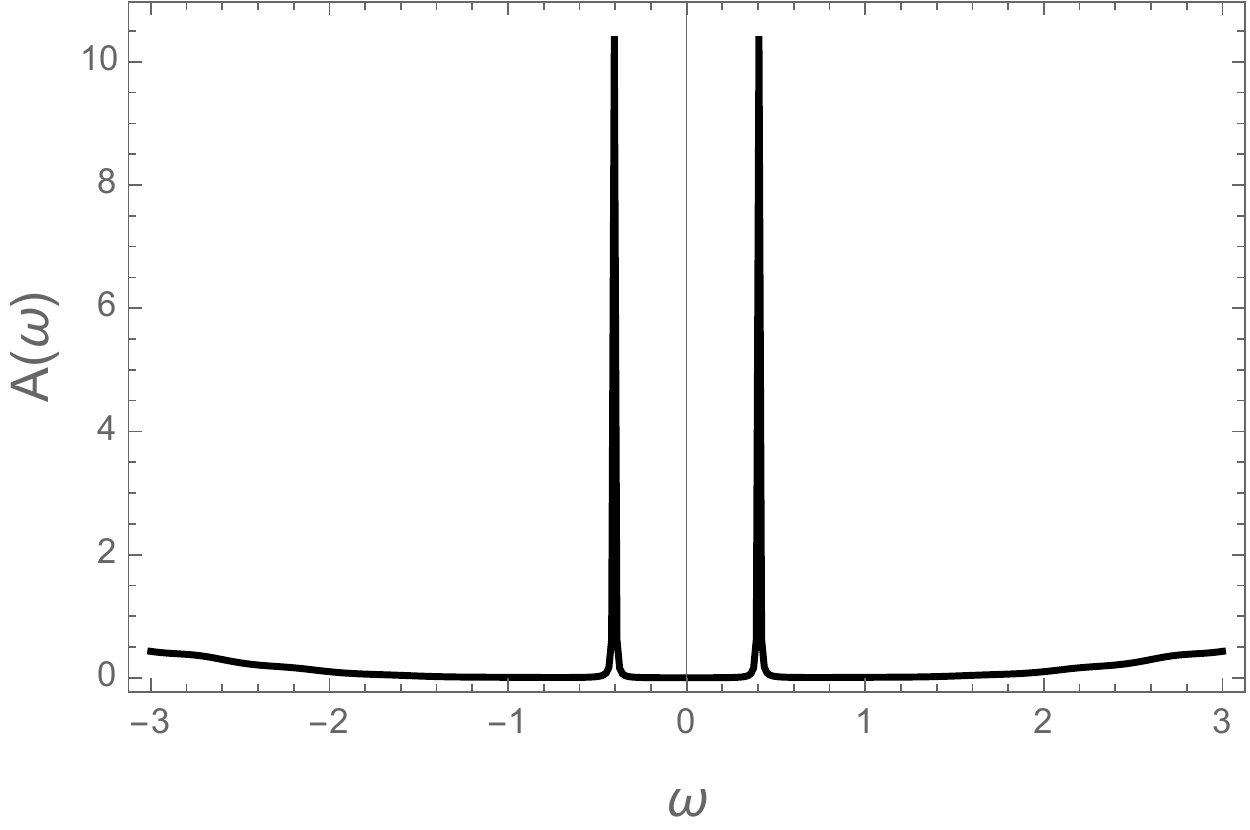}  }
   %
         \caption{ $U$-evolutions   for   linear Embedding.
         (a)-(d) are at marked point of embedding diagram for upper (Yellow) line, (e)-(h) for   for lower (Red) line.        } \label{fig:emb1}
\end{figure}
\begin{figure}[ht!]
\centering 
       \subfigure[]
   {\includegraphics[width=3.7cm]{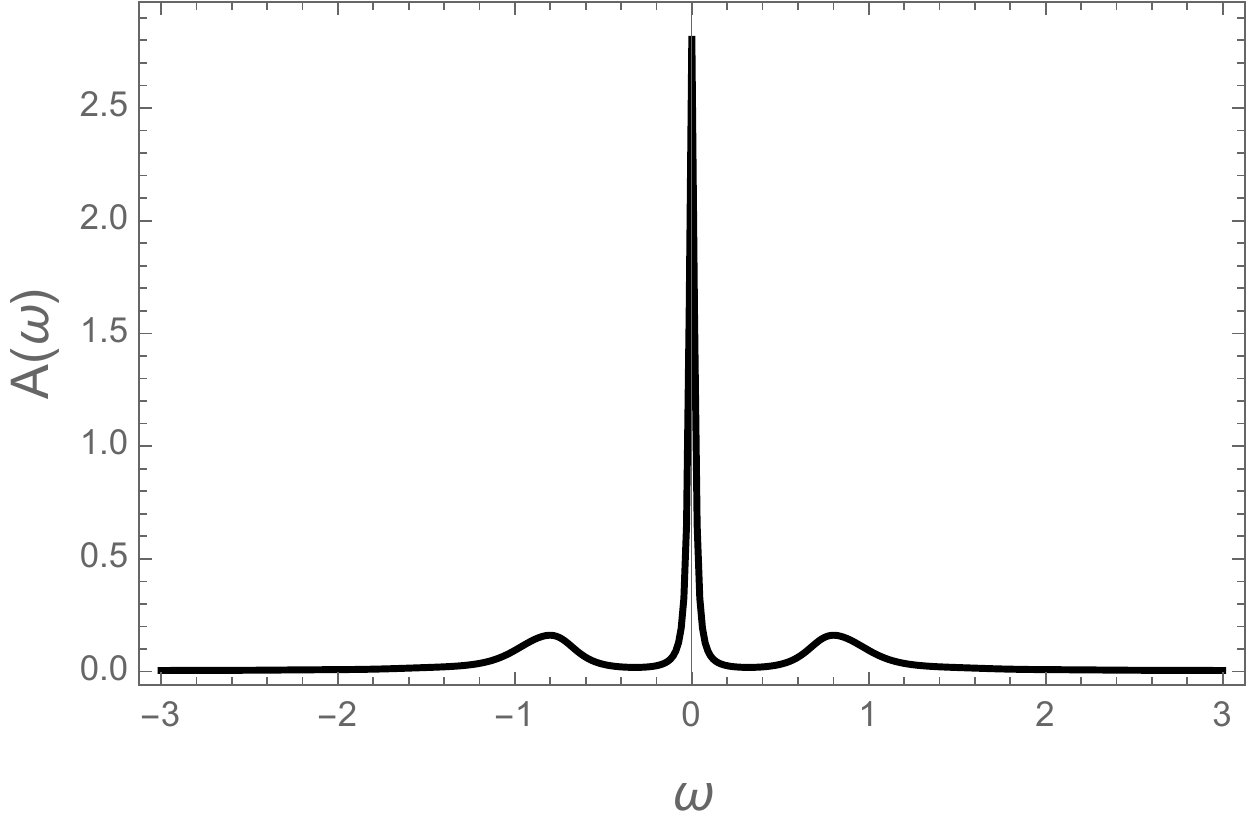}  }
       \subfigure[ ]
   {\includegraphics[width=3.7cm]{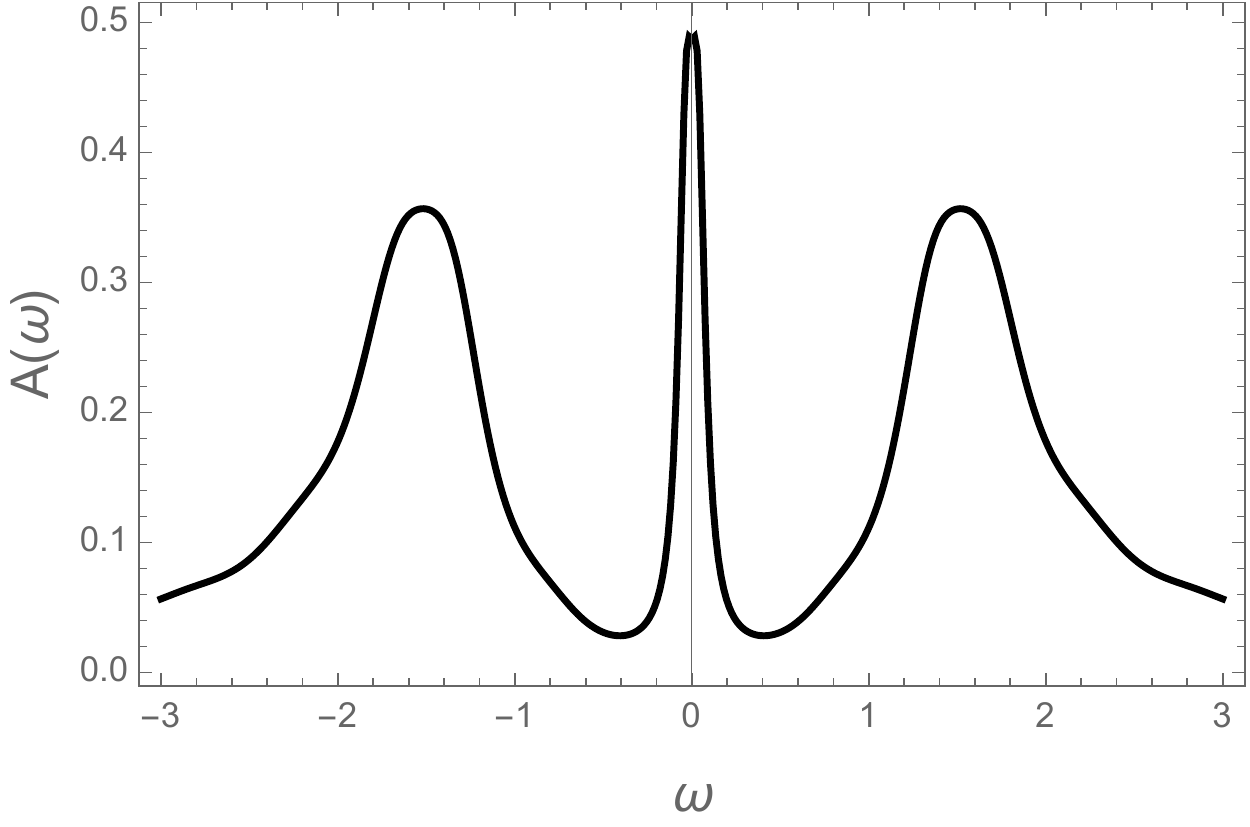}  }
         \subfigure[]
   {\includegraphics[width=3.7cm]{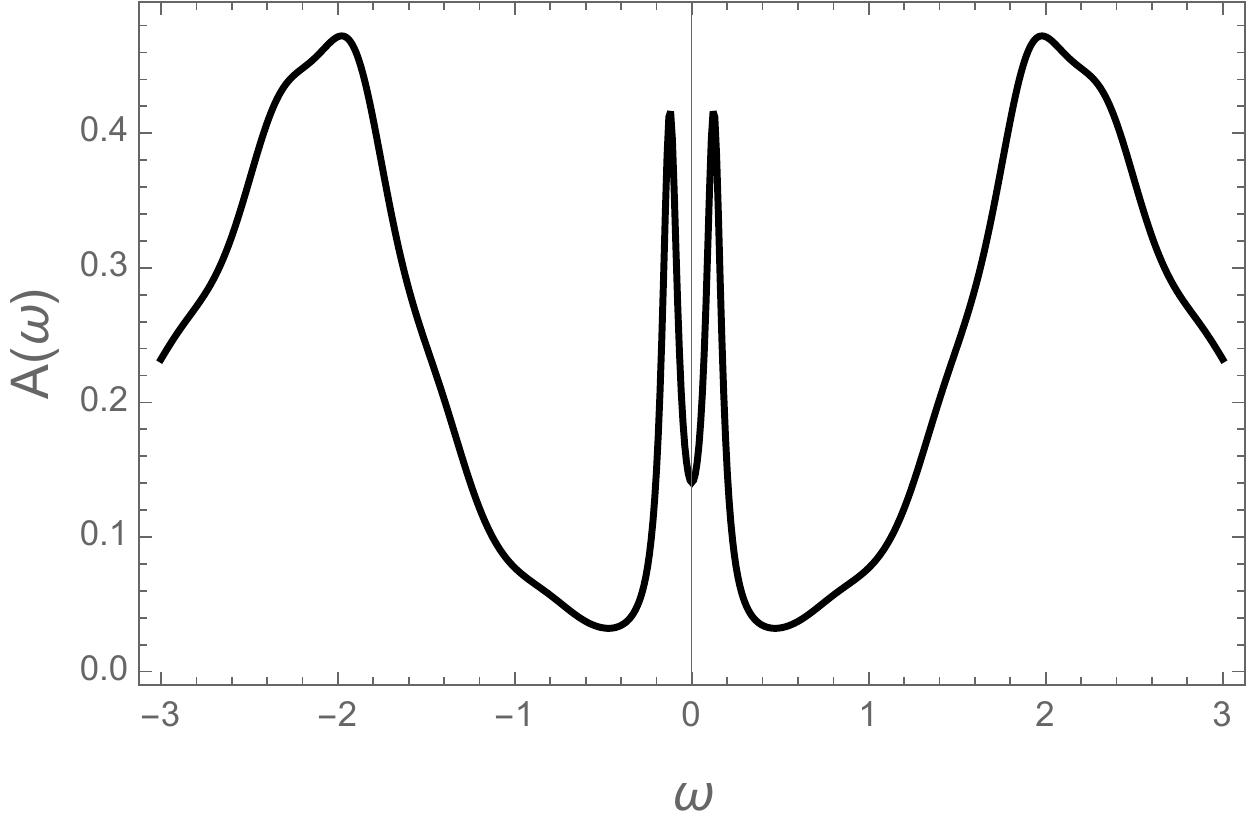}  }
         \subfigure[ ]
   {\includegraphics[width=3.7cm]{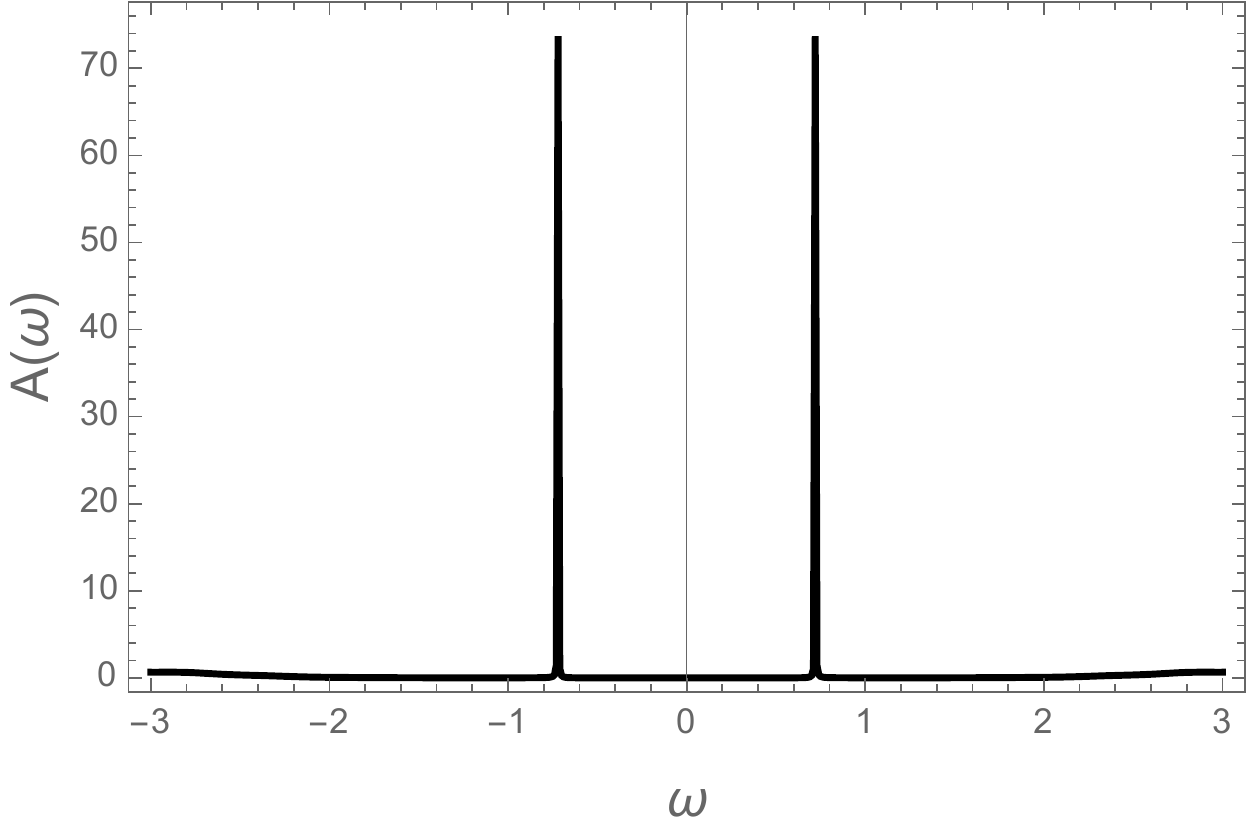}  }
    \subfigure[]
   {\includegraphics[width=3.7cm]{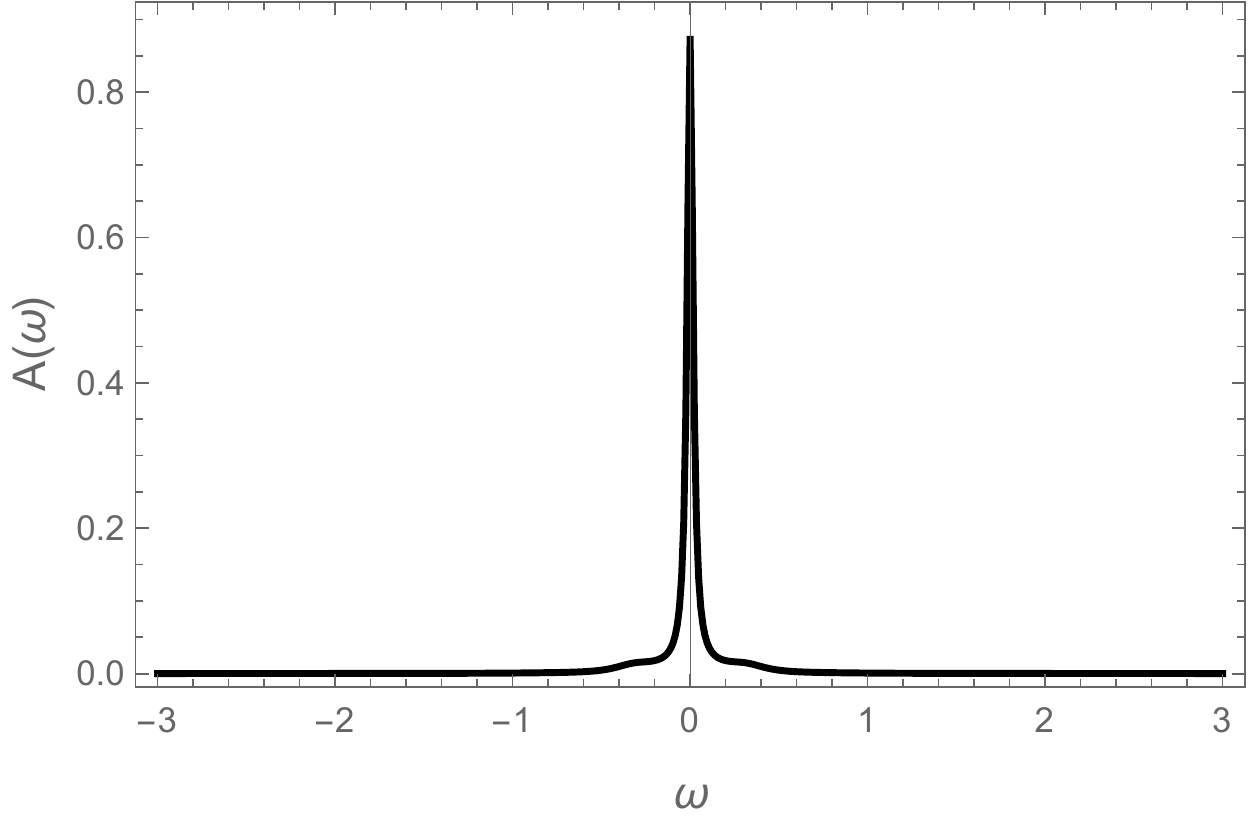}  }
       \subfigure[ ]
   {\includegraphics[width=3.7cm]{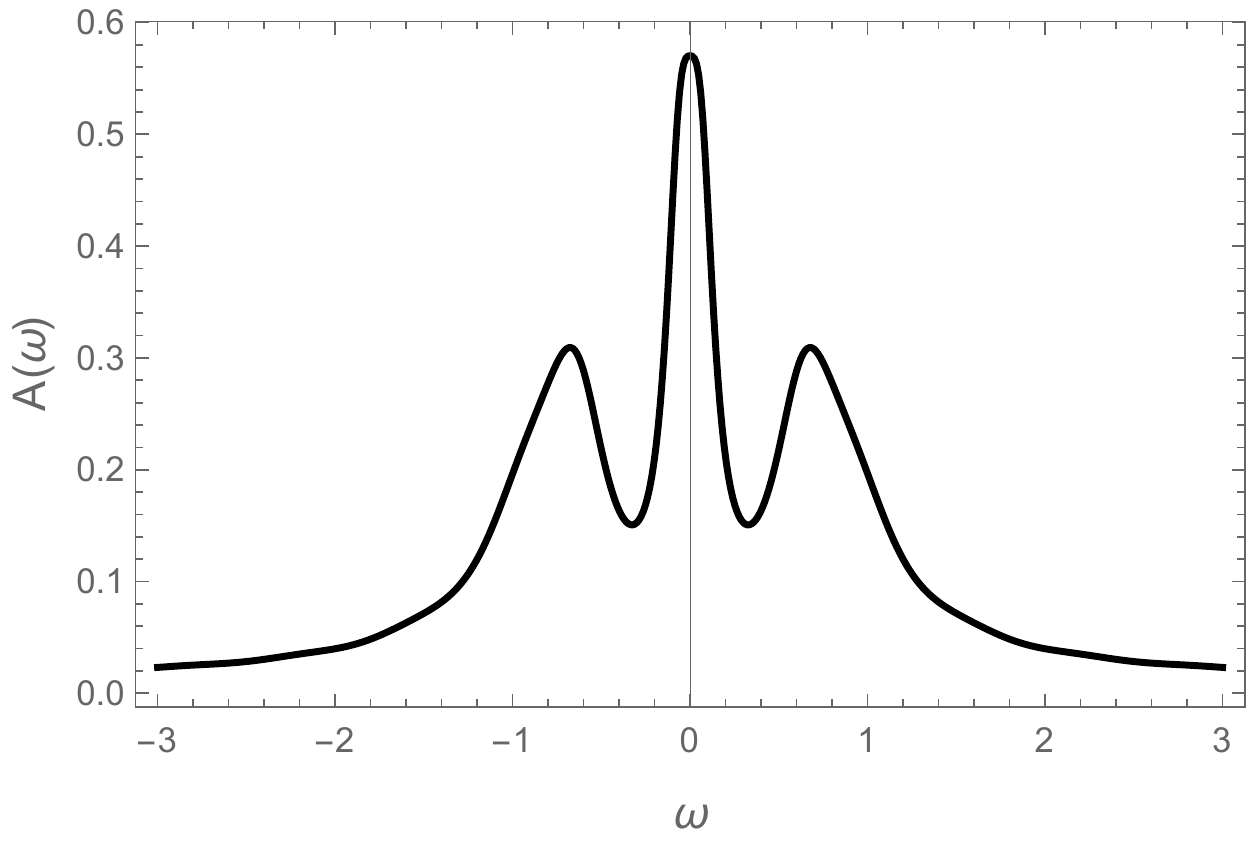}  }
         \subfigure[]
   {\includegraphics[width=3.7cm]{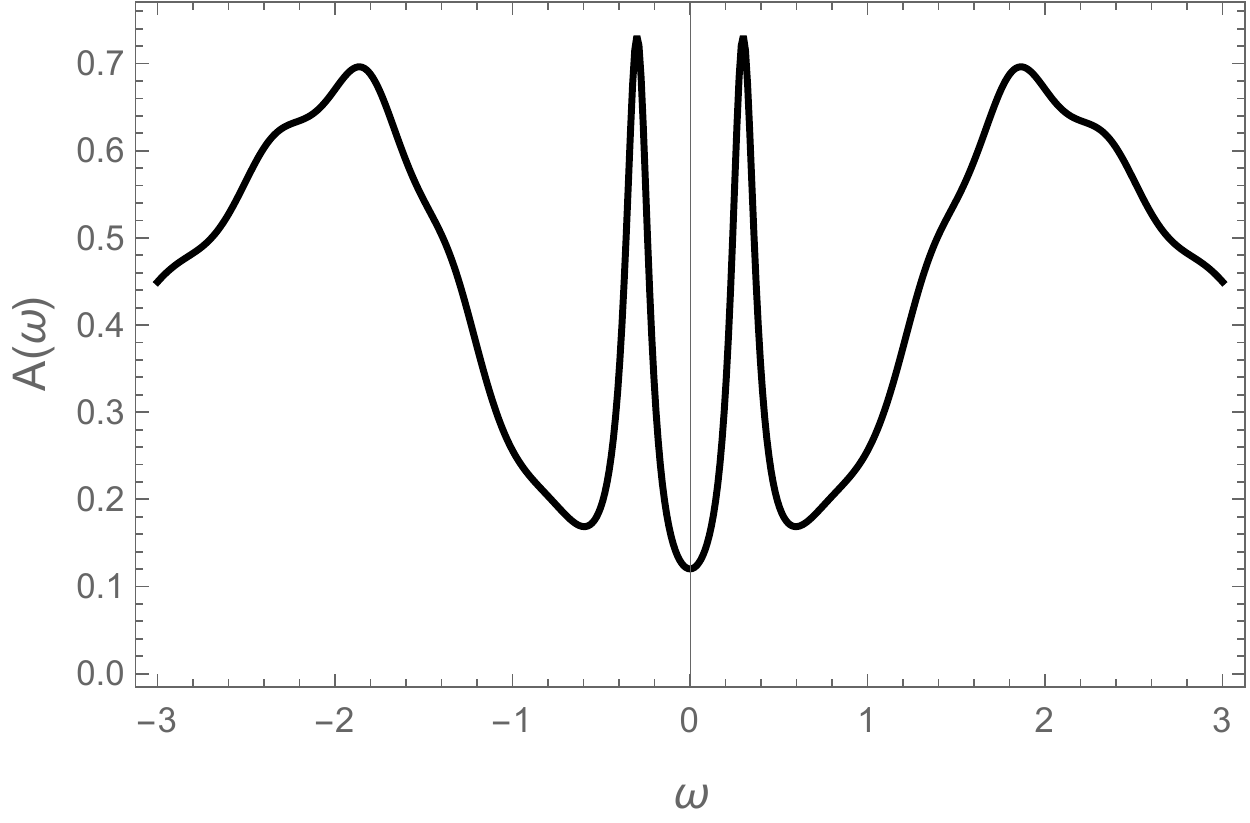}  }
         \subfigure[ ]
   {\includegraphics[width=3.7cm]{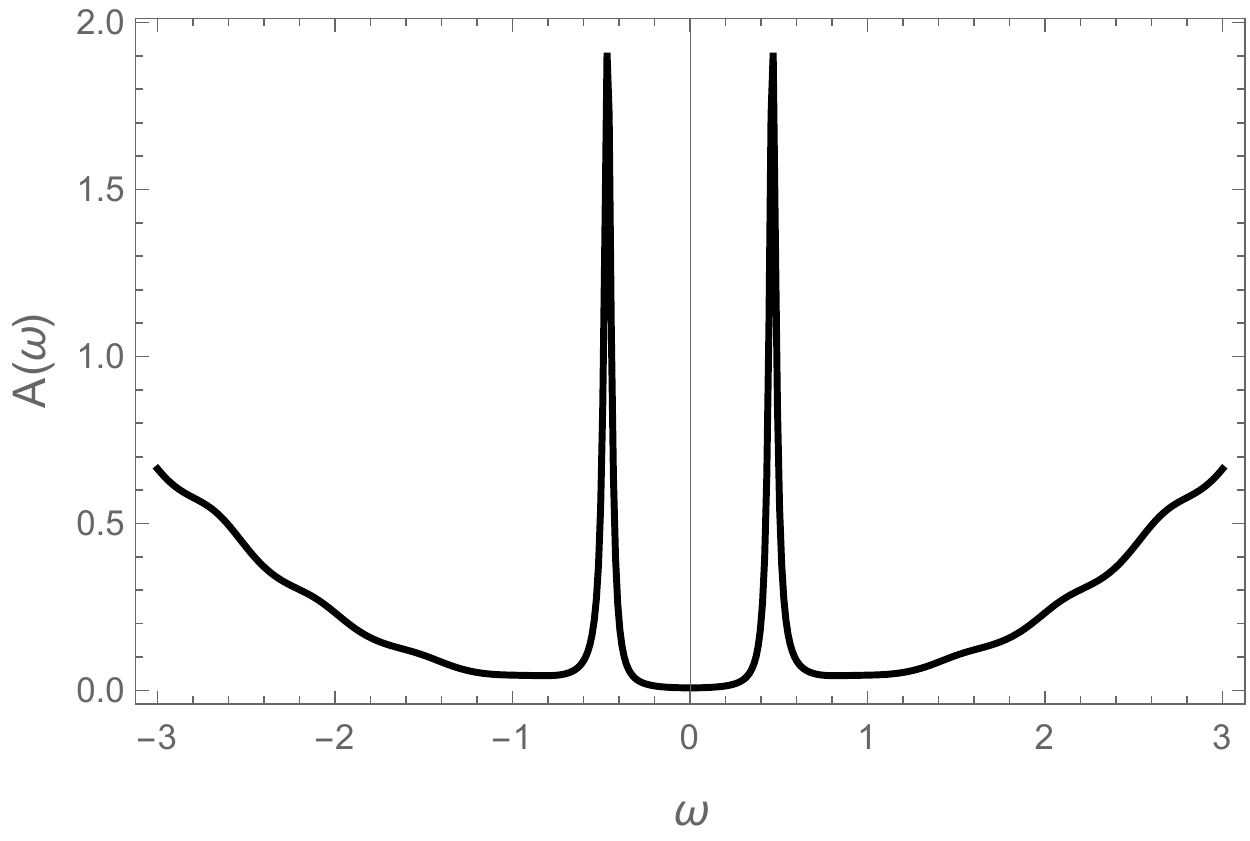}  } 
                   \caption{ $U$-evolutions   for Hyperbolic Embedding ($V\neq 0$). 
         (a)-(d) for upper (Blue) curve, (e)-(h) for  lower (Green) curve,    
         Spectral functions at $k_c$ for $V \neq 0$ (Green Curve).     } \label{fig:emb2}
\end{figure}
 
\vfill
    
\acknowledgments
 We thank   Ara Go, Myeong-Joon Han,  Ki-Seok Kim and  Hunpyo Lee   for useful discussions. This  work is supported by Mid-career Researcher Program through the National Research Foundation of Korea grant No. NRF-2016R1A2B3007687.  YS is  supported  by Basic Science Research Program through NRF grant No. NRF-2016R1D1A1B03931443.

\bibliographystyle{JHEP}
 \bibliography{Refs_spectrum.bib}
\end{document}